\documentclass[10pt,aps,prd,twocolumn,superscriptaddress,nofootinbib,floatfix,raggedbottom]{revtex4-2}

\usepackage[T1]{fontenc}
\usepackage[english]{babel}

\usepackage{amsmath,amssymb}
\usepackage{bm}
\usepackage{bbm}
\usepackage{braket}
\usepackage{cancel}
\usepackage{physics}
\usepackage{extpfeil}
\usepackage{hepnames}
\usepackage{csquotes}
\usepackage{xparse}
\usepackage{array}
\usepackage{multirow}
\usepackage{makecell}

\usepackage[table,xcdraw]{xcolor}
\definecolor{darkgreen}{rgb}{0.0, 0.5, 0.0}
\definecolor{blue}{rgb}{0.0, 0.0, 1.0}
\usepackage[most]{tcolorbox}
\newtcbox{\redbox}{on line, box align=base, colframe=red, colback=white, arc=3pt}
\newtcbox{\bluebox}{on line, box align=base, colframe=blue, colback=white, arc=3pt}

\usepackage{hyperref}
\hypersetup{
    colorlinks = true,
    citecolor = darkgreen,
    linkcolor = blue,
    urlcolor = blue,
}
\usepackage{cleveref}

\usepackage{graphicx}
\graphicspath{{images/}{../images/}}
\usepackage{adjustbox}

\usepackage{subfiles}
\usepackage{comment}

\begin{document}

\title{Coherent State Path Integral Reveals Unexpected Vacuum Structure in Thermal Field Theory}

\author{Rens Roosenstein}
\email{rens.roosenstein@gmail.com}
\affiliation{Department of Physics, University of Cape Town, Private Bag X3, Rondebosch 7701, South Africa}
\affiliation{Institute for Theoretical Physics, University of Amsterdam, PO Box 94485, 1090 GL Amsterdam, The Netherlands}
\affiliation{Nikhef, Theory Group, Science Park 105, 1098 XG, Amsterdam, The Netherlands}

\author{Maximilian Attems}
\email{m.attems@uva.nl}
\affiliation{Institute for Theoretical Physics, University of Amsterdam, PO Box 94485, 1090 GL Amsterdam, The Netherlands}
\affiliation{Nikhef, Theory Group, Science Park 105, 1098 XG, Amsterdam, The Netherlands}

\author{W.\ A.\ Horowitz}
\email{wa.horowitz@uct.ac.za}
\affiliation{Department of Physics, University of Cape Town, Private Bag X3, Rondebosch 7701, South Africa}
\affiliation{Department of Physics, New Mexico State University, Las Cruces, New Mexico, 88003, USA}

\date{\today}

\begin{abstract}
We construct the path integral formulation of the partition function for a free scalar thermal field theory using coherent states, first in the ladder operator basis and then in the field operator basis. In so doing, we provide for the first time a mapping in quantum field theory between the field-basis coherent states and ladder-basis coherent states. Using either basis, one finds terms missed in the usual path integral derivation, which we identify as the vacuum energy contribution to the partition function. We then extend the field-basis coherent state method to the interacting $\phi^4$ theory. In addition to the vacuum energy contribution, one finds a coupling of a vacuum expectation value to the mass term that is absent in the existing literature.
\end{abstract}

\maketitle

\section{Introduction}\label{Introduction}
Thermal Field Theory (TFT) extends Quantum Field Theory (QFT) to systems at finite temperature and density with wide-ranging applications; see the lectures and reviews \cite{Landsman:1986uw,Smilga:1996cm,Das:1997gg,Carrington:1997sq,LeBellac:2000,Thoma:2000dc,KapustaGale,Rebhan:2001wt,Arnold:2007pg,Coleman2015,LaineVuorinen,Rammer:2007zz,Zagoskin2014,Strickland:2019pdt,Strickland:2019tnd,Ghiglieri:2020dpq,Mustafa2022,Salvio:2024upo}. 
TFTs provide insight into understanding systems ranging from the very large to the very small --- e.g.\ cosmological phase transitions \cite{Linde:1978px,Linde:1981zj,Kolb:1990vq,Cohen:1991iu,Yamaguchi:1994yt,Croon:2020cgk}, thermal particle production in the early universe and in the Quark-Gluon Plasma (QGP) \cite{Braaten:1989mz,Rischke:2003mt,Giudice:2003jh,Turbide:2005fk,Anisimov:2010gy,vanHees:2011vb,Gorda:2022fci} produced in high-energy heavy-ion collisions at facilities such as the Relativistic Heavy Ion Collider (RHIC) at Brookhaven National Lab (BNL) \cite{Gyulassy:2004zy,PHOBOS:2004zne,BRAHMS:2004adc,PHENIX2005,STAR2005} and the Large Hadron Collider (LHC) at CERN \cite{ALICE2010,ATLAS2010,CMS:2011iwn,CMS2012,LHCb:2015coe,Dean:2021rlo}. Moreover, TFT plays a central role in modeling the thermodynamical properties of neutron stars \cite{Friman:1979ecl,Alford:2008xm,Schmitt:2010pn,Gorda:2022lsk,Moore:2023glb,Gorda:2023mkk} and their mergers \cite{Most:2018eaw,Fields:2023bhs,Vuorinen:2024qws}, and in simulations of lattice Quantum Chromo-Dynamics (QCD) \cite{Aoki:2006br,Aoki:2009sc,Borsanyi:2010bp,Rothkopf:2011db}, essential for the calculation of the QCD Equation of State (EoS) \cite{Cheng:2007jq,Borsanyi:2011zzb,Borsanyi:2013bia}, the phase diagram of QCD \cite{Soltz:2015ula,Busza:2018rrf}, QCD susceptibilities \cite{Bazavov:2013uja,Datta:2016ukp}, and the properties of quarkonia \cite{Aarts:2007pk,Aarts:2010ek,Aarts:2015tyj} and heavy quark diffusion constants \cite{Brambilla:2020siz,Altenkort:2023oms} in a thermal bath, to name a few. A recent application of TFT is to systems of finite-size \cite{Svaiter:2004ad,Ghisoiu:2010fga,Fister:2015eca}, which are relevant to examine the Casimir effect at finite temperature and to compute the properties of the QGP in small systems \cite{Karabali:2018ael,Kitazawa:2019otp,Mogliacci:2018oea,Horowitz:2021dmr}.

This paper is the result of a careful rederivation of the path integral representation of the partition function in TFT, in which we include an explanation of all logical steps and provide detailed proofs of all intermediate results. (We do not consider the real-time formalism in this treatment \cite{Schwinger:1960qe,Keldysh:1964ud,Chou:1984es,Kobes:1990kr}.) A major concern in our investigation arose from the observation that the standard treatments of the transition from free to interacting theories present this step in a condensed manner, simply replacing the free theory Euclidean action in the path integral with the interacting theory Euclidean action. Moreover, in the usual scattering treatment of vacuum QFT there are a number of subtleties, e.g., the picture the field operators reside in, the asymptotic states of the theory, the renormalization and LSZ reduction \cite{Itzykson:1980rh,PeskinSchroeder1995,weinberg1995quantum,Coleman:2011xi,Schwartz:2014sze}, which demand consideration.

Central to our derivation of the path integral is the use of coherent states. The standard treatments in TFT derive the path integral representation of the partition function by inserting decompositions of unity $\mathbbm{1} \equiv \int d\phi \ketbra{\phi}{\phi}$ and $\mathbbm{1} \equiv \int d\pi \ketbra{\pi}{\pi}$ using eigenstates of the field operator and its canonically conjugate momentum, $\hat{\phi}\ket{\phi} = \phi\ket{\phi}$, $\hat{\pi}\ket{\pi} = \pi\ket{\pi}$. While the use of these unities makes the derivation of the path integral straightforward, it is unclear how these eigenkets and, especially, the integrals over them, are defined. In our derivation, we rather use well-defined coherent states. Our final result is an explicit expression for the path integral as a well-defined finite product of integrals over real numbers.

Of special mention in our derivation is the connection between coherent states in the field basis with coherent states in the ladder basis; deferring the details for later,
\begin{align}
   \ket{\boldsymbol{\alpha}(\tau,t)} &\sim e^{i\int d^n x \, \phi(\tau,\mathbf{x}) \hat \pi(t,\mathbf{x}) - \pi(\tau, \mathbf{x}) \hat\phi(t,\mathbf{x})}\ket{0}\nonumber\\
   & \sim e^{\int\frac{d^np}{(2\pi)^n2 E_{\mathbf{p}}}\left(\alpha(\tau,\mathbf{p})\hat a^\dagger(t,\mathbf{p}) - \alpha^*(\tau,\mathbf{p})\hat a(t,\mathbf{p})\right)}\ket{0},
\end{align}
where
\begin{align}
   \hat{a}(t,\mathbf{p}) \ket{\boldsymbol{\alpha}(\tau,t)} = \alpha(\tau, \mathbf{p})\ket{\boldsymbol{\alpha}(\tau,t)}.
\end{align}
We are unaware of any other reference that shows such a connection.

As a result of our careful derivation of the path integral formulation of the partition function, we see the emergence of the vacuum energy contribution to the Euclidean action in the path integral of the free theory. This vacuum energy contribution appears naturally when writing the Hamiltonian in terms of ladder operators. We then show how the path integral formulation of the partition function is identical, including the vacuum energy contribution to the Euclidean action, when the Hamiltonian is written in terms of the field operator and its canonically conjugate momentum operator. In the field operator basis, the Hamiltonian is simple and does not include any vacuum energy contribution. Rather, the vacuum energy contribution arises from the path integral measure in this basis. We then extend the use of the coherent states in the field-operator basis to the interacting theory case. (We are unaware of an easy way to derive the path integral representation of the partition function for the interacting theory when the Hamiltonian is expressed in terms of ladder operators.) For the interacting theory, we find in addition to the vacuum energy a novel contribution to the Euclidean action, a vacuum bubble that acts as a mass correction. 

This paper is organized as follows. In \cref{Free Theory Partition Function} we introduce the theoretical framework and fix our conventions. We define the free scalar field theory and present two distinct computations of the free theory path integral partition function. We extend the derivation to the interacting theory in \cref{Interacting Theory Partition Function}. Finally, in \cref{Discussion}, we discuss our main results and provide an outlook.

\section{Free Theory Partition Function}
\label{Free Theory Partition Function}

\subsection{Setting Up the Scalar Field Theory}\label{Setting Up the Scalar Field Theory}
For the sake of simplicity, we start to examine a free massive single component scalar field theory in $n + 1$ dimensions. The weak $\lambda \phi^4$ interactions for this classical real scalar theory will follow later on. Due to the thermal system, the discretization even in equilibrium is non-trivial. The Lagrangian for the considered theory\footnote{We work with the mostly-minus metric in $D=n+1$ Minkowskian spacetime; $\eta^M_{\mu \nu} = \text{diag}(+1,-1,\cdots,-1)$, use natural units throughout this work, $\hbar = c = k_B = 1$, and denote $\mathrm{i} \equiv \sqrt{-1}$.} is
\begin{align}
    \mathcal{L} &= \frac{1}{2}\partial_\mu\phi\partial^\mu\phi - \frac{1}{2}m^2\phi^2,\label{eq:real scalar field lagrangian}\\
    & \phi^*(t,\mathbf{x})=\phi(t,\mathbf{x}).\nonumber
\end{align}
In our scalar field theory, we assume that the vacuum expectation value of the scalar field is zero, $\langle \phi(t,\mathbf{x}) \rangle_0 = 0$, as is typical for scalar theories in thermal equilibrium without spontaneous symmetry breaking.

The Hamiltonian is obtained via a Legendre transform, 
\begin{align} 
    \mathcal{H} &\equiv \pi \dot{\phi} - \mathcal{L}, \quad \pi \equiv \frac{\partial \mathcal{L}}{\partial \dot{\phi}} = \dot{\phi}\nonumber\\
    \therefore \mathcal{H} &= \frac{1}{2} \left( \pi^2 + (\nabla \cdot \phi)^2 + m^2 \phi^2 \right). \nonumber 
\end{align} 

Our derivations take place in a discretized spacetime with periodic boundary conditions imposed over a finite-sized box, $x_j \in [0, L_j]$. Throughout this work, we adopt the notation for the momentum and spacetime discretization indices as given in \cref{tab:discretized_vs_continuous}.
\begin{table}[htbp!]
\centering
\begin{tabular}{l|l|l}
 &  Continuous  &  Discretized   \\ \hline
Spacetime \rule{0pt}{10pt} & $\phi(x) \equiv \phi(t, \mathbf{x})$  &  $\phi_{x_r} \equiv \phi_{\mathbf{x}_{\mathbf{r},a}}$ \\ \hline
 Momentum \rule{0pt}{10pt} & $\Tilde{\phi}(p) \equiv \Tilde{\phi}(\omega, \mathbf{p})$ & $\Tilde{\phi}_{p_k} \equiv \Tilde{\phi}_{\mathbf{p}_{\mathbf{k},m}}$
\end{tabular}
\caption{Continuous vs discretized representations of spacetime and 4-momentum.}
\label{tab:discretized_vs_continuous}
\end{table}

In \cref{tab:discretized_vs_continuous}, time is divided into \( K_0 + 1 \) steps and space into \( K_j + 1 \) steps along each direction \( j \). Due to the periodic discretization of spacetime, the momentum modes are restricted to lie within the first Brillouin zone as described in \cite{MontvayMunster}; i.e. \( -\frac{\pi}{\Delta x_j} \leq p_j < \frac{\pi}{\Delta x_j} \), where \( \Delta x_j \) is the lattice spacing. We thus have that

\begin{align}
    t_a &\equiv t_\text{initial} + a\,\Delta t, 
    \quad \Delta t \equiv \frac{t_\text{final}-t_\text{initial}}{K_0+1}, \nonumber \\
    &\quad a \in \left[0, K_0\right], \label{eq:Time discretization} \\[0.2cm]
    x_{r_j} &\equiv r_j\,\Delta x, 
    \quad \Delta x \equiv \frac{L_j}{K_j+1}, \nonumber \\
    &\quad r_j \in \left[0, K_j\right], \label{eq:Space discretization} \\[0.2cm]
    \omega_m &\equiv m\,\Delta\omega, 
    \quad \Delta\omega \equiv \frac{2\pi}{t_\text{final}-t_\text{initial}}, \nonumber \\
    &\quad m \in 
    \left\{
    \begin{array}{ll}
        \left[-\tfrac{K_0}{2}, \tfrac{K_0}{2}\right] & \text{for even } K_0 \\[0.2cm]
        \left[-\tfrac{K_0 + 1}{2}, \tfrac{K_0 - 1}{2}\right] & \text{for odd } K_0
    \end{array}
    \right. , \label{eq:Frequency discretization} \\[0.2cm]
    p_{k_j} &\equiv k_j\,\Delta p, 
    \quad \Delta p \equiv \frac{2\pi}{L_j}, \nonumber \\
    &\quad k_j \in 
    \left\{
    \begin{array}{ll}
        \left[-\tfrac{K_j}{2}, \tfrac{K_j}{2}\right] & \text{for even } K_j \\[0.2cm]
        \left[-\tfrac{K_j + 1}{2}, \tfrac{K_j -1}{2}\right] & \text{for odd } K_j
    \end{array}
    \right. . \label{eq:Momentum discretization}
\end{align}
The \cref{eq:Time discretization,eq:Space discretization,eq:Frequency discretization,eq:Momentum discretization} lead to the following correspondence between continuous integrals and discrete Riemann sums:
\begin{align} 
    \int \frac{d^n \mathbf{p}}{(2\pi)^n2E_\mathbf{p}} &\longleftrightarrow \Delta p^n \sum_{\mathbf{k}=-\frac{\mathbf{K}}{2}}^\frac{\mathbf{K}}{2} \frac{1}{(2 \pi)^n 2E_\mathbf{k}} \label{eq:discretized momentum integral}\\
    \int d^n \mathbf{x} &\longleftrightarrow \Delta x^n \sum_{\mathbf{r}=\mathbf{0}}^\mathbf{K}.\label{eq:discretized position integral}
\end{align}
One gets
\begin{align}
    \left(\frac{\Delta p}{2\pi}\right)^n &= \prod_{j=1}^n \frac{\Delta p_j}{2\pi} = \prod_{j=1}^n \frac{1}{L_j} = \frac{1}{V},\label{eq:Definition of Volume}
\end{align}
where $V$ is the volume of our spatial $n$ dimensional box.

The discretized formalism sketched in this section leads to the following form of the Kronecker delta function (as proven in \cref{Kronecker Delta Function Proof}),
\begin{align}
    \delta_{\mathbf{k},\mathbf{q}} &= \frac{1}{(K+1)^n} \sum_{\mathbf{r}=\mathbf{0}}^\mathbf{K} e^{\frac{2\pi \mathrm{i}(\mathbf{k} - \mathbf{q})\cdot \mathbf{r}}{K+1} }.\label{eq:Kronecker delta function}
\end{align}
In the continuum limit, the Dirac delta function, as defined in \cite{griffiths2018introduction}, is given by
\begin{align}
    \int_{-\infty}^\infty \frac{d^n \mathbf{p}}{(2\pi)^n} e^{\mathrm{i} \mathbf{p} \cdot (\mathbf{x} - \mathbf{y})} = \delta^{(n)}(\mathbf{x} - \mathbf{y}).\label{eq:Continuum Dirac delta function}
\end{align} 
The discrete analogue of \cref{eq:Continuum Dirac delta function} involving the Kronecker delta reads
\small{
\begin{align}
    \left(\frac{\Delta p}{2\pi}\right)^n \sum_{\mathbf{k}=-\frac{\mathbf{K}}{2}}^{\frac{\mathbf{K}}{2}} e^{\mathrm{i} \mathbf{p_k} \cdot (\mathbf{x_r} - \mathbf{x_s})}&= \prod_{j=1}^n \frac{1}{L_j} \sum_{k_j=-\frac{K_j}{2}}^{\frac{K_j}{2}} e^{\frac{2\pi \mathrm{i} k_j}{L_j} \frac{(r_j - s_j) L_j}{K_j+1}}\nonumber\\
    &= \frac{1}{\Delta x^n}\delta_{\mathbf{r},\mathbf{s}},\label{eq:series representation of Dirac delta}
\end{align}}
with its inverse given by
\begin{align}
    \Delta x^n \sum_{\mathbf{r}=\mathbf{0}}^\mathbf{K} e^{\mathrm{i}(\mathbf{p}_\mathbf{k} - \mathbf{p}_\mathbf{q})\cdot \mathbf{x}_\mathbf{r} } &= \prod_{j=1}^n \frac{L_j}{K_j+1} \sum_{r_j=0}^{K_j}  e^{\frac{2\pi \mathrm{i}(k_j - q_j)}{L_j} \frac{r_j L_j}{K_j+1}}\nonumber\\
    &=\left( \frac{2\pi}{\Delta p} \right)^{n}\delta_{\mathbf{k},\mathbf{q}}.\label{eq:Series representation of inverse Dirac delta function}
\end{align}
Hence
\begin{align}
    \delta^{(n)}\left(\mathbf{x}-\mathbf{y}\right) \longleftrightarrow \frac{1}{\Delta x^n}\delta_{\mathbf{r},\mathbf{s}}.
\end{align}
This sketched discretization leads to the following discretized form of the Lagrangian and Hamiltonian:
\begin{align}
     \mathcal{L} &= \frac{1}{2}\dot{\phi}^2\left(t_{a},\mathbf{x_r}\right) -\frac{1}{2}\left(\nabla_{\mathbf{x_r}}\cdot \phi\left(t_{a},\mathbf{x_r}\right)\right)^2\nonumber\\
     & \quad - \frac{1}{2}m^2\phi^2\left(t_{a},\mathbf{x_r}\right)\label{eq:real scalar field lagrangian discrete}\\
     \pi\left(t_{a},\mathbf{x_r}\right) &\equiv \frac{\partial \mathcal{L}}{\partial \dot{\phi}\left(t_{a},\mathbf{x_r}\right)} = \dot{\phi}\left(t_{a},\mathbf{x_r}\right)\label{eq:Discrete conjugate field}\\
    \therefore H &= \Delta x^n \sum_{\mathbf{r}=\mathbf{0}}^\mathbf{K} \mathcal{H}\nonumber\\
     &= \frac{\Delta x^n}{2} \sum_{\mathbf{r}=\mathbf{0}}^\mathbf{K} \left[\pi^2\left(t_{a},\mathbf{x_r}\right) +\left(\nabla_{\mathbf{x_r}}\cdot \phi\left(t_{a},\mathbf{x_r}\right)\right)^2 \right. \nonumber\\
     & \left. \quad + m^2\phi^2\left(t_{a},\mathbf{x_r}\right)\right].\label{eq:real scalar field hamiltonian discrete}
\end{align}
To simplify notation, we denote the spacetime dependence of the fields as follows:
\begin{align}
    \phi_{\mathbf{x}_{\mathbf{r},a}} \equiv \phi\left(t_{a},\mathbf{x_r}\right).\nonumber
\end{align}
Moreover, we adopt the familiar derivative notation from the continuum limit to express our discrete derivatives in a compact form. Specifically, we define
\begin{align}
    \partial_{t_{a}} \phi_{\mathbf{x}_{\mathbf{r},a}} &\equiv \dot{\phi}_{\mathbf{x_r},t_{a}} \nonumber\\
    &\equiv \frac{\phi_{\mathbf{x}_{\mathbf{r},a}} - \phi_{\mathbf{x}_{\mathbf{r},a-1}}}{\Delta t}, \label{eq:Discrete temporal derivative compact notation}\\
    \nabla_\mathbf{x_r} \cdot \phi_{\mathbf{x}_{\mathbf{r},a}} & \equiv \frac{\phi_{\mathbf{x}_{\mathbf{r},a}} - \phi_{\mathbf{x_{r-1}},a}}{\Delta x} \nonumber\\
    &\equiv \sum_{j=1}^n \frac{\phi_{\mathbf{x}_{\mathbf{r},a}} - \phi_{\mathbf{x_r} - \Delta x\, \hat{e}_j, a}}{\Delta x}.\label{eq:Discrete spatial derivative compact notation}
\end{align}
In \cref{eq:Discrete spatial derivative compact notation}, the notation $\mathbf{r-1}$ is used as a shorthand for notational simplicity and should not be interpreted as a literal multi-index shift; the notation serves to make the expressions more compact. From \cref{eq:real scalar field hamiltonian discrete} we deduce that the dimensions of our discretized scalar fields are
\begin{align}
    \left[\phi_{\mathbf{x}_{\mathbf{r},a}}\right]&=E^{\frac{n-1}{2}}, \quad \left[\pi_{\mathbf{x}_{\mathbf{r},a}}\right]=E^{\frac{n+1}{2}}.\label{eq:Field dimensions}
\end{align}
We now show that the classical equations of motion retain their form in the discretized setting, with all derivatives replaced by finite differences. We start from the discretized action,
\begin{align}
    S =& \Delta t \Delta x^n \sum_{a=1}^{K_0} \sum_{\mathbf{r}=\mathbf{0}}^\mathbf{K} \frac{1}{2} \left[\dot{\phi}_{\mathbf{x}_{\mathbf{r},a}}^2-\left(\nabla_\mathbf{x_r} \cdot \phi_{\mathbf{x}_{\mathbf{r},a}}\right)^2 \right. \nonumber\\
    & \left. \qquad -m^2 \phi_{\mathbf{x}_{\mathbf{r},a}}^2 \right].\label{eq:Discretized action}
\end{align}

Note that we have adjusted the lower limit of the temporal sum according to the discrete temporal derivative to ensure that we never evaluate field configurations outside the domain of our temporal lattice. In contrast, the spatial lattice is defined on a torus. Since each spatial direction is periodic, the indices $-1$ and $K_j + 1$ are identified with $K_j$ and $0$, respectively. As a result, we do not need to adjust the limits of the sum over our spatial lattice sites.

To obtain the discrete equations of motion, we compute the variation of the action with respect to the field values,
\begin{widetext}
\begin{align}
    \delta S &= \frac{\Delta t \Delta x^n}{2} \sum_{a,b=1}^{K_0} \sum_{\mathbf{r},\mathbf{s}=\mathbf{0}}^{\mathbf{K}} \frac{\partial}{\partial \phi_{\mathbf{x}_{\mathbf{r},a}}}\left[\left(\frac{\phi_{\mathbf{x}_{\mathbf{s},b}}-\phi_{\mathbf{x}_{\mathbf{s},b-1}}}{\Delta t}\right)^2 -\left(\frac{\phi_{\mathbf{x}_{\mathbf{s},b}}-\phi_{\mathbf{x}_{\mathbf{s-1},b}}}{\Delta x}\right)^2-m^2 \phi_{\mathbf{x}_{\mathbf{s},b}}^2 \right]\delta \phi_{\mathbf{x}_{\mathbf{r},a}}\nonumber\\
    &= \Delta t \Delta x^n \sum_{a,b=1}^{K_0} \sum_{\mathbf{r},\mathbf{s}=\mathbf{0}}^{\mathbf{K}} \left[\frac{\phi_{\mathbf{x}_{\mathbf{s},b}}\delta_{\mathbf{s},\mathbf{r}}\delta_{b,a}-\phi_{\mathbf{x}_{\mathbf{s},b}}\delta_{\mathbf{s},\mathbf{r}}\delta_{b-1,a}-\phi_{\mathbf{x}_{\mathbf{s},b-1}}\delta_{\mathbf{s},\mathbf{r}}\delta_{b,a}+\phi_{\mathbf{x}_{\mathbf{s},b-1}}\delta_{\mathbf{s},\mathbf{r}}\delta_{b-1,a}}{\Delta t^2}\right.\nonumber\\
    & \left. \qquad -\frac{\phi_{\mathbf{x}_{\mathbf{s},b}}\delta_{\mathbf{s},\mathbf{r}}\delta_{b,a}-\phi_{\mathbf{x}_{\mathbf{s},b}}\delta_{\mathbf{s-1},\mathbf{r}}\delta_{b,a}-\phi_{\mathbf{x}_{\mathbf{s-1},b}}\delta_{\mathbf{s},\mathbf{r}}\delta_{b,a}+\phi_{\mathbf{x}_{\mathbf{s-1},b}}\delta_{\mathbf{s-1},\mathbf{r}}\delta_{b,a}}{\Delta x^2} -m^2 \phi_{\mathbf{x}_{\mathbf{s},b}}\delta_{\mathbf{s},\mathbf{r}}\delta_{b,a} \right]\delta \phi_{\mathbf{x}_{\mathbf{r},a}}\nonumber\\
    &= -\Delta t \Delta x^n \sum_{a=1}^{K_0-1} \sum_{\mathbf{r}=\mathbf{0}}^{\mathbf{K}} \left[\frac{\phi_{\mathbf{x}_{\mathbf{r},a+1}}-2\phi_{\mathbf{x}_{\mathbf{r},a}}+\phi_{\mathbf{x}_{\mathbf{r},a-1}}}{\Delta t^2} -\frac{\phi_{\mathbf{x}_{\mathbf{r+1},a}}-2\phi_{\mathbf{x}_{\mathbf{r},a}}+\phi_{\mathbf{x}_{\mathbf{r-1},a}}}{\Delta x^2} + m^2 \phi_{\mathbf{x}_{\mathbf{r},a}} \right]\delta \phi_{\mathbf{x}_{\mathbf{r},a}}.\label{eq:variation of action}
\end{align}
\end{widetext}
It is significant when applying the variational principle, the variations vanish at the temporal boundaries, i.e., $\delta \phi_{\mathbf{x}_{\mathbf{r},0}} = \delta \phi_{\mathbf{x}_{\mathbf{r},K_0}} = 0$. Since $\delta \phi_{\mathbf{x}_{\mathbf{r},K_0}} = 0$, the upper limit of the sums over $a$ and $b$ in \cref{eq:variation of action} reduces from $K_0$ to $K_0 -1$ from the second to third steps.

Setting $\delta S = 0$ and assuming that all $\delta \phi_{\mathbf{x}_{\mathbf{r},a}}$'s are completely arbitrary and independent, we derive the equations of motion. Using the notation defined in \cref{eq:Discrete temporal derivative compact notation,eq:Discrete spatial derivative compact notation}, we obtain
\begin{align}
   \ddot{\phi}_{\mathbf{x}_{\mathbf{r},a}}-\nabla_\mathbf{x_r}^2 \cdot \phi_{\mathbf{x}_{\mathbf{r},a}}+m^2 \phi_{\mathbf{x}_{\mathbf{r},a}} = 0\label{eq:Classical equations of motion}
\end{align}
One derives the dispersion relation in the discrete limit by performing a Discrete Fourier Transform (DFT) on \cref{eq:Classical equations of motion}. The DFT of our field is defined by
\begin{align}
    \phi_{\mathbf{x}_{\mathbf{r},a}} &\equiv \frac{\Delta \omega}{2\pi}\left( \frac{\Delta p}{2\pi} \right)^n  \sum_{m=-\frac{K_0}{2}}^{\frac{K_0}{2}} \sum_{\mathbf{k}=-\mathbf{K}/2}^{\mathbf{K}/2} e^{\mathrm{i} \mathbf{p_k} \cdot \mathbf{x_r}-\mathrm{i}\omega_{m} t_a} \, \Tilde{\phi}_{\mathbf{p}_{\mathbf{k},m}},\label{eq:Real scalar field DFT}\\
    \pi_{\mathbf{x}_{\mathbf{r},a}} &\equiv \frac{\Delta \omega}{2\pi}\left( \frac{\Delta p}{2\pi} \right)^n \sum_{m=-\frac{K_0}{2}}^{\frac{K_0}{2}} \sum_{\mathbf{k}=-\mathbf{K}/2}^{\mathbf{K}/2} e^{\mathrm{i} \mathbf{p_k} \cdot \mathbf{x_r}-\mathrm{i}\omega_{m} t_a} \, \Tilde{\pi}_{\mathbf{p}_{\mathbf{k},m}},\label{eq:Real conjugate scalar field DFT}
\end{align}
with the inverse DFT
\begin{align}
    \Tilde{\phi}_{\mathbf{p}_{\mathbf{k},m}} &=  \Delta t \Delta x^n \sum_{a=0}^{K_0} \sum_{\mathbf{r}=\mathbf{0}}^\mathbf{K}e^{-\mathrm{i}\mathbf{p_k} \cdot \mathbf{x_r}+\mathrm{i}\omega_{m} t_a}\phi_{\mathbf{x}_{\mathbf{r},a}},\label{eq:Real scalar field inverse DFT}\\
    \Tilde{\pi}_{\mathbf{p}_{\mathbf{k},m}} &= \Delta t \Delta x^n \sum_{a=0}^{K_0} \sum_{\mathbf{r}=\mathbf{0}}^\mathbf{K}e^{-\mathrm{i}\mathbf{p_k} \cdot \mathbf{x_r}+\mathrm{i}\omega_{m} t_a}\pi_{\mathbf{x}_{\mathbf{r},a}}.\label{eq:Real scalar field inverse conjugate DFT}
\end{align}
From the normalizations of \cref{eq:Real scalar field DFT,eq:Real conjugate scalar field DFT}, we deduce that the dimensions of our momentum space fields are
\begin{align}
    \left[ \Tilde{\phi}_{\mathbf{p}_{\mathbf{k},m}}\right]=E^{-\frac{1+n}{2}}, \quad \left[\Tilde{\pi}_{\mathbf{p}_{\mathbf{k},m}}\right]=E^{\frac{1-n}{2}}.
\end{align}
As proven in \cref{Discrete Spatial Derivative Fourier Decomposition}, in Fourier space, the discrete derivatives turn into multiplicative factors, just like in the continuum limit. We can thus substitute $\nabla_\mathbf{x_r}^2 \rightarrow -\mathbf{p}_\mathbf{k}^2$ and $\partial_{t_a}^2 \rightarrow -\omega_{m}^2$. Applying the DFT from \cref{eq:Real scalar field DFT} to the classical equations of motion and replacing the derivatives with their corresponding multiplicative factors in Fourier space yields
\begin{align}
    \left(-\omega_{m}^2 + \mathbf{p}_\mathbf{k}^2 + m^2\right) \Tilde{\phi}_{\mathbf{p}_{\mathbf{k},m}} &= 0,\nonumber
\end{align}
which leads to the dispersion relation
\begin{align}
    E_\mathbf{k} &\equiv \omega_{m} = \mathbf{p}_\mathbf{k}^2 + m^2.\label{eq:Dispersion relation}
\end{align}
This \cref{eq:Dispersion relation} is analogous to the continuum limit dispersion relation, and allows us to write down the classical mode expansion in the discretized theory,
\begin{align}
    \phi_{\mathbf{x}_{\mathbf{r},a}} =& \Delta p^n \sum_{\mathbf{k}=-\mathbf{K}/2}^{\mathbf{K}/2} \frac{1}{(2\pi)^n 2E_\mathbf{k}}\Big[a_{\mathbf{p_k}} e^{\mathrm{i} \mathbf{p_k} \cdot \mathbf{x_r} - \mathrm{i} E_\mathbf{k} t_a}\nonumber\\
    & + a^*_{\mathbf{p_k}} e^{-\mathrm{i} \mathbf{p_k} \cdot \mathbf{x_r} + \mathrm{i} E_\mathbf{k} t_a} \Big]\bigg|_{p^0_k = E_\mathbf{k}=\sqrt{\mathbf{p}_\mathbf{k}^2+m^2}}.\label{eq:Classical mode expansion}
\end{align}

We quantize the system by promoting the fields $\phi_{\mathbf{x}_{\mathbf{r},a}}, \pi_{\mathbf{x}_{\mathbf{r},a}}$ to operators $\hat{\phi}_{\mathbf{x}_{\mathbf{r},a}}, \hat{\pi}_{\mathbf{x}_{\mathbf{r},a}}$ and imposing the equal-time canonical commutation relations
\begin{align}
    \left[ \hat{\phi}_{\mathbf{x}_{\mathbf{r},a}}, \hat{\pi}_{\mathbf{x}_{\mathbf{s},a}}\right] &= \frac{\mathrm{i}}{\Delta x^n}\delta_{\mathbf{r},\mathbf{s}}.\label{eq:Canonical equal time commutation relations}
\end{align}
To construct a quantum field theory consistent with \cref{eq:Classical mode expansion,eq:Canonical equal time commutation relations}, we express the field operators in terms of a discrete Fourier mode expansion involving creation and annihilation operators:
\begin{align}
    \hat{\phi}_{\mathbf{x}_{\mathbf{r},a}} =& \Delta p^n \sum_{\mathbf{k}=-\mathbf{K}/2}^{\mathbf{K}/2} \frac{1}{(2\pi)^n 2E_\mathbf{k}}\Big[\hat{a}_{\mathbf{p_k}} e^{\mathrm{i} \mathbf{p_k} \cdot \mathbf{x_r} - \mathrm{i} E_\mathbf{k} t_a}\nonumber\\
    & + \hat{a}^{\dagger}_{\mathbf{p_k}} e^{-\mathrm{i} \mathbf{p_k} \cdot \mathbf{x_r} + \mathrm{i} E_\mathbf{k} t_a} \Big]\bigg|_{p^0_k=E_\mathbf{k}=\sqrt{\mathbf{p}_\mathbf{k}^2+m^2}},\label{eq:Free Heisenberg picture field mode expansion}\\
    \hat{\pi}_{\mathbf{x}_{\mathbf{r},a}} =& \Delta p^n \sum_{\mathbf{k}=-\mathbf{K}/2}^{\mathbf{K}/2} \frac{-\mathrm{i}}{(2\pi)^n 2E_\mathbf{k}} E_\mathbf{k}\Big[\hat{a}_{\mathbf{p_k}} e^{\mathrm{i} \mathbf{p_k} \cdot \mathbf{x_r} - \mathrm{i} E_\mathbf{k} t_a}\nonumber\\
    & - \hat{a}^{\dagger}_{\mathbf{p_k}} e^{-\mathrm{i} \mathbf{p_k} \cdot \mathbf{x_r} + \mathrm{i} E_\mathbf{k} t_a} \Big]\bigg|_{p^0_k=E_\mathbf{k}=\sqrt{\mathbf{p}_\mathbf{k}^2+m^2}}.\label{eq:Free Heisenberg picture conjugate field mode expansion}
\end{align}
From \cref{eq:Free Heisenberg picture field mode expansion}, we deduce that the creation and annihilation operators have mass dimension
\begin{align}
    \left[\hat{a}_{\mathbf{p_k}}\right]&=\left[\hat{a}^\dagger_{\mathbf{p_k}}\right]=E^{\frac{1-n}{2}}.\nonumber
\end{align}
In \cref{Discretized Canonical Commutation relations}, we demonstrate that, for the mode expansions \cref{eq:Free Heisenberg picture field mode expansion,eq:Free Heisenberg picture conjugate field mode expansion} to satisfy the canonical commutation relations \cref{eq:Canonical equal time commutation relations}, the ladder operators must obey the commutation relations
\newpage
\begin{align}
    \left[ \hat{a}_{\mathbf{p_k}}, \hat{a}^\dagger_{\mathbf{p_q}} \right] &= (2\pi)^n 2E_{\mathbf{k}} \frac{1}{\Delta p^n} \delta_{\mathbf{k}, \mathbf{q}}, \label{eq:real scalar ladder operator commutation relation 1}\\
    [\hat{a}_{\mathbf{p_k}}, \hat{a}_{\mathbf{p_q}}] &= [\hat{a}_{\mathbf{p_k}}^\dagger, \hat{a}_{\mathbf{p_q}}^\dagger] = 0.\label{eq:real scalar ladder operator commutation relation 2}
\end{align}

\subsection{Thermal Trace Trotter Formula}
\label{Thermal Trace Trotter Formula}
Before we start with computing the partition function, we show the coherent state relations (proven in \cref{Coherent State Relations}) that will be necessary in order to transition from the trace of the partition function via the operator formalism into a path integral representation:
 \begin{align}
    \ket{\boldsymbol{\alpha}} \equiv& e^{\Delta p^n\sum_{\mathbf{k}=-\mathbf{K}/2}^{\mathbf{K}/2}\frac{1}{(2\pi)^n 2E_\mathbf{k}}\alpha_{\mathbf{p_k}}\hat{a}^\dagger_{\mathbf{p_k}}}\ket{0}, \ \alpha_{\mathbf{p_k}} \in \mathbbm{C}, \label{eq:Coherent state definition}\\
    \hat{a}_{\mathbf{p_k}}\ket{\boldsymbol{\alpha}}=&\alpha_{\mathbf{p_k}}\ket{\boldsymbol{\alpha}},\label{eq:Eigenstate property of coherent state}\\ 
    \braket{\tilde{\boldsymbol{\alpha}}}{\boldsymbol{\alpha}}=&e^{\Delta p^n\sum_{\mathbf{k}=-\mathbf{K}/2}^{\mathbf{K}/2}\frac{1}{(2\pi)^n 2E_\mathbf{k}}\tilde{\alpha}^*_{\mathbf{p_k}}\alpha_{\mathbf{p_k}}},\label{eq:Overcompleteness of coherent state}\\ 
    \mathbbm{1}=&\left(\prod_{\mathbf{k}=-\frac{\mathbf{K}}{2}}^\frac{\mathbf{K}}{2} \int_{-\infty}^\infty \frac{da_{\mathbf{p_k}} db_{\mathbf{p_k}}}{2\pi VE_\mathbf{k}} \right) \nonumber\\
    & e^{\Delta p^n\sum_{\mathbf{k}=-\mathbf{K}/2}^{\mathbf{K}/2}\frac{1}{(2\pi)^n 2E_\mathbf{k}}\alpha^*_{\mathbf{p_k}}\alpha_{\mathbf{p_k}}}\ketbra{\boldsymbol{\alpha}}{\boldsymbol{\alpha}}.\label{eq:coherent state identity element}
\end{align}
(All results related to coherent states are grouped into \cref{Coherent states}.) In \cref{eq:coherent state identity element}, we define  $a_{\mathbf{p_k}}, b_{\mathbf{p_k}} \in \mathbbm{R}$ as the real and imaginary parts of $\alpha_{\mathbf{p_k}}$, respectively,
\begin{align}
    \alpha_{\mathbf{p_k}}=a_{\mathbf{p_k}}+\mathrm{i}b_{\mathbf{p_k}}, \quad \alpha_{\mathbf{p_k}}^*=a_{\mathbf{p_k}}-\mathrm{i}b_{\mathbf{p_k}}.\label{eq:Complex decomposition of alpha}
\end{align} 
The complex decomposition in \cref{eq:Complex decomposition of alpha} ensures that the integration measure in \cref{eq:coherent state identity element} is real and that the integrals are well-defined over the real domain of these decomposed variables, $a_{\mathbf{p_k}}$ and $b_{\mathbf{p_k}}$.

Let us now consider the partition function in the canonical ensemble. Suppose that $\big\{\ket{n}\big\}$ provides a complete set of states for our system, such that $\mathbbm{1} = \sum_n \ketbra{n}{n}$. Then
\begin{align}
    Z &\equiv \Tr\left[e^{-\beta \hat{H}}\right]\label{eq:First principles partition function}\\
    &= \sum_{n}\bra{n}e^{-\beta \hat{H}}\ket{n}.\nonumber
\end{align}
Inserting a complete set of coherent states gives
\begin{align}
    Z =& \sum_{n}\bra{n}\left(\prod_{\mathbf{k}=-\frac{\mathbf{K}}{2}}^\frac{\mathbf{K}}{2} \int_{-\infty}^\infty \frac{da_{\mathbf{p_k}} db_{\mathbf{p_k}}}{2\pi VE_\mathbf{k}}\right) \nonumber\\ 
    & \ e^{-\Delta p^n \sum_{\mathbf{k}=-\mathbf{K}/2}^{\mathbf{K}/2} \frac{1}{(2\pi)^n2E_\mathbf{k}}\alpha_{\mathbf{p_k}}^* \alpha_{\mathbf{p_k}}} \ket{\boldsymbol{\alpha}}\bra{\boldsymbol{\alpha}} e^{-\beta\hat{H}}\ket{n}\nonumber\\
    =& \left(\prod_{\mathbf{k}=-\frac{\mathbf{K}}{2}}^\frac{\mathbf{K}}{2} \int_{-\infty}^\infty \frac{da_{\mathbf{p_k}} db_{\mathbf{p_k}}}{2\pi VE_\mathbf{k}}\right) \nonumber\\ 
    & \ e^{-\Delta p^n \sum_{\mathbf{k}=-\mathbf{K}/2}^{\mathbf{K}/2} \frac{1}{(2\pi)^n2E_\mathbf{k}}\alpha_{\mathbf{p_k}}^* \alpha_{\mathbf{p_k}}} \bra{\boldsymbol{\alpha}} e^{-\beta\hat{H}}\sum_{n} \ket{n} \braket{n}{\boldsymbol{\alpha}} \nonumber\\
    =& \left(\prod_{\mathbf{k}=-\frac{\mathbf{K}}{2}}^\frac{\mathbf{K}}{2} \int_{-\infty}^\infty \frac{da_{\mathbf{p_k}} db_{\mathbf{p_k}}}{2\pi VE_\mathbf{k}}\right) \nonumber\\ 
    & \ e^{-\Delta p^n \sum_{\mathbf{k}=-\mathbf{K}/2}^{\mathbf{K}/2} \frac{1}{(2\pi)^n2E_\mathbf{k}}\alpha_{\mathbf{p_k}}^* \alpha_{\mathbf{p_k}}} \bra{\boldsymbol{\alpha}} e^{-\beta\hat{H}} \ket{\boldsymbol{\alpha}}.\label{eq:Intermediate free real partition function 1}
\end{align}
Applying trotterization \cite{Trotter:1959}, we divide the partition function in \cref{eq:Intermediate free real partition function 1} into $N \rightarrow \infty$ segments and insert \cref{eq:coherent state identity element} in between each segment to find the thermal trace trotter formula
\begin{align}
    Z =& \left(\prod_{\mathbf{k}=-\frac{\mathbf{K}}{2}}^\frac{\mathbf{K}}{2} \int_{-\infty}^\infty \frac{da_{\mathbf{p_k}} db_{\mathbf{p_k}}}{2\pi VE_\mathbf{k}}\right) \ e^{-\Delta p^n \sum_{\mathbf{k}=-\mathbf{K}/2}^{\mathbf{K}/2}\frac{1}{(2\pi)^n2E_\mathbf{k}}\alpha_\mathbf{p_k}^* \alpha_\mathbf{p_k}}\nonumber\\ 
    & \ \bra{\boldsymbol{\alpha}} e^{-\frac{\beta}{N}\hat{H}} \cdots e^{-\frac{\beta}{N}\hat{H}}\ket{\boldsymbol{\alpha}}\nonumber\\
    =& \prod_{i=1}^{N} \left(\prod_{\mathbf{k}=-\frac{\mathbf{K}}{2}}^\frac{\mathbf{K}}{2} \int_{-\infty}^\infty \frac{da_{\mathbf{p}_{\mathbf{k},i}} db_{\mathbf{p}_{\mathbf{k},i}}}{2\pi VE_\mathbf{k}}\right)\nonumber\\ 
    & \ e^{-\Delta p^n\sum_{\mathbf{k}=-\mathbf{K}/2}^{\mathbf{K}/2}\frac{1}{(2\pi)^n 2E_\mathbf{k}}\alpha_{\mathbf{p}_{\mathbf{k},i}}^* \alpha_{\mathbf{p}_{\mathbf{k},i}}}\bra{\boldsymbol{\alpha}} e^{-\frac{\beta}{N}\hat{H}} \ket{\boldsymbol{\alpha}_{N-1}}\nonumber\\
    & \ \bra{\boldsymbol{\alpha}_{N-1}} e^{-\frac{\beta}{N}\hat{H}} \ket{\boldsymbol{\alpha}_{N-2}} \cdots \bra{\boldsymbol{\alpha}_{1}} e^{-\frac{\beta}{N}\hat{H}} \ket{\boldsymbol{\alpha}}.\label{eq:Intermediate free real partition function 2}
\end{align}
We emphasize that the trace has been discretized into segments of length $\frac{\beta}{N} \equiv \Delta\tau$ and is periodic at the boundaries, such that $\ket{\boldsymbol{\alpha}} \equiv \ket{\boldsymbol{\alpha}_N} \equiv \ket{\boldsymbol{\alpha}_0}$. In this trotterization process, our coherent state eigenvalues picked up an additional index $i$, $\alpha_{\mathbf{p}_{\mathbf{k},i}}$.

Consider a single matrix element from \cref{eq:Intermediate free real partition function 2}. By expanding in the infinitesimal parameter $\Delta\tau$ and inserting the relativistic free real Klein-Gordon Hamiltonian, we can derive the path integral. Such a matrix element can be computed in two distinct ways. Comparing the two results reveals a crucial subtlety related to the vacuum energy, an insight that becomes particularly relevant when interactions are introduced in \cref{Interacting Theory Partition Function}.

\subsection{Path Integral from the Number Operator Hamiltonian}\label{Path Integral from the Number Operator Hamiltonian}

\noindent
For the first calculation, we use the coherent state $\ket{\boldsymbol{\alpha}_i}$ from \cref{eq:Coherent state definition} with an additional index $i$, and insert the Klein-Gordon Hamiltonian expressed in terms of creation and annihilation operators.

In \cref{Klein Gordon Hamiltonian}, we derive the Hamiltonian operator in the ladder operator basis using \cref{eq:real scalar field hamiltonian discrete,eq:real scalar ladder operator commutation relation 2,eq:real scalar ladder operator commutation relation 1,eq:Free Heisenberg picture field mode expansion,eq:Canonical equal time commutation relations},
\begin{align}
    \hat{H} &=\Delta p^n \sum_{\mathbf{k}=-\mathbf{K}/2}^{\mathbf{K}/2} \frac{1}{(2\pi)^n 2E_\mathbf{k}}E_\mathbf{k}\left[\hat{a}^\dagger_\mathbf{p_k}\hat{a}_\mathbf{p_k}+ \left(\frac{2\pi}{\Delta p}\right)^n E_\mathbf{k}\right].\label{eq:Free scalar ladder operator Hamiltonian}
\end{align}
We retain the vacuum energy term explicitly in \cref{eq:Free scalar ladder operator Hamiltonian}, as it is precisely this contribution that gives rise to the subtlety we aim to highlight. Substituting \cref{eq:Free scalar ladder operator Hamiltonian} into a single matrix element from \cref{eq:Intermediate free real partition function 2} yields
\begin{align}
    \bra{\boldsymbol{\alpha}_i} e^{-\Delta\tau\hat{H}} & \ket{\boldsymbol{\alpha}_{i-1}} \nonumber\\
    =& \bra{\boldsymbol{\alpha}_i} (1-\Delta\tau\hat{H}+\order{\Delta\tau^2} \ket{\boldsymbol{\alpha}_{i-1}}\nonumber\\
    =& \braket{\boldsymbol{\alpha}_i}{\boldsymbol{\alpha}_{i-1}} e^{-\Delta\tau H\left(\alpha^*_{\mathbf{p}_{\mathbf{k},i}},\alpha_{\mathbf{p}_{\mathbf{k},i-1}}\right)} +\order{\Delta\tau^2}\nonumber\\
    =& e^{\Delta p^n \sum_{\mathbf{k}=-\mathbf{K}/2}^{\mathbf{K}/2} \frac{1}{(2\pi)^n 2E_\mathbf{k}}\alpha^*_{\mathbf{p}_{\mathbf{k},i}}\alpha_{\mathbf{p}_{\mathbf{k},i-1}}}\nonumber\\
    & \ e^{-\Delta\tau H\left(\alpha^*_{\mathbf{p}_{\mathbf{k},i}},\alpha_{\mathbf{p}_{\mathbf{k},i-1}}\right)}+\order{\Delta\tau^2},
\end{align}

where we used the function $H\big(\alpha^*_{\mathbf{p}_{\mathbf{k},i}},\alpha_{\mathbf{p}_{\mathbf{k},i-1}}\big)$, defined by
\begin{align}
    H\big(\alpha^*_{\mathbf{p}_{\mathbf{k},i}},\alpha_{\mathbf{p}_{\mathbf{k},i-1}}\big) \equiv& \Delta p^n \sum_{\mathbf{k}=-\mathbf{K}/2}^{\mathbf{K}/2} \frac{1}{(2\pi)^n 2E_\mathbf{k}}E_\mathbf{k}\nonumber\\
    & \left[ \alpha^*_{\mathbf{p}_{\mathbf{k},i}}\alpha_{\mathbf{p}_{\mathbf{k},i-1}}+ \left(\frac{2\pi}{\Delta p}\right)^n E_\mathbf{k}\right].\label{eq:Hamiltonian in terms of alpha}
\end{align}
Neglecting corrections of order $\order{\Delta\tau^2}$, the partition function takes the form
\begin{align}
    Z =& \left(\prod_{i=1}^{N} \prod_{\mathbf{k}=-\frac{\mathbf{K}}{2}}^\frac{\mathbf{K}}{2}\frac{1}{2\pi VE_\mathbf{k}}\int_{-\infty}^\infty da_{\mathbf{p}_{\mathbf{k},i}}\int_{-\infty}^\infty db_{\mathbf{p}_{\mathbf{k},i}}\right) \nonumber\\
    & \exp\bigg\{-\Delta \tau \Delta p^n  \sum_{i=1}^{N}  \sum_{\mathbf{k}=-\mathbf{K}/2}^{\mathbf{K}/2} \left[\frac{1}{(2\pi)^n 2E_\mathbf{k}} \left(\alpha^*_{\mathbf{p}_{\mathbf{k},i}}\alpha'_{\mathbf{p}_{\mathbf{k},i}} \right. \right. \nonumber\\
    & \left. \left. + E_\mathbf{k} \alpha^*_{\mathbf{p}_{\mathbf{k},i}}\alpha_{\mathbf{p}_{\mathbf{k},i-1}}\right)+ \frac{1}{\Delta p^n} \frac{E_\mathbf{k}}{2}\right]\bigg\}.\label{eq:Finite size discrete real free scalar field partitition function alpha}
\end{align}

It is important to note that this trotterization procedure involves ambiguities at higher orders in the discretization parameter, $\Delta\tau$. These subtleties are closely related to operator ordering issues in the corresponding canonical quantization and may, in more general contexts, affect the precise form of the action or the path integral measure. However, in the present work we restrict ourselves to leading-order considerations in $\Delta\tau$, for which these effects are negligible, and thus we do not address these technical subtleties further. For more detailed discussions, see \cite{Kleinert:2009,Schulman:1981,Trotter:1959}.

We should emphasize the vacuum energy term in \cref{eq:Finite size discrete real free scalar field partitition function alpha} appears with a dimensionful prefactor. This prefactor arises naturally from the discretized nature of spacetime in our formulation, where summations over discrete lattice points replace integrals, leading to additional factors of the lattice spacing.

In analogy with the finite-difference notation introduced in \cref{eq:Discrete temporal derivative compact notation,eq:Discrete spatial derivative compact notation}, we introduced a compact notation for finite differences along the discrete index $i$. Specifically, we write
\begin{align}
    \alpha'_{\mathbf{p}_{\mathbf{k},i}} & \equiv \frac{\alpha_{\mathbf{p}_{\mathbf{k},i}}-\alpha_{\mathbf{p}_{\mathbf{k},i-1}}}{\Delta \tau}.\label{eq:Discrete beta space derivative}
\end{align}

Unlike real-time derivatives, denoted by a dot, we use a prime to emphasize that the evolution here occurs in “$\beta$-space” (i.e., along the inverse temperature direction). This distinction helps avoid confusion between real-time derivatives and derivatives along the $i$ direction.

To proceed, we apply a change of variables. Motivated by the mode decomposition of the field, we define 
\begin{align}
    \phi_{\mathbf{p}_{\mathbf{k},i}} &\equiv \frac{1}{2E_\mathbf{k}} \left(\alpha_{\mathbf{p}_{\mathbf{k},i}} + \alpha_{-\mathbf{p}_{\mathbf{k},i}}^*\right),\label{eq:Field operator change of variables}\\
    \pi_{\mathbf{p_k},i} &\equiv \frac{\mathrm{i}}{2}\left(\alpha_{\mathbf{p}_{\mathbf{k},i}}-\alpha_{-\mathbf{p}_{\mathbf{k},i}}^*\right).\label{eq:Conjugate field operator change of variables}
\end{align}
The inverse of the transform in \cref{eq:Field operator change of variables,eq:Conjugate field operator change of variables} is given by
\begin{align}
    \alpha_{\mathbf{p}_{\mathbf{k},i}} &= E_\mathbf{k}\phi_{\mathbf{p}_{\mathbf{k},i}}+\mathrm{i}\pi_{\mathbf{p}_{\mathbf{k},i}},\label{eq:Coherent eigenvalue change of variables}\\
    \alpha^*_{\mathbf{p}_{\mathbf{k},i}} &= E_\mathbf{k}\phi_{-\mathbf{p}_{\mathbf{k},i}}-\mathrm{i}\pi_{-\mathbf{p}_{\mathbf{k},i}}.\label{eq:Conjugate coherent eigenvalue change of variables}
\end{align}
Applying the change of variables from \cref{eq:Coherent eigenvalue change of variables,eq:Conjugate coherent eigenvalue change of variables} to the partition function in \cref{eq:Finite size discrete real free scalar field partitition function alpha} introduces a Jacobian, which exactly cancels the energy-dependent prefactors and numerical constants in the integration measure. After changing variables, we apply the identities established in \cref{Partition Function Reality Condition,Summation by Parts with Periodic Boundaries}, specifically \cref{eq:Partition function reality condition,eq:Total field derivative vanish}, to obtain the final expression
\begin{align}
    Z =& \prod_{i=1}^N \int_{-\infty}^\infty \frac{da^\phi_{\mathbf{0},i}da^\pi_{\mathbf{0},i}}{2 \pi V} 
    \left(\prod'_{\mathbf{k}}\frac{da^\phi_{\mathbf{p}_{\mathbf{k},i}} da^\pi_{\mathbf{p}_{\mathbf{k},i}} db^\phi_{\mathbf{p}_{\mathbf{k},i}}db^\pi_{\mathbf{p}_{\mathbf{k},i}}}{(\pi V)^2}\right) \nonumber\\
    & \exp\bigg\{-\frac{\Delta\tau}{2}\left(\frac{\Delta p}{2\pi}\right)^n  \sum_{\mathbf{k}=-\mathbf{K}/2}^{\mathbf{K}/2} \left[E^2_\mathbf{k}\abs{\phi_{\mathbf{p}_{\mathbf{k},i}}}^2 + \abs{\pi_{\mathbf{p}_{\mathbf{k},i}}}^2 \right. \nonumber\\
    & \left. -\mathrm{i} \left(\pi_{\mathbf{p}_{\mathbf{k},i}}\left(\phi^*_{\mathbf{p}_{\mathbf{k},i}}\right)'+\pi^*_{\mathbf{p}_{\mathbf{k},i}}\phi'_{\mathbf{p}_{\mathbf{k},i}}\right)+\left(\frac{2\pi}{\Delta p}\right)^n E_\mathbf{k}\right]\bigg\}.\label{eq:Finite size discrete real free scalar field partitition function phi_p}
\end{align}
In order to make sure we integrate over the real domain, we have decomposed the fields into their real and imaginary components,
\begin{align}
    \phi_{\mathbf{p}_{\mathbf{k},i}} &\equiv a^\phi_{\mathbf{p}_{\mathbf{k},i}}+\mathrm{i}b^\phi_{\mathbf{p}_{\mathbf{k},i}},\nonumber\\
    \pi_{\mathbf{p}_{\mathbf{k},i}} &\equiv a^\pi_{\mathbf{p}_{\mathbf{k},i}}+\mathrm{i}b^\pi_{\mathbf{p}_{\mathbf{k},i}}.\label{eq:Complex decomposition of scalar fields}
\end{align}
The reality condition of the position-space fields $\phi_{\mathbf{x}_{\mathbf{r},i}}^*=\phi_{\mathbf{x}_{\mathbf{r},i}}$ imposes a symmetry in momentum space: $\phi_{\mathbf{p}_{\mathbf{k},i}}^* = \phi_{-\mathbf{p_k},i}$. This symmetry follows from the DFT in \cref{eq:Real scalar field DFT} and is reflected in the measure used in \cref{eq:Finite size discrete real free scalar field partitition function phi_p}, where we integrate only over the positive momentum modes, and treat the zero-modes separately.

To obtain the path integral partition function in position space, we perform an inverse DFT on the fields in \cref{eq:Finite size discrete real free scalar field partitition function phi_p} and apply the dispersion relation $E^2_{\mathbf{k}} = \mathbf{p}_\mathbf{k}^2 + m^2$. The Fourier decomposition then allows the substitution $\mathbf{p}^2_\mathbf{k} \rightarrow -\nabla_\mathbf{x_r}^2$, where $\nabla_\mathbf{x_r}$ is used as shorthand for the discrete spatial derivative, defined explicitly in \cref{eq:Discrete spatial derivative compact notation}. This substitution is more subtle than in the continuum limit due to the discrete nature of the derivative, as explicitly shown in \cref{Discrete Spatial Derivative Fourier Decomposition}. We obtain
\begin{align}
    \left(\frac{\Delta p}{2\pi}\right)^n & \sum_{\mathbf{k}=-\mathbf{K}/2}^{\mathbf{K}/2} E_\mathbf{k}^2 \abs{\phi_{\mathbf{p}_{\mathbf{k},i}}}^2 \nonumber\\
    =&\left(\frac{\Delta p}{2\pi}\right)^n \sum_{\mathbf{k}=-\mathbf{K}/2}^{\mathbf{K}/2} E_\mathbf{k}^2 \left(\Delta x^n \sum_{\mathbf{r}=\mathbf{0}}^\mathbf{K} e^{\mathrm{i} \mathbf{p_k} \cdot \mathbf{x_r}} \phi_{\mathbf{x}_{\mathbf{r},i}} \right) \nonumber\\
    & \ \left( \Delta x^n \sum_{\mathbf{s}=\mathbf{0}}^{\mathbf{K}} e^{-\mathrm{i} \mathbf{p_k} \cdot \mathbf{x_s}} \phi_{\mathbf{x}_{\mathbf{s},i}} \right)\nonumber\\
    =&\left(\frac{\Delta p}{2\pi}\right)^n \Delta x^{2n}  \sum_{\mathbf{r},\mathbf{s}=\mathbf{0}}^\mathbf{K} \phi_{\mathbf{x}_{\mathbf{r},i}} \phi_{\mathbf{x}_{\mathbf{s},i}} \nonumber\\
    & \ \sum_{\mathbf{k}=-\mathbf{K}/2}^{\mathbf{K}/2}\left(\mathbf{p}_\mathbf{k}^2 +m^2\right) e^{\mathrm{i} \mathbf{p_k} \cdot (\mathbf{x_r} - \mathbf{x_s})} \nonumber\\
    =&\frac{\Delta x^n}{(K+1)^n} \sum_{\mathbf{r},\mathbf{s}=\mathbf{0}}^\mathbf{K} \phi_{\mathbf{x}_{\mathbf{r},i}} \phi_{\mathbf{x}_{\mathbf{s},i}} \nonumber\\
    & \ \sum_{\mathbf{k}=-\mathbf{K}/2}^{\mathbf{K}/2}\left(-\nabla_\mathbf{x_s}^2 +m^2\right) e^{\mathrm{i} \mathbf{p_k} \cdot (\mathbf{x_r} - \mathbf{x_s})} \nonumber
    \end{align}
    \begin{align}
    =& \Delta x^{n}  \sum_{\mathbf{r},\mathbf{s}=\mathbf{0}}^\mathbf{K} \phi_{\mathbf{x}_{\mathbf{r},i}} \left(-\nabla_\mathbf{x_s}^2 +m^2\right) \phi_{\mathbf{x}_{\mathbf{s},i}} \delta_{\mathbf{r},\mathbf{s}} \nonumber\\
    =& \Delta x^{n}  \sum_{\mathbf{r}=\mathbf{0}}^\mathbf{K}\left(\nabla_\mathbf{x_r} \cdot \phi_{\mathbf{x}_{\mathbf{r},i}}\right)^2  +m^2 \phi_{\mathbf{x}_{\mathbf{r},i}}^2,
\end{align}
where we repeatedly used summation by parts with periodic boundary conditions (see \cref{Summation by Parts with Periodic Boundaries}) to rewrite the discrete spatial derivative.

\bigskip
\noindent
Subsequently performing the DFT on the remaining terms in \cref{eq:Finite size discrete real free scalar field partitition function phi_p}, and keeping track of the Jacobian associated with the change of measure induced by the DFT, we obtain
\begin{align}
    Z =& \prod_{i=1}^{N} \left(\prod_{\mathbf{r}=\mathbf{0}}^\mathbf{K} \int_{-\infty}^\infty \frac{\Delta x^n}{2\pi}d\phi_{\mathbf{x}_{\mathbf{r},i}}d\pi_{\mathbf{x}_{\mathbf{r},i}} \right)  \nonumber\\
    & \exp\bigg\{-\frac{\Delta\tau \Delta x^n}{2} \sum_{\mathbf{r}=\mathbf{0}}^\mathbf{K}\left[\pi_{\mathbf{x}_{\mathbf{r},i}}^2 -2\mathrm{i}\pi_{\mathbf{x}_{\mathbf{r},i}}\phi_{\mathbf{x}_{\mathbf{r},i}}' \right. \nonumber\\
    & \left. + \left(\nabla_\mathbf{x_r} \cdot \phi_{\mathbf{x}_{\mathbf{r},i}}\right)^2+m_r^2\phi_{\mathbf{x}_{\mathbf{r},i}}^2\right]-\Delta\tau\sum_{\mathbf{k}=-\mathbf{K}/2}^{\mathbf{K}/2}\frac{E_\mathbf{k}}{2}\bigg\}.\label{Finite size discrete real free scalar field partitition function phi_x}
\end{align}
Integrating out the $\pi_{\mathbf{x}_{\mathbf{r},i}}$ fields then yields
\begin{align}
    Z =& \prod_{i=1}^{N}\left( \prod_{\mathbf{r}=\mathbf{0}}^\mathbf{K} \sqrt{\frac{\Delta x^n}{2\pi\Delta\tau}} \int_{-\infty}^\infty d\phi_{\mathbf{x}_{\mathbf{r},i}}\right) \nonumber\\
    & \ \exp\bigg\{-\frac{\Delta\tau \Delta x^n}{2} \sum_{\mathbf{r}=\mathbf{0}}^\mathbf{K}\left[ \phi_{\mathbf{x}_{\mathbf{r},i}}'^2  + \left( \mathbf{\nabla}_{\mathbf{x_r}}\phi_{\mathbf{x}_{\mathbf{r},i}}\right)^2+m^2\phi^2_{\mathbf{x}_{\mathbf{r},i}}\right] \nonumber\\
    & \ -\Delta\tau\sum_{\mathbf{k}=-\mathbf{K}/2}^{\mathbf{K}/2}\frac{E_\mathbf{k}}{2}\bigg\}.\label{eq:Final real free scalar field partitition function 1}
\end{align}
Here, the final term involving $\frac{E_\mathbf{k}}{2}$ represents the vacuum energy contribution. This term diverges in the continuum limit, and is typically discarded. The vacuum energy term in \cref{eq:Final real free scalar field partitition function 1} is retained as it is precisely this contribution that gives rise to the subtlety we aim to highlight in the next section.

\subsection{Path Integral from Field Operator Hamiltonian}\label{Path Integral from Field Operator Hamiltonian}
Although the coherent states were originally defined using time-independent ladder operators, our goal in this section is to rewrite the coherent states in terms of field operators to enable a seamless transition to the interacting theory. In the presence of interactions, the time dependence of field operators becomes more intricate, and it is therefore crucial to implement the change of variables to the field basis in a way that remains consistent with the time evolution of the full theory. To this end, we promote the coherent states in \cref{eq:Coherent state definnition displacement operator} to a time-dependent form by expressing them in terms of Heisenberg picture ladder operators.

We begin by defining the time evolution of the creation and annihilation operators using the standard Heisenberg time evolution:
\begin{align}
    \hat{a}_H(t_a,\mathbf{p_k}) &\equiv e^{\mathrm{i}\hat{H}t_a}\hat{a}_S(\mathbf{p_k})e^{-\mathrm{i}\hat{H}t_a}\nonumber\\
    &= e^{-\mathrm{i}E_\mathbf{k}t_a}\hat{a}_S(\mathbf{p_k})e^{\mathrm{i}\hat{H}t_a}e^{-\mathrm{i}\hat{H}t_a} \nonumber\\
    &= e^{-\mathrm{i}E_\mathbf{k}t_a}\hat{a}_S(\mathbf{p_k}),\nonumber\\
    \hat{a}^\dagger_H(t_a,\mathbf{p_k}) &\equiv e^{\mathrm{i}\hat{H}t_a}\hat{a}^\dagger_S(\mathbf{p_k})e^{-\mathrm{i}\hat{H}t_a} \nonumber\\
    &= e^{\mathrm{i}E_\mathbf{k}t_a}\hat{a}^\dagger_S(\mathbf{p_k})e^{\mathrm{i}\hat{H}t_a}e^{-\mathrm{i}\hat{H}t_a} \nonumber\\
    &= e^{\mathrm{i}E_\mathbf{k}t_a}\hat{a}^\dagger_S(\mathbf{p_k}).\nonumber
\end{align}
(Recall that we are still working with the Hamiltonian operator in the free theory from \cref{eq:Free scalar ladder operator Hamiltonian}.) Using discretized notation, we write
\begin{align}
    \hat{a}_H(t_a,\mathbf{p_k}) &\equiv \hat{a}_{\mathbf{p}_{\mathbf{k},a}}, \quad
    \hat{a}^\dagger_H(t_a,\mathbf{p_k}) \equiv \hat{a}^\dagger_{\mathbf{p}_{\mathbf{k},a}}.\label{eq:Time dependent annihalation operator}
\end{align}
Since the statistical properties of the system are time-independent, we may fix $t_a$ to any value and work on a constant-time slice. On any particular time-slice, the coherent state properties that we have derived in \cref{Coherent State Relations} continue to hold.

For the second computation, we thus use a generalized coherent state defined as
\begin{align}
    \ket{\boldsymbol{\alpha}_{i,a}} \equiv& N^{+}_{i,i}\hat{D}(\boldsymbol{\alpha}_{i,a})\ket{0}\nonumber\\
    =& e^{\frac{1}{2} \Delta p^n \sum_{\mathbf{k}=-\mathbf{K}/2}^{\mathbf{K}/2} \frac{\alpha^*_{\mathbf{p}_{\mathbf{k},i}}\alpha_{\mathbf{p}_{\mathbf{k},i}}}{(2\pi)^n 2E_\mathbf{k}}}\nonumber\\
    & \ e^{\Delta p^n \sum_{\mathbf{k}=-\mathbf{K}/2}^{\mathbf{K}/2} \frac{1}{(2\pi)^n 2E_\mathbf{k}} \left[\alpha_{\mathbf{p}_{\mathbf{k},i}}\hat{a}^\dagger_{\mathbf{p}_{\mathbf{k},a}}-\alpha^*_{\mathbf{p}_{\mathbf{k},i}}\hat{a}_{\mathbf{p}_{\mathbf{k},a}}\right]}\ket{0},\label{eq:Coherent state definnition displacement operator}
\end{align}
where we introduced a generalized normalization constant
\begin{align}
    N^{\pm}_{i,j} &\equiv e^{\pm \frac{1}{2}\Delta p^n \sum_{\mathbf{k}=-\mathbf{K}/2}^{\mathbf{K}/2} \frac{\alpha^*_{\mathbf{p}_{\mathbf{k},i}}\alpha_{\mathbf{p_k},j}}{(2\pi)^n 2E_\mathbf{k}}},\label{eq:Coherent state generalized normalization}
\end{align}
and the displacement operator
\begin{align}
    \hat{D}(\boldsymbol{\alpha}_{i,a}) &\equiv e^{\Delta p^n \sum_{\mathbf{k}=-\mathbf{K}/2}^{\mathbf{K}/2} \frac{1}{(2\pi)^n 2E_\mathbf{k}} \left[\alpha_{\mathbf{p}_{\mathbf{k},i}}\hat{a}^\dagger_{\mathbf{p}_{\mathbf{k},a}}-\alpha^*_{\mathbf{p}_{\mathbf{k},i}}\hat{a}_{\mathbf{p}_{\mathbf{k},a}}\right]}.\label{eq:Displacement operator}
\end{align}
We retain the index $i$ to label the evolution along $\beta$, whereas the index $a$ indicates real-time evolution. This distinction between the evolution along $\beta$ and the evolution along $t$ allows us to clearly separate the thermal evolution of the system from its dynamical time evolution.

In \cref{Coherent State Representation using Displacement Operator}, we demonstrate the equivalence between the coherent states defined in \cref{eq:Coherent state definnition displacement operator,eq:Coherent state definition}. In \cref{Coherent State in the Field Basis}, we further show how the coherent state can be rewritten in terms of the field operators and classical field configurations $\hat{\phi}_{\mathbf{x}_{\mathbf{r},a}}, \ \hat{\pi}_{\mathbf{x}_{\mathbf{r},a}}, \ \phi_{\mathbf{x}_{\mathbf{r},i}}, \text{ and } \pi_{\mathbf{x}_{\mathbf{r},i}}$. The resulting expression provides an equivalent form of the coherent state in the field basis,
\begin{align}
    \hat{D}(\phi_{i,a}) &= e^{-\mathrm{i} (\Delta x)^n\sum_{\mathbf{r}=\mathbf{0}}^\mathbf{K} \left[\phi_{\mathbf{x}_{\mathbf{r},i}}\hat{\pi}_{\mathbf{x}_{\mathbf{r},a}}-\pi_{\mathbf{x}_{\mathbf{r},i}}\hat{\phi}_{\mathbf{x}_{\mathbf{r},a}}\right]}\nonumber\\
    \therefore \ket{\phi_{i,a}}&= N^{+}_{i,i}\hat{D}(\phi_{i,a})\ket{0}\nonumber\\
    &= N^{+}_{i,i}\hat{D}(\boldsymbol{\alpha}_{i,a})\ket{0}\nonumber\\
    &=\ket{\boldsymbol{\alpha}_{i,a}},\label{eq:Coherent state field basis}
\end{align}
where $\phi_{\mathbf{p}_{\mathbf{k},i}}$ and $\alpha_{\mathbf{p}_{\mathbf{k},i}}$ are related via \cref{eq:Field operator change of variables,eq:Conjugate field operator change of variables}. We emphasize that $\ket{\phi_{i,a}}$ remains an eigenstate of the annihilation operator. $\ket{\phi_{i,a}}$ is \emph{not} an eigenstate of the field operator: $\hat{\phi}_{\mathbf{x}_{\mathbf{r},a}}\ket{\phi_{i,a}} \neq \phi_{\mathbf{x}_{\mathbf{r},i}}\ket{\phi_{i,a}}$.

Note that the notation $\hat{D}(\phi_{i,a})$ and $\ket{\phi_{i,a}}$ does not imply dependence on the field $\phi$ alone: both the displacement operator and the corresponding coherent state depend on the full phase space configuration, including the conjugate momentum $\pi$; we have suppressed the dependence on $\pi$ for readability. Additionally, there is no explicit spatial dependence in the notation, as the spatial coordinates are summed over in the exponentials.

Key commutation relations involving the displacement operator in the field basis, which will be important for evaluating matrix elements from the partition function of \cref{eq:Intermediate free real partition function 2}, are listed below and proven explicitly in \cref{Field Basis Displacement Operator Commutation Relations}:
\begin{align}
    \left[\hat{\phi}_{\mathbf{x}_{\mathbf{r},a}},\hat{D}(\phi_{i,a})\right]&=\phi_{\mathbf{x}_{\mathbf{r},i}}\hat{D}(\phi_{i,a}),\label{eq:Commutator field and displacement operator}\\
    \left[\hat{D}^\dagger(\phi_{i,a}),\hat{\phi}_{\mathbf{x}_{\mathbf{r},a}}\right]&=\phi_{\mathbf{x}_{\mathbf{r},i}}\hat{D}^\dagger(\phi_{i,a}),\label{eq:Commutator field and conjugate displacement operator}\\
    \left[\hat{\pi}_{\mathbf{x}_{\mathbf{r},a}},\hat{D}(\phi_{i,a})\right]&=\pi_{\mathbf{x}_{\mathbf{r},i}}\hat{D}(\phi_{i,a}),\label{eq:Commutator conjugate field and displacement operator}\\
    \left[\hat{D}^\dagger(\phi_{i,a}),\hat{\pi}_{\mathbf{x}_{\mathbf{r},a}}\right]&=\pi_{\mathbf{x}_{\mathbf{r},i}}\hat{D}^\dagger(\phi_{i,a}).\label{eq:Commutator conjugate field and conjugate displacement operator}
\end{align}

We now compute the matrix element of the Hamiltonian between two coherent states at consecutive points in the thermal trace. The Klein-Gordon Hamiltonian in the field basis from \cref{eq:real scalar field hamiltonian discrete} in the operator formalism takes the form
\begin{align}
     \hat{H}& = \frac{\Delta x^n}{2} \sum_{\mathbf{r}=\mathbf{0}}^\mathbf{K} \left(\hat{\pi}_{\mathbf{x}_{\mathbf{r},a}}^2 + \left(\mathbf{\nabla}_\mathbf{x_r}\cdot\hat{\phi}_{\mathbf{x}_{\mathbf{r},a}}\right)^2 + m^2\hat{\phi}_{\mathbf{x}_{\mathbf{r},a}}^2\right).\nonumber
\end{align}
Inserting this Hamiltonian between two consecutive coherent states, we find
\begin{align}
    \bra{\phi_{i,a}} & \Delta\tau \hat{H}[\hat{\phi}_{\mathbf{x}_{\mathbf{r},a}}] \ket{\phi_{i-1,a}} \nonumber\\
    =& \bra{\phi_{i,a}} \frac{\Delta\tau \Delta x^n}{2} \sum_{\mathbf{r}=\mathbf{0}}^\mathbf{K} \left[\hat{\pi}_{\mathbf{x}_{\mathbf{r},a}}^2 \right. \nonumber\\
    & \left. \ + \left( \nabla_\mathbf{x_r} \cdot \hat{\phi}_{\mathbf{x}_{\mathbf{r},a}}\right)^2 + m^2\hat{\phi}_{\mathbf{x}_{\mathbf{r},a}}^2\right]\ket{\phi_{i-1,a}}\nonumber\\
    =&  N^+_{i,i}N^+_{i-1,i-1} \bra{0}\hat{D}^\dagger(\phi_{i,a})\frac{\Delta\tau \Delta x^n}{2} \sum_{\mathbf{r}=\mathbf{0}}^\mathbf{K} \left[\hat{\pi}_{\mathbf{x}_{\mathbf{r},a}}^2 \right. \nonumber\\
    &\left. \  + \left( \nabla_\mathbf{x_r} \cdot \hat{\phi}_{\mathbf{x}_{\mathbf{r},a}}\right)^2+m^2\hat{\phi}_{\mathbf{x}_{\mathbf{r},a}}^2\right]\hat{D}(\phi_{i-1,a})\ket{0}.\label{eq:Intermediate free matrix element}
\end{align}

In order to evaluate the matrix element in \cref{eq:Intermediate free matrix element}, we consider each of the terms in the square brackets individually. Our general strategy will be to use \cref{eq:Commutator field and conjugate displacement operator} and \cref{eq:Commutator conjugate field and conjugate displacement operator} to commute the $\hat{D}^\dagger(\phi_{i,a})$ across the operators $\hat{\mathcal{O}}=\left\{\hat{\pi}_{\mathbf{x}_{\mathbf{r},a}},\nabla_\mathbf{x_r} \cdot \hat{\phi}_{\mathbf{x}_{\mathbf{r},a}},\hat{\phi}_{\mathbf{x}_{\mathbf{r},a}}\right\}$. 

Starting with the momentum conjugate fields $\pi_{\mathbf{x}_{\mathbf{r},i}}$,
\begin{align}
    \Delta\tau\bra{\phi_{i,a}}& \hat{\pi}_{\mathbf{x}_{\mathbf{r},a}}^2\ket{\phi_{i-1,a}} \nonumber\\
    =& \Delta\tau N^+_{i,i}N^+_{i-1,i-1} \bra{0}\hat{D}^\dagger(\phi_{i,a}) \hat{\pi}_{\mathbf{x}_{\mathbf{r},a}}^2 \hat{D}(\phi_{i-1,a}) \ket{0}\nonumber\\
    =& \Delta\tau N^+_{i,i}N^+_{i-1,i-1} \bra{0} \left(\hat{\pi}_{\mathbf{x}_{\mathbf{r},a}}+\pi_{\mathbf{x}_{\mathbf{r},i}}\right)^2 \nonumber\\
    & \hat{D}^\dagger(\phi_{i,a})\hat{D}(\phi_{i-1,a}) \ket{0}.\label{eq:Pi partition function matrix element}
\end{align}
For the discrete spatial derivative of the field operators, we find
\begin{align}
    \Delta\tau\bra{\phi_{i,a}} & \left(\nabla_\mathbf{x_r} \cdot \hat{\phi}_{\mathbf{x}_{\mathbf{r},a}}\right)^2\ket{\phi_{i-1,a}} \nonumber\\
    =& \Delta\tau N^+_{i,i}N^+_{i-1,i-1} \bra{0} \hat{D}^\dagger(\phi_{i,a}) \left(\frac{\hat{\phi}_{\mathbf{x}_{\mathbf{r},a}}-\hat{\phi}_{\mathbf{x_{r-1}},a}}{\Delta x} \right)^2 \nonumber\\
    & \ \hat{D}(\phi_{i-1,a}) \ket{0}\nonumber\\
    =& \Delta\tau N^+_{i,i}N^+_{i-1,i-1} \bra{0} \left( \frac{\hat{\phi}_{\mathbf{x}_{\mathbf{r},a}}-\hat{\phi}_{\mathbf{x}_{\mathbf{r-1}},a}}{\Delta x} \right.  \nonumber\\
    & \ \left. + \frac{\phi_{\mathbf{x}_{\mathbf{r},i}}-\phi_{\mathbf{x}_{\mathbf{r-1},i}}}{\Delta x} \right)^2 \hat{D}^\dagger(\phi_{i,a})\hat{D}(\phi_{i-1,a})\ket{0}\nonumber\\
    =& \Delta\tau N^+_{i,i}N^+_{i-1,i-1} \bra{0} \left( \nabla_\mathbf{x_r} \cdot \hat{\phi}_{\mathbf{x}_{\mathbf{r},a}}+\nabla_\mathbf{x_r} \cdot\phi_{\mathbf{x}_{\mathbf{r},i}}\right)^2 \nonumber\\
    & \ \hat{D}^\dagger(\phi_{i,a})\hat{D}(\phi_{i-1,a}) \ket{0}.\label{eq:Del phi partition function matrix element}
\end{align}
Consider next $\Delta\tau\bra{\phi_{i,a}} \hat{\phi}_{\mathbf{x}_{\mathbf{r},a}}^n\ket{\phi_{i-1,a}}$, where we keep $n$ general in anticipation of the interacting theory derivation:
\begin{align}
    \Delta\tau\bra{\phi_{i,a}} & \hat{\phi}_{\mathbf{x}_{\mathbf{r},a}}^n\ket{\phi_{i-1,a}} \nonumber\\
    =& \Delta\tau N^+_{i,i}N^+_{i-1,i-1} \bra{0}\hat{D}^\dagger(\phi_{i,a}) \hat{\phi}_{\mathbf{x}_{\mathbf{r},a}}^n \hat{D}(\phi_{i-1,a}) \ket{0}\nonumber\\
    =& \Delta\tau N^+_{i,i}N^+_{i-1,i-1} \bra{0} \left(\hat{\phi}_{\mathbf{x}_{\mathbf{r},a}}+\phi_{\mathbf{x}_{\mathbf{r},i}}\right)^n \nonumber\\
    & \ \hat{D}^\dagger(\phi_{i,a}) \hat{D}(\phi_{i-1,a}) \ket{0}. \label{eq:General partition function matrix element}
\end{align}

In each expression, we encounter the term $\Delta \tau \hat{D}^\dagger(\phi_{i,a})\hat{D}(\phi_{i-1,a})$, which, as shown in \cref{Commutation of Displacement Operators in the Path Integral}, can be rewritten as
\begin{align}
    \Delta\tau  \hat{D}^\dagger&(\phi_{i,a}) \hat{D}(\phi_{i-1,a}) \nonumber\\
    &= \Delta \tau N^-_{i-1,i-1}N^+_{i,i-1}N^-_{i,i}N^+_{i,i-1} \nonumber\\
    & \quad e^{-\left(\frac{\Delta p}{2\pi}\right)^n \sum_{\mathbf{k}=-\mathbf{K}/2}^{\mathbf{K}/2} \frac{1}{2E_\mathbf{k}}\left(\alpha_{\mathbf{p}_{\mathbf{k},i}}-\alpha_{\mathbf{p}_{\mathbf{k},i-1}}\right)\hat{a}^\dagger_{\mathbf{p_k}}}.\label{eq:Displacement operator normalization}
\end{align}
Recall that the index $i \in [1,N]$ labels discrete points on a circle due to the trace of the partition function. We define $\tau_i \equiv \frac{i \beta}{N} = i\Delta\tau$, such that $\tau_0 = \tau_N$; $\tau_i$ parametrizes the circle with uniform spacing. The periodicity in $\tau$ permits a discrete Fourier transform over $\tau_i$, with corresponding frequencies given by $\nu_\ell \equiv \frac{2\pi \ell}{\beta} = \ell\Delta\nu$. These frequencies are known as Matsubara frequencies, named after Takeo Matsubara, who first formulated thermal field theory using the imaginary-time formalism \cite{Matsubara:1955ws}. As discussed in \cite{KapustaGale,LaineVuorinen}, these Matsubara frequencies are not tied to a dispersion relation but arise purely from the finite periodic lattice on the circle. The allowed frequency modes lie within the first Brillouin zone as discussed in \cite{MontvayMunster}, so that $\ell \in [-\frac{N-1}{2},\frac{N-1}{2}]$. The DFT is defined by
\begin{align}
    f(\tau_i) &= \frac{\Delta\nu}{2\pi}\sum_{\ell = -\frac{N-1}{2}}^{\frac{N-1}{2}} e^{-\mathrm{i}\nu_\ell \tau_i} \tilde{f}(\nu_\ell),\label{eq:DFT Matsubara frequency}\\
    \tilde{f}(\nu_\ell) &= \Delta\tau \sum_{i = 1}^N e^{\mathrm{i}\nu_\ell \tau_i} f(\tau_i).\label{eq:Inverste DFT Matsubara frequency}
\end{align}
The DFTs in \cref{eq:DFT Matsubara frequency,eq:Inverste DFT Matsubara frequency} allow us to express discrete finite differences on the circle, such as between coherent state eigenvalues, in frequency space:
\begin{align}
    \alpha_{\mathbf{p}_{\mathbf{k},i}} & -\alpha_{\mathbf{p}_{\mathbf{k},i-1}}\nonumber\\
    &= \alpha_{\mathbf{p}_{\mathbf{k},i}} \left(1-e^{-\mathrm{i}\nu_\ell \Delta\tau}\right)\nonumber\\
    &=\alpha_{\mathbf{p}_{\mathbf{k},i}}\left(1-\left(1-\mathrm{i}\nu_\ell \Delta\tau+\frac{1}{2}\left(-\mathrm{i}\nu_\ell \Delta\tau\right)^2+\cdots\right)\right)\nonumber\\
    &=\order{\Delta\tau}.\label{eq:Discrete finite difference}
\end{align}
Returning to \cref{eq:Displacement operator normalization}, we observe the exponential contains a discrete difference $\alpha_{\mathbf{p}_{\mathbf{k},i}} - \alpha_{\mathbf{p}_{\mathbf{k},i-1}}$, which is itself already of $\order{\Delta\tau}$. When expanding the exponential, the leading term is simply 1, while the first nontrivial contribution appears at $\order{\Delta\tau}$. Since \cref{eq:Intermediate free matrix element} already includes a factor of $\Delta\tau$, these exponential corrections contribute at $\order{\Delta\tau^2}$. We have previously chosen to ignore $\order{\Delta\tau^2}$ terms and may therefore rewrite \cref{eq:Displacement operator normalization} to obtain the result
\begin{align}
    \Delta \tau \hat{D}^\dagger(\phi_{i,a})\hat{D}(\phi_{i-1,a}) &= \Delta \tau N^-_{i-1,i-1}N^+_{i,i-1}N^-_{i,i}N^+_{i,i-1}.\label{eq:Displacement operator normalization 2}
\end{align}
After substituting \cref{eq:Displacement operator normalization 2} into
\cref{eq:Pi partition function matrix element,eq:Del phi partition function matrix element,eq:General partition function matrix element}, and rewriting the various normalization factors in each of these expressions, \cref{eq:Intermediate free matrix element} becomes
\begin{align}
    \bra{\phi_{i,a}} & \Delta\tau \hat{H}[\hat{\phi}_{\mathbf{x}_{\mathbf{r},a}}] \ket{\phi_{i-1,a}} \nonumber\\
    =& N^{+2}_{i,i-1} \bra{0}\frac{\Delta\tau \Delta x^n}{2} \sum_{\mathbf{r}=\mathbf{0}}^\mathbf{K} \left[\left(\hat{\pi}_{\mathbf{x}_{\mathbf{r},a}}+\pi_{\mathbf{x}_{\mathbf{r},i}}\right)^2 \right.\nonumber\\
    & + \left(\nabla_\mathbf{x_r} \cdot \hat{\phi}_{\mathbf{x}_{\mathbf{r},a}}+\nabla_\mathbf{x_r} \cdot \phi_{\mathbf{x}_{\mathbf{r},i}}\right)^2 + \nonumber\\
    & \left. m^2 \left(\hat{\phi}_{\mathbf{x}_{\mathbf{r},a}}+\phi_{\mathbf{x}_{\mathbf{r},i}}\right)^2 \right]\ket{0}.\label{eq:Expanded thingy}
\end{align}
Expanding the product of fields and field operators, we obtain the following structure
\begin{align}
    \bra{0} \left(\hat{\phi}_{\mathbf{x}_{\mathbf{r},a}}+\phi_{\mathbf{x}_{\mathbf{r},i}}\right)^2 \ket{0} =& \phi_{\mathbf{x}_{\mathbf{r},i}}^2 + 2 \phi_{\mathbf{x}_{\mathbf{r},i}} \cancelto{0}{\bra{0}\hat{\phi}_{\mathbf{x}_{\mathbf{r},a}}\ket{0}} \nonumber\\
    & \ + \bra{0}\hat{\phi}_{\mathbf{x}_{\mathbf{r},a}}^2\ket{0},\label{eq:Squared field operator VEV expansion}
\end{align}
and similarly for the expansion of the other terms in \cref{eq:Expanded thingy}. \Cref{eq:Expanded thingy,eq:Squared field operator VEV expansion} lead to the following matrix elements to be inserted in the partition function
\begin{align}
    \bra{\phi_{i,a}} & e^{-\Delta\tau \hat{H}[\hat{\phi}_{\mathbf{x}_{\mathbf{r},a}}]}\ket{\phi_{i-1,a}} \nonumber\\
    =& N^{+2}_{i,i-1} \exp\Bigg\{-\frac{\Delta\tau \Delta x^n}{2} \sum_{\mathbf{r}=\mathbf{0}}^\mathbf{K} \left[\pi_{\mathbf{x}_{\mathbf{r},i}}^2 +\left(\nabla_\mathbf{x_r} \cdot \phi_{\mathbf{x}_{\mathbf{r},i}}\right)^2 \right.\nonumber\\
    & \ + m^2\phi_{\mathbf{x}_{\mathbf{r},i}}^2 + \bra{0}\hat{\pi}_{\mathbf{x}_{\mathbf{r},a}}^2\ket{0}+ \bra{0}\left(\nabla_\mathbf{x_r} \cdot \hat{\phi}_{\mathbf{x}_{\mathbf{r},a}}\right)^2\ket{0} \nonumber\\
    &\left. \ + m^2\bra{0}\hat{\phi}_{\mathbf{x}_{\mathbf{r},a}}^2\ket{0} \right]\Bigg\}.\label{eq:Matrix element with VEV corrections}
\end{align}
Substituting the matrix element in \cref{eq:Matrix element with VEV corrections} into the the partition function, we find
\begin{align}
    Z &= \prod_{i=1}^{N} \left(\prod_{\mathbf{r}=\mathbf{0}}^\mathbf{K} \int_{-\infty}^\infty \frac{\Delta x^n}{2\pi}d\phi_{\mathbf{x}_{\mathbf{r},i}}d\pi_{\mathbf{x}_{\mathbf{r},i}} N^{-2}_{i,i}N^{+2}_{i,i-1}\right) \nonumber\\
    &\exp\Bigg\{-\frac{\Delta\tau \Delta x^n}{2} \sum_{\mathbf{r}=\mathbf{0}}^\mathbf{K} \left[\pi_{\mathbf{x}_{\mathbf{r},i}}^2 +\left(\nabla_\mathbf{x_r} \cdot \phi_{\mathbf{x}_{\mathbf{r},i}}\right)^2 + m^2\phi_{\mathbf{x}_{\mathbf{r},i}}^2 \right. \nonumber\\
    & \left. + \bra{0}\hat{\pi}_{\mathbf{x}_{\mathbf{r},a}}^2 + \left(\nabla_\mathbf{x_r} \cdot \hat{\phi}_{\mathbf{x}_{\mathbf{r},a}}\right)^2 + m^2 \hat{\phi}_{\mathbf{x}_{\mathbf{r},a}}^2\ket{0} \right]\Bigg\}.\label{eq:Free partition function field basis phi and pi}
\end{align}
In \cref{Coherent State Normalization in the Field Basis}, we show that we can rewrite the normalization constants as
\begin{align}
    \prod_{i=1}^N N^{-2}_{i,i}N^{+2}_{i,i-1} &= e^{\mathrm{i}\Delta\tau \Delta x^n \sum_{i=1}^N \sum_{\mathbf{r}=\mathbf{0}}^\mathbf{K}\pi_{\mathbf{x}_{\mathbf{r},i}}\phi_{\mathbf{x}_{\mathbf{r},i}}'}.\label{eq:Coherent state normalization rewrite}
\end{align}
After evaluating the Gaussian integrals in terms of the conjugate momentum fields $\pi_{\mathbf{x}_{\mathbf{r},i}}$, we obtain
\begin{align}
    Z &= \prod_{i=1}^{N} \left(\prod_{\mathbf{r}=\mathbf{0}}^\mathbf{K} \int_{-\infty}^\infty \sqrt{\frac{\Delta x^n}{2\pi \Delta\tau}} d\phi_{\mathbf{x}_{\mathbf{r},i}}\right)\nonumber\\
    & \exp\Bigg\{-\frac{\Delta\tau \Delta x^n}{2} \sum_{\mathbf{r}=\mathbf{0}}^\mathbf{K} \bigg[\phi_{\mathbf{x}_{\mathbf{r},i}}'^2 +\left(\nabla_\mathbf{x_r} \cdot \phi_{\mathbf{x}_{\mathbf{r},i}}\right)^2+m^2\phi_{\mathbf{x}_{\mathbf{r},i}}^2 \nonumber\\
    & + \bra{0}\underbrace{\hat{\pi}_{\mathbf{x}_{\mathbf{r},a}}^2+\left(\nabla_\mathbf{x_r} \cdot \hat{\phi}_{\mathbf{x}_{\mathbf{r},a}}\right)^2+m^2\hat{\phi}_{\mathbf{x}_{\mathbf{r},a}}^2}_{\hat{\mathcal{H}}}\ket{0} \bigg]\Bigg\}.\label{eq:Final real free scalar field partitition function 2}
\end{align}
The Vacuum Expectation Values (VEVs) that appear in \cref{eq:Final real free scalar field partitition function 2} correspond precisely to the vacuum energy we found in the result of the previous computation, \cref{eq:Final real free scalar field partitition function 1}. This vacuum energy is typically ignored. When interactions are included, we will show that some VEVs couple to the other terms in the Lagrangian and modify the structure of the counterterms, as shown in \cref{Interacting Theory Partition Function}.

\section{Interacting Theory Partition Function}\label{Interacting Theory Partition Function}
\subsection{Quantization and Picture of Interacting Fields}\label{Quantization and Picture of Interacting Fields}
We start with the bare scalar $\phi^4$ Lagrangian in $D=n+1$ dimensions and with the coupling constant $\lambda_0$,
\begin{align}
    \mathcal{L}_0 &\equiv \frac{1}{2} \partial_\mu \phi_0 \, \partial^\mu \phi_0 - \frac{1}{2} m_0^2 \phi_0^2 - \frac{\lambda_0}{4!} \phi_0^4,
    \label{eq:Bare Lagrangian}
\end{align}
and renormalize using multiplicative renormalization. We define
\begin{align}
    \phi_0 &\equiv \sqrt{Z_\phi}\phi_r,  \quad Z_\phi \equiv 1 + \delta_\phi, \nonumber\\
    m_0 &\equiv \frac{\sqrt{Z_m}}{\sqrt{Z_\phi}}m_r,  \quad Z_m \equiv 1+\frac{\delta_m}{m_r},\label{eq:Renormalized parameters}\\
    \lambda_0 &\equiv \frac{Z_\lambda}{Z_\phi^2}\lambda_r,  \quad Z_\lambda \equiv 1 + \frac{\delta_\lambda}{\lambda_r}.\nonumber
\end{align}
Inserting the renormalized parameters from \cref{eq:Renormalized parameters} into the bare Lagrangian, \cref{eq:Bare Lagrangian} yields
\begin{align}
    \mathcal{L}_r &= \mathcal{L}_\text{free} +  \mathcal{L}_\text{int},\nonumber\\
    \mathcal{L}_\text{free} &= \frac{1}{2} \partial_\mu \phi_r \,  \partial^\mu \phi_r - \frac{1}{2} m_r^2 \phi_r^2,\nonumber\\
    \mathcal{L}_\text{int} &= - \frac{\lambda_r}{4!} \phi_r^4 + \frac{1}{2}\delta_\phi\partial_\mu \phi_r \, \partial^\mu \phi_r - \frac{1}{2} \delta_m \phi_r^2 - \frac{\delta_\lambda}{4!} \phi_r^4.\label{eq:Renormalized Lagrangian}
\end{align}
Performing a Legendre transform then gives the renormalized Hamiltonian,
\begin{align}
    \mathcal{H}_r =& \mathcal{H}_\text{free} + \mathcal{H}_\text{int},\nonumber\\
    \mathcal{H}_\text{free} =& \frac{1}{2} \pi_r^2 + \frac{1}{2} (\nabla \cdot \phi_r)^2 + \frac{1}{2} m_r^2 \phi_r^2,\nonumber\\
    \mathcal{H}_\text{int} =& \frac{\lambda_r}{4!} \phi_r^4 + \frac{1}{2}\delta_\phi \left(\pi_r^2 + \frac{1}{2} (\nabla \cdot \phi_r)^2\right) + \frac{1}{2} \delta_m \phi_r^2\nonumber\\
    & \ + \frac{\delta_\lambda}{4!} \phi_r^4.\label{eq:Renormalized Hamiltonian}
\end{align}

Before quantizing the fields and imposing commutation relations, we first highlight a crucial subtlety. The choice of picture (such as the Heisenberg or Interaction picture) determines how the time evolution is implemented. By working in the Interaction picture, the evolution of our operators is governed by the free Hamiltonian, which means we may use the mode expansion of the free theory in the Heisenberg picture. Recognizing this subtlety is essential for formulating a consistent description of real-time interactions within a thermal framework. It ensures that all expressions remain well defined in terms of the free theory creation and annihilation operators.

We may transition from the Heisenberg picture to the Interaction picture via the Interaction picture unitary time evolution operator
\begin{align}
    \hat{U}_I(t,t_0) &\equiv e^{
    \mathrm{i}\hat{H}_\text{free}(t-t_0)}e^{-\mathrm{i}\hat{H}(t-t_0)}\nonumber\\
    &= \hat{U}^\dagger_\text{free}(t,t_0)\hat{U}(t,t_0).\label{eq:Shorthand unitary operator}
\end{align}
The Interaction picture unitary time evolution operator $\hat{U}_I(t,t_0)$ of \cref{eq:Shorthand unitary operator} satisfies the following identities:
\begin{align}
    \hat{U}_I^\dagger(t,t_0)&=\hat{U}_I^{-1}(t,t_0)\nonumber\\
    &= \hat{U}_I(t_0,t) ; \label{eq:Dyson series identities}\\
    \hat{U}_I(t_0,t)\hat{U}_I(t,t_0) &= \hat{U}_I^{-1}(t,t_0)\hat{U}_I(t,t_0)\nonumber\\
    &= \hat{U}_I(t_0,t_0)\nonumber\\
    &= \mathbbm{1}.
\end{align}
Consequently, our quantized fields in the Interaction picture can be expressed in terms of the Heisenberg picture fields as
\begin{align}
    \hat{\phi}_I\left(t,\mathbf{x_r}\right) \equiv& \hat{U}_I(t,t_0) \hat{\phi}_H \left(t,\mathbf{x_r}\right)  \hat{U}^\dagger_I(t,t_0)\nonumber\\
    =& \hat{U}^\dagger_\text{free}(t,t_0)\hat{U}(t,t_0) \hat{\phi}_H\left(t,\mathbf{x_r}\right) \nonumber\\
    & \ \hat{U}^\dagger(t,t_0) \hat{U}_\text{free}(t,t_0).
\end{align}
These unitary operators, $\hat{U}_I(t,t_0)$, represent continuous-time evolution. In contrast to the derivation of the free theory partition function, where both space and time were discretized, we now work with discretized space but continuous time. We work in continuous time for simplicity: first, the generator of continuous time evolution is simply the Hamiltonian; second, in the equilibrium thermal field theory we are considering, the final answer cannot depend on any continuous time variable, and we will see that the only effect of including the (continuous) real time evolution is changing the Heisenberg picture fields to Interaction picture fields. This change to Interaction picture fields is crucial for us as, as mentioned previously, these fields are decomposed into the Heisenberg picture creation and annihilation operators of the free theory, which are under full theoretical control. 

As a result all the derivations in the appendices, which assumed discretized space \emph{and} time, remain valid for the fields in this section, which use continuous time and are in the Interaction picture. We will thus continue to refer to the same appendices in this section as we referred to in the free theory derivation in \cref{Free Theory Partition Function}. Note further that while the field operators such as $\hat{\phi}(t,\mathbf{x_r})$ have both time and space as an argument, the non-operator fields $\phi_{\mathbf{x}_{\mathbf{r},i}}$ and $\alpha_{\mathbf{x}_{\mathbf{r},i}}$ associated with the coherent states iterate over evolution in $\beta$ but have no real time dependence at all.

The next step in the quantization procedure is to impose equal time canonical commutation relations on the renormalized Interaction picture fields,
\begin{align}
    \left[\hat{\phi}_{I,r}\left(t,\mathbf{x_r}\right),\hat{\pi}_{I,r}\left(t,\mathbf{x_s}\right)\right] &= \frac{\mathrm{i}}{\Delta x^n}\delta_{\mathbf{r},\mathbf{s}},\label{eq:Renormalized interaction picture scalar field commutation relations}\\
    \hat{\pi}_{I,r}(t,\mathbf{x_r}) \equiv \frac{\partial \mathcal{L}_\text{free}}{\partial \dot{\hat{\phi}}_{I,r}(t,\mathbf{x_r})} &= \dot{\hat{\phi}}_{I,r}(t,\mathbf{x_r}).
\end{align}
The renormalized Interaction picture field operators in continuous time and finite, discretized space are defined via the mode expansion with free theory creation and annihilation operators as
\begin{align}
    \hat{\phi}_\mathbf{x_r}(t) =& \Delta p^n \sum_{\mathbf{k}=\mathbf{K}/2}^{\mathbf{K}/2}  \frac{1}{(2\pi)^n 2E_\mathbf{k}} \left[ \hat{a}_{\mathbf{p_k}} e^{-\mathrm{i} \mathbf{p_k} \cdot \mathbf{x_r} + \mathrm{i} E_\mathbf{k} t} \right. \nonumber\\
    & \left. \ + \hat{a}^{\dagger}_{\mathbf{p_k}} e^{\mathrm{i} \mathbf{p_k} \cdot \mathbf{x_r} - \mathrm{i} E_\mathbf{k} t} \right]\bigg|_{E_\mathbf{k}=\sqrt{\mathbf{p}_\mathbf{k}^2+m^2}}.\label{eq:Interatction picture field mode expansion}
\end{align}
In \cref{eq:Interatction picture field mode expansion} and from now on, for notational simplicity, we suppress the indices related to the renormalization and the picture the field is in.

\subsection{Trotterization of the Interacting Theory Path Integral Partition Function}
\label{Phi4 Path Integral Partition Function}
We start the construction of the path integral representation of the partition function by inserting a complete set of coherent states into the thermal trace. For notational simplicity, we adopt a shorthand for the identity operator in terms of coherent states, rewriting \cref{eq:coherent state identity element} as
\begin{align}
    \mathbbm{1} &= \prod_{\mathbf{k} =-\frac{\mathbf{K}}{2}}^\frac{\mathbf{K}}{2} \int d\alpha^*_{\mathbf{p}_{\mathbf{k},i}}\, d\alpha_{\mathbf{p}_{\mathbf{k},i}}\, \ketbra{\boldsymbol{\alpha}_i(t)}{\boldsymbol{\alpha}_i(t)},\label{eq:Shorthand coherent state identity element}
\end{align}
where $\ket{\boldsymbol{\alpha}_i(t)}$ is the continuous-time analogue of the coherent state, given below both in terms of ladder operators and in the field basis
\begin{align}
    \ket{\boldsymbol{\alpha}_i(t)} \equiv& N^{+}_{i,i}e^{\Delta p^n \sum_{\mathbf{k}=\mathbf{K}/2}^{\mathbf{K}/2} \frac{1}{(2\pi)^n 2E_\mathbf{k}} \left[\alpha_{\mathbf{p}_{\mathbf{k},i}}\hat{a}^\dagger_\mathbf{p_k}(t)-\alpha^*_{\mathbf{p}_{\mathbf{k},i}}\hat{a}_\mathbf{p_k}(t)\right]}\nonumber\\
    & \ \ket{0},\label{eq:Coherent state definition continuous time alpha}\\
    \ket{\phi_i(t)} \equiv& N^{+}_{i,i} e^{-\mathrm{i} (\Delta x)^n\sum_{\mathbf{r}=\mathbf{0}}^\mathbf{K} \left[\phi_{\mathbf{x}_{\mathbf{r},i}}\hat{\pi}_\mathbf{x_r}(t)-\pi_{\mathbf{x}_{\mathbf{r},i}}\hat{\phi}_\mathbf{x_r}(t)\right]}\ket{0}.\label{eq:Coherent state definition continuous time phi}
\end{align}
Recall that $i$ iterates over the $\beta$ evolution of the thermal trace. At this stage $t$ is arbitrary. Recall that the operators appearing in the coherent states are in the Interaction picture for the full theory, which is the same as the Heisenberg picture for the free theory, and thus the operators carry time dependence. Crucially, but extremely subtly, the coherent states $\ket{\alpha_i(t)}=\ket{\phi_i(t)}$ defined in \cref{eq:Coherent state definition continuous time alpha,eq:Coherent state definition continuous time phi} are \emph{not} the coherent states at time $t$ in the Interaction picture. Rather, $\ket{\alpha_i(t)}$ is created by the displacement operator in the Interaction picture acting on the free theory vacuum. To wit,
\begin{align}
   \ket{\alpha_i(t)}
   & \equiv N^{+}_{i,i}\,\hat{D}_I(\alpha_i(t))\ket{0},\label{eq:Interaction picture Coherent state 1}
\end{align}
where $\hat{D}_I(\alpha_i)$ is the displacement operator composed of creation and annihilation operators (or, similarly, fields $\hat \phi$ and $\hat \pi$) in the Interaction picture.  On the other hand, the coherent state in the Interaction picture at time $t$ is given by the time evolution of the coherent state at time $t_0$ evolved with the Interaction picture time evolution operator:
\begin{align}
   \ket{\alpha_i(t)}_I
   & \equiv \hat{U}_I(t,t_0)\ket{\alpha_i(t_0)} \nonumber\\
   & = N^{+}_{i,i}\,\hat{U}_I(t,t_0) \hat{D}(\alpha_i(t_0)\ket{0} \nonumber\\
   & = N^{+}_{i,i}\,\hat{U}_{H_\text{free}}(t,t_0)\hat{U}^\dagger_{H}(t,t_0)\hat{D}(\alpha_i(t_0))\ket{0}.\label{eq:Interaction picture Coherent state 2}
\end{align}
In \cref{eq:Interaction picture Coherent state 2}, the displacement operator is evaluated at time $t_0$, and thus the operators in $\hat{D}(\alpha_i(t_0))$ can be in any of the pictures. We may explicitly see that $\ket{\alpha_i(t)} \ne \ket{\alpha_i(t)}_I$ by slightly manipulating \cref{eq:Interaction picture Coherent state 1}:
\begin{align}
    \ket{\alpha_i(t)}
    & = N^{+}_{i,i}\,\hat{D}_I(\alpha_i(t))\ket{0} \nonumber\\
    & = N^{+}_{i,i}\,\hat{U}^\dagger_\text{free}(t,t_0)\hat{D}(\alpha_i(t_0))\hat{U}_\text{free}(t,t_0)\ket{0} \nonumber\\
    & = N^{+}_{i,i}\,\hat{U}^\dagger_\text{free}(t,t_0)\hat{D}(\alpha_i(t_0))\ket{0},\label{eq:Interaction picture Coherent state 3}
\end{align}
where in the last line we set the ground state energy of the free theory to 0.  Unless the full Hamiltonian is trivial, \cref{eq:Interaction picture Coherent state 3} can never equal \cref{eq:Interaction picture Coherent state 2}. 

We now compute the partition function from first principles. Starting from
\begin{align}
    Z &= \Tr\left[e^{-\beta H_\text{full}[\hat{\phi}_H(t,\mathbf{x_r})]}\right],
\end{align}
where the full Hamiltonian in the Heisenberg picture is given by
\begin{align}
    \hat{H}_\text{full}[\hat{\phi}_H(t,\mathbf{x_r})] &= \Delta x^n \sum_{\mathbf{r} = \mathbf{0}}^\mathbf{K} \hat{\mathcal{H}}_r[\hat{\phi}_H(t,\mathbf{x_r})],
\end{align}
and $\hat{\mathcal{H}}_r$ is the quantized renormalized Hamiltonian density, defined classically in \cref{eq:Renormalized Hamiltonian}. At this point, the time $t$ is arbitrary.

To evaluate the trace, we insert a complete set of states $\mathbbm{1} = \sum_n \ketbra{n}{n}$ and proceed as outlined above:
\begin{align}
    Z =& \sum_{n}\bra{n}e^{-\beta \hat{H}_\text{full}[\hat{\phi}_H(t,\mathbf{x_r})]}\ket{n}\nonumber\\
    =& \sum_{n}\bra{n}\hat{U}_I(t_0,t)\hat{U}_I(t,t_0)e^{-\beta \hat{H}_\text{full}[\hat{\phi}_H(t,\mathbf{x_r})]}\ket{n}\nonumber\\
    =& \sum_{n} \bra{n}\hat{U}_I(t_0,t) \prod_{\mathbf{k} =-\frac{\mathbf{K}}{2}}^\frac{\mathbf{K}}{2} \int d\alpha^*_\mathbf{p_k} d\alpha_\mathbf{p_k} \ketbra{\boldsymbol{\alpha}(t)}{\boldsymbol{\alpha}(t)}\nonumber\\
    & \quad \hat{U}_I(t,t_0)e^{-\beta \hat{H}_\text{full}[\hat{\phi}_H(t,\mathbf{x_r})]}\ket{n}\nonumber\\
    =& \prod_{\mathbf{k} =-\frac{\mathbf{K}}{2}}^\frac{\mathbf{K}}{2} \int d\alpha^*_\mathbf{p_k} d\alpha_\mathbf{p_k} \bra{\boldsymbol{\alpha}(t)}\hat{U}_I(t,t_0)e^{-\beta \hat{H}_\text{full}[\hat{\phi}_H(t,\mathbf{x_r})]}\nonumber\\
    & \quad \sum_{n}\ketbra{n}{n}\hat{U}_I(t_0,t)\ket{\boldsymbol{\alpha}(t)}\nonumber\\
    =& \prod_{\mathbf{k} =-\frac{\mathbf{K}}{2}}^\frac{\mathbf{K}}{2} \int d\alpha^*_\mathbf{p_k} d\alpha_\mathbf{p_k}\bra{\boldsymbol{\alpha}(t)}\hat{U}_I(t,t_0)e^{-\beta \hat{H}_\text{full}[\hat{\phi}_H(t,\mathbf{x_r})]}\nonumber\\
    & \quad \hat{U}_I(t_0,t) \ket{\boldsymbol{\alpha}(t)} \nonumber\\
    =& \prod_{\mathbf{k} =-\frac{\mathbf{K}}{2}}^\frac{\mathbf{K}}{2} \int d\alpha^*_\mathbf{p_k} d\alpha_\mathbf{p_k}\bra{\boldsymbol{\alpha}(t)}e^{-\beta \hat{H}_\text{full}[\hat{\phi}_I(t,\mathbf{x_r})]}\ket{\boldsymbol{\alpha}(t)}.\label{eq:intermediate Interaction picture partition function 1}
\end{align}
Note that the derivation is completely insensitive to the time at which the eigenstates of the full Hamiltonian $\ket{n}$ are evaluated. Notice especially that from the second to last to the last step, the fields have gone from the Heisenberg picture to the Interaction picture. The final \cref{eq:intermediate Interaction picture partition function 1} is under complete control: the kets are created by displacement operators whose operators are in the Interaction picture acting on the free theory vacuum, and the Hamiltonian is also a function of Interaction picture operators.

A subtlety arises should one choose to take $t_0 \to t$. When $t_0 = t$, $\hat{U}_I = \mathbbm{1}$ and the Heisenberg and Interaction pictures coincide. When the Heisenberg and the Interaction pictures coincide, $\hat{H}_\text{full}[\hat{\phi}_H] = \hat{H}_\text{full}[\hat{\phi}_I]$, suggesting that we have, in fact, exactly solved the fully interacting problem. In the context of scattering theory, this apparent coincidence between the different pictures is avoided by working in the in--out formalism, where $t_0 \to -\infty$ and $t \to +\infty$, and asymptotic states are evolved with the full Hamiltonian via the LSZ reduction procedure.

We may resolve this seeming contradiction as follows.  In order to capture the effects of the interactions, we must include enough real time evolution to include several scatterings; one may think of this requirement as requiring that should the system become slightly out of equilibrium, we allow the system enough time to re-equilibrate. Thus we require that $t - t_0 \gg t_\text{int}$, where $t_\text{int}$ is the timescale associated with the interactions in the system. We will simplify our calculations by taking $t_0 \to -\infty$, which simultaneously ensures that 1) $t - t_0 \gg t_\text{int}$ and 2) two scales are removed from the problem: $t_0$, since $t_0 \to -\infty$, and also $t$, since the final result of the partition function must be $t$ independent.

We next divide the partition function into $N \rightarrow \infty$ segments, inserting the identity element of \cref{eq:coherent state identity element} between each segment. Note that all identity insertions happen at a fixed real-time slice $t$. Returning to the discretized space notation for the renormalized Interaction picture field operators $\hat{\phi}_\mathbf{x_r}(t)$, we obtain
\begin{align}
    Z =& \prod_{\mathbf{k} =-\frac{\mathbf{K}}{2}}^\frac{\mathbf{K}}{2} \int d\alpha^*_{\mathbf{p}_{\mathbf{k}}} d\alpha_{\mathbf{p}_{\mathbf{k}}}e^{-\Delta p^n \sum_\mathbf{k} \frac{1}{(2\pi)^n 2E_{\mathbf{k}}}\alpha_{\mathbf{p}_{\mathbf{k}}}^* \alpha_{\mathbf{p}_{\mathbf{k}}}}\nonumber\\
    & \quad \bra{\boldsymbol{\alpha}(t)}e^{-\Delta\tau \hat{H}_\text{full}[\hat{\phi}_\mathbf{x_r}(t)]}\cdots e^{-\Delta\tau \hat{H}_\text{full}[\hat{\phi}_\mathbf{x_r}(t)]}\ket{\boldsymbol{\alpha}(t)}\nonumber\\
    =& \prod_{i=1}^N \prod_{\mathbf{k} =-\frac{\mathbf{K}}{2}}^\frac{\mathbf{K}}{2} \int d\alpha^*_{\mathbf{p}_{\mathbf{k},i}} d\alpha_{\mathbf{p}_{\mathbf{k},i}}e^{- \Delta p^n \sum_\mathbf{k} \frac{1}{(2\pi)^n 2E_\mathbf{k}}\alpha_{\mathbf{p}_{\mathbf{k},i}}^* \alpha_{\mathbf{p}_{\mathbf{k},i}}}\nonumber\\
    & \quad \bra{\boldsymbol{\alpha}_{N}(t)} e^{-\Delta\tau \hat{H}_\text{full}[\hat{\phi}_\mathbf{x_r}(t)]}\ket{\boldsymbol{\alpha}_{N-1}(t)} \cdots \nonumber\\
    & \quad \bra{\boldsymbol{\alpha}_1(t)} e^{-\Delta\tau \hat{H}_\text{full}[\hat{\phi}_\mathbf{x_r}(t)]}\ket{\boldsymbol{\alpha}_{0}(t)}.\label{eq:intermediate Interaction picture partition function 2}
\end{align}

We now turn our attention to a single matrix element from \cref{eq:intermediate Interaction picture partition function 2}. By expanding in the infinitesimal parameter $\Delta\tau$ and inserting the full Hamiltonian in terms of the Interaction picture fields, we find
\begin{align}
    \bra{\boldsymbol{\alpha}_{i}(t)} & e^{-\Delta\tau \hat{H}_\text{full}[\hat{\phi}_\mathbf{x_r}(t)]}\ket{\boldsymbol{\alpha}_{i-1}(t)}\nonumber\\
    =&  \bra{\boldsymbol{\alpha}_{i}(t)}\left(1-\Delta\tau\hat{H}_\text{full}[\hat{\phi}_\mathbf{x_r}(t)]+\order{\Delta\tau^2}\right)\ket{\boldsymbol{\alpha}_{i-1}(t)}\nonumber\\
    =&  \bra{\boldsymbol{\alpha}_{i}(t)}\left(1-\Delta\tau\hat{H}_\text{free}[\hat{\phi}_\mathbf{x_r}(t)]-\Delta\tau\hat{H}_I[\hat{\phi}_\mathbf{x_r}(t)] \right. \nonumber\\
    & \left. \quad +\order{\Delta\tau^2}\right)\ket{\boldsymbol{\alpha}_{i-1}(t)}.\nonumber
\end{align}
Using the commutation relations of the displacement operator from \cref{Field Basis Displacement Operator Commutation Relations} and the results from \cref{Commutation of Displacement Operators in the Path Integral}, we compute the matrix element of the full Hamiltonian between two consecutive coherent states in exactly the same way as in \cref{Path Integral from Field Operator Hamiltonian}:
\begin{widetext}
\begin{align}
    \bra{\boldsymbol{\alpha}_i(t)} \Delta\tau & \hat{H}_\text{full}[\hat{\phi}_\mathbf{x_r}(t)]\ket{\boldsymbol{\alpha}_{i-1}(t)}\nonumber\\
    =& N^+_{i,i} N^+_{i-1,i-1} \bra{0} \hat{D}^\dagger(\boldsymbol{\alpha}_{i}(t)) \left(\Delta\tau\Delta x^n \sum_{\mathbf{r} = \mathbf{0}}^\mathbf{K} \left[ \frac{1}{2} \left( \hat{\pi}_\mathbf{x_r}^2(t) + \left(\nabla_\mathbf{x_r} \cdot \hat{\phi}_\mathbf{x_r}(t) \right)^2 +m_r^2 \hat{\phi}_\mathbf{x_r}^2(t)\right) \right. \right. \nonumber\\
    & \quad \left. \left. + \frac{\lambda_r}{4!}\hat{\phi}_\mathbf{x_r}^4(t) + \frac{\delta_\phi}{2} \left( \hat{\pi}_\mathbf{x_r}^2(t) + \left(\nabla_\mathbf{x_r} \cdot \hat{\phi}_\mathbf{x_r}(t) \right)^2 \right) + \delta_m \hat{\phi}_\mathbf{x_r}^2(t) + \frac{\delta_\lambda}{4!}\hat{\phi}_\mathbf{x_r}^4(t) \right] \right) \hat{D}(\boldsymbol{\alpha}_{i-1}(t))\ket{0} \nonumber\\
    =& N^+_{i,i} N^+_{i-1,i-1} \bra{0} \left(\Delta\tau\Delta x^n \sum_{\mathbf{r} = \mathbf{0}}^\mathbf{K} \left[\frac{1}{2} \left( \hat{\pi}_\mathbf{x_r}(t)+\pi_{\mathbf{x}_{\mathbf{r},i}} \right)^2 + \frac{1}{2} \left( \nabla_\mathbf{x_r} \cdot \hat{\phi}_\mathbf{x_r}(t) + \nabla_\mathbf{x_r} \cdot \phi_{\mathbf{x}_{\mathbf{r},i}} \right)^2 + \frac{m_r^2}{2} \left( \hat{\phi}_\mathbf{x_r}(t)+\phi_{\mathbf{x}_{\mathbf{r},i}} \right)^2 \right. \right. \nonumber\\
    & \quad   + \frac{\lambda_r}{4!} \left( \hat{\phi}_\mathbf{x_r}(t)+\phi_{\mathbf{x}_{\mathbf{r},i}} \right)^4 + \frac{\delta_\phi}{2} \left( \hat{\pi}_\mathbf{x_r}(t)+\pi_{\mathbf{x}_{\mathbf{r},i}} \right)^2 + \frac{\delta_\phi}{2}\left( \nabla_\mathbf{x_r} \cdot \hat{\phi}_\mathbf{x_r}(t) + \nabla_\mathbf{x_r} \cdot \phi_{\mathbf{x}_{\mathbf{r},i}} \right)^2\nonumber\\
    &\left. \left. \quad  + \frac{\delta_m}{2} \left( \hat{\phi}_\mathbf{x_r}(t)+\phi_{\mathbf{x}_{\mathbf{r},i}} \right)^2 + \frac{\delta_\lambda}{4!} \left( \hat{\phi}_\mathbf{x_r}(t)+\phi_{\mathbf{x}_{\mathbf{r},i}} \right)^4 \right] \right) \hat{D}^\dagger(\boldsymbol{\alpha}_i(t))\hat{D}(\boldsymbol{\alpha}_{i-1}(t)) \ket{0} \nonumber\\
    =& N^{+2}_{i,i-1} \bra{0} \left(\Delta\tau\Delta x^n \sum_{\mathbf{r} = \mathbf{0}}^\mathbf{K} \left[\frac{1}{2}\left( \hat{\pi}_\mathbf{x_r}(t)+\pi_{\mathbf{x}_{\mathbf{r},i}} \right)^2 + \frac{1}{2}\left( \nabla_\mathbf{x_r} \cdot \hat{\phi}_\mathbf{x_r}(t) + \nabla_\mathbf{x_r} \cdot \phi_{\mathbf{x}_{\mathbf{r},i}} \right)^2 + \frac{m_r^2}{2} \left( \hat{\phi}_\mathbf{x_r}(t)+\phi_{\mathbf{x}_{\mathbf{r},i}} \right)^2 \right. \right. \nonumber\\
    & \quad + \frac{\lambda_r}{4!} \left( \hat{\phi}_\mathbf{x_r}(t)+\phi_{\mathbf{x}_{\mathbf{r},i}} \right)^4 + \frac{\delta_\phi}{2}\left( \hat{\pi}_\mathbf{x_r}(t)+\pi_{\mathbf{x}_{\mathbf{r},i}} \right)^2 + \frac{\delta_\phi}{2} \left( \nabla_\mathbf{x_r} \cdot \hat{\phi}_\mathbf{x_r}(t) + \nabla_\mathbf{x_r} \cdot \phi_{\mathbf{x}_{\mathbf{r},i}} \right)^2 \nonumber\\
    & \left. \left. \quad + \frac{\delta_m}{2} \left( \hat{\phi}_\mathbf{x_r}(t)+\phi_{\mathbf{x}_{\mathbf{r},i}} \right)^2 + \frac{\delta_\lambda}{4!} \left( \hat{\phi}_\mathbf{x_r}(t)+\phi_{\mathbf{x}_{\mathbf{r},i}} \right)^4 \right] \right) \ket{0}. \label{eq:Interacting coherent state Hamiltonian matrix element}
\end{align}
\end{widetext}
In the first line, we simply write out the various expressions fully; in the second line, we commute $\hat{D}^\dagger(\boldsymbol{\alpha}_i(t))$ to the right of the expression;  and in the final line we evaluate $\hat{D}^\dagger(\boldsymbol{\alpha}_i(t))\hat{D}(\boldsymbol{\alpha}_{i-1}(t))$ and simplify.

Next, we expand the shifted interaction term
\begin{align}
    \bra{0}(\hat{\phi}_\mathbf{x_r}(t) & + \phi_{\mathbf{x}_{\mathbf{r}_i}} \hat{\mathbbm{1}})^4\ket{0}\nonumber\\
    =& \phi_{\mathbf{x}_{\mathbf{r}_i}}^4 + 4 \phi_{\mathbf{x}_{\mathbf{r}_i}}^3\cancelto{0}{\bra{0}\hat{\phi}_\mathbf{x_r}(t)\ket{0}}+6\phi_{\mathbf{x}_{\mathbf{r}_i}}^2\bra{0}\hat{\phi}_\mathbf{x_r}^2(t)\ket{0}\nonumber\\
    & \ + 4 \phi_{\mathbf{x}_{\mathbf{r}_i}}\cancelto{0}{\bra{0}\hat{\phi}_\mathbf{x_r}^3(t)\ket{0}}+\bra{0}\hat{\phi}_\mathbf{x_r}^4(t)\ket{0}\nonumber\\
    =& \phi_{\mathbf{x}_{\mathbf{r}_i}}^4 + 6 \phi_{\mathbf{x}_{\mathbf{r}_i}}^2 \bra{0} \hat{\phi}_\mathbf{x_r}^2(t) \ket{0} + \bra{0}\hat{\phi}_\mathbf{x_r}^4(t) \ket{0},\nonumber
\end{align}
which reveals an important structural feature: unlike in the free theory, the interaction term now couples the vacuum fluctuations to the fields, thereby modifying the effective mass term in the action.

\bigskip
\noindent
Substituting \cref{eq:Interacting coherent state Hamiltonian matrix element} into the partition function \cref{eq:intermediate Interaction picture partition function 2} and rewriting the normalization constants as shown in \cref{Coherent State Normalization in the Field Basis}, yields
\begin{widetext}
\begin{align}
    Z =& \prod_{i=1}^{N} \left( \prod_{\mathbf{r} = \mathbf{0}}^\mathbf{K} \int_{-\infty}^\infty \frac{\Delta x^n}{2\pi}d\phi_{\mathbf{x}_{\mathbf{r},i}}d\pi_{\mathbf{x}_{\mathbf{r},i}} e^{\mathrm{i}\Delta\tau \Delta x^n\sum_{\mathbf{r}=\mathbf{0}}^{\mathbf{K}}\pi_{\mathbf{x}_{\mathbf{r},i}}\phi_{\mathbf{x}_{\mathbf{r},i}}'}\right)\nonumber\\
    & \quad \exp\Bigg\{-\Delta\tau \Delta x^n \sum_{\mathbf{r} = \mathbf{0}}^\mathbf{K} \left[\frac{1}{2}\pi_{\mathbf{x}_{\mathbf{r},i}}^2+ \frac{1}{2}\left(\nabla_\mathbf{x_r} \cdot \phi_{\mathbf{x}_{\mathbf{r},i}}\right)^2+\frac{m_r^2}{2}\phi_{\mathbf{x}_{\mathbf{r},i}}^2 \right.\nonumber\\
    & \quad +\frac{\lambda_r}{4!}\phi_{\mathbf{x}_{\mathbf{r},i}}^4 + \frac{\delta_\phi}{2} \pi_{\mathbf{x}_{\mathbf{r},i}}^2 +\frac{\delta_\phi}{2} \left(\nabla_\mathbf{x_r} \cdot \phi_{\mathbf{x}_{\mathbf{r},i}} \right)^2 + \frac{1}{2}\delta_m\phi_{\mathbf{x}_{\mathbf{r},i}}^2  + \frac{\delta_\lambda}{4!}\phi_{\mathbf{x}_{\mathbf{r},i}}^4 \nonumber\\
    & \left. \quad +\frac{\lambda_r + \delta_\lambda}{2} \bra{0}\hat{\phi}_\mathbf{x_r}^2(t)\ket{0}\phi_{\mathbf{x}_{\mathbf{r},i}}^2 + \bra{0}\hat{\mathcal{H}}_r\ket{0}\right] \Bigg\}.\label{eq:Interacting scalar field partition function with pi fields}
\end{align}
\end{widetext}
All vacuum expectation values (VEVs) appearing in the partition function yield divergent contributions in the continuum limit. Those VEVs that sum to the vacuum expectation value of the renormalized Hamiltonian operator are typically omitted.

A notable exception from the literature arises from the coupling between the VEV $\bra{0} \hat{\phi}_{\mathbf{x}_\mathbf{r}}^2(t) \ket{0}$ and the classical field configurations $\phi_{\mathbf{x}_{\mathbf{r},i}}^2$. This coupling introduces a nontrivial correction to the mass counterterm and marks a key distinction from the standard path integral formulations in the literature where such couplings between operator VEVs and classical fields are absent.

We proceed by integrating out the $\pi_{\mathbf{x}_{\mathbf{r},i}}$ fields, yielding the final form of the interacting path integral partition function:
\begin{widetext}
\begin{align}
    Z =& \prod_{i=1}^{N} \left( \prod_{\mathbf{r} = \mathbf{0}}^\mathbf{K} \int_{-\infty}^\infty \sqrt{\frac{\Delta x^n}{2\pi \Delta\tau\left(1+\delta_\phi\right)}} d\phi_{\mathbf{x}_{\mathbf{r},i}}\right)\nonumber\\
    & \quad \exp\Bigg\{-\Delta\tau \Delta x^n \sum_{\mathbf{r} = \mathbf{0}}^\mathbf{K} \bigg[\frac{1}{2}\phi_{\mathbf{x}_{\mathbf{r},i}}'^2+ \frac{1}{2}\left(\nabla_\mathbf{x_r} \cdot \phi_{\mathbf{x}_{\mathbf{r},i}}\right)^2+\frac{m_r^2}{2}\phi_{\mathbf{x}_{\mathbf{r},i}}^2 + \frac{\lambda_r}{4!}\phi_{\mathbf{x}_{\mathbf{r},i}}^4 \nonumber\\
    & \quad + \frac{1}{2} \delta_\phi \phi_{\mathbf{x}_{\mathbf{r},i}}'^2 +\frac{1}{2} \delta_\phi \left(\nabla_\mathbf{x_r} \cdot \phi_{\mathbf{x}_{\mathbf{r},i}} \right)^2 + \frac{1}{2}\left(\delta_m +\frac{\lambda_r + \delta_\lambda}{2} \bra{0}\hat{\phi}_\mathbf{x_r}^2(t)\ket{0}\right)\phi_{\mathbf{x}_{\mathbf{r},i}}^2 + \frac{\delta_\lambda}{4!} \phi_{\mathbf{x}_{\mathbf{r},i}}^4 \nonumber\\
    & \quad \bra{0}\hat{\mathcal{H}}_r\ket{0} \bigg] \Bigg\}.\label{eq:Final interacting scalar field partition function}
\end{align}
\end{widetext}

\leavevmode\thispagestyle{empty}\clearpage
\section{Conclusions and Outlook}\label{Discussion}
In this work, we systematically derived the path integral formulation of the thermal partition function of a scalar quantum field theory from first principles. We started by deriving the path integral formulation of the partition function for a free scalar field theory in two ways. First, we wrote the Hamiltonian in terms of creation and annihilation operators, making sure to keep the vacuum energy contribution. Second, we kept the Hamiltonian in terms of the field operators, where the vacuum energy contributions are not explicit at the level of the Hamiltonian. Crucial for these derivations, we used well-defined field theoretic coherent states, as opposed to the ill-defined unities comprised of the eigenstates of the field operators used in all other derivations existing in the literature. For the former derivation, we made use of coherent states in the ladder operator basis; for the latter derivation, we made use of coherent states in the field operator basis. We provided an explicit mapping between these two bases for coherent states; we are unaware of any other reference that shows such a connection. In the ladder operator basis, no additional terms appear from the matrix element of the Hamiltonian with the coherent states. In the field operator basis, the vacuum energy contribution emerges from the matrix element of the Hamiltonian with the coherent states. Thus both methods yielded the identical well-defined expression of a finite product of integrals over real numbers for the path integral representation of the partition function in scalar TFT. This formula included the usual Euclidean action seen in the literature, but also included the explicit VEV of the Hamiltonian.

We then computed the partition function in $\phi^4$ theory using the field-basis coherent state method we developed for the free theory case. We found the usual Euclidean action, the VEV of the Hamiltonian, and yet another VEV that, this time, coupled to the mass term of the theory. Both of these vacuum terms are absent in the literature.

Since the usual derivation in the literature misses these VEVs, one wonders where the usual derivation breaks down. We expect that the culprit lies in the assumption that unity can be decomposed as $\mathbbm{1} \equiv \int d\phi \ketbra{\phi}{\phi}$ and $\mathbbm{1} \equiv \int d\pi \ketbra{\pi}{\pi}$. In particular, the integral over $\phi$ appears ill-defined; we have not been able to find in the literature any rigorous construction of the identity operator in terms of these field eigenstates.

It would be interesting to examine whether the additional VEVs we discovered in the partition function have physical consequences. First we discuss the VEV of the Hamiltonian. Naively, one might expect that such a VEV will not have any physical consequences. However, a common understanding of the Casimir effect \cite{Casimir:1948dh,Mogliacci:2018oea} is that the effect is due to a difference in the vacuum energy of the confined theory compared to the theory in infinite volume; thus the VEV of the Hamiltonian might be important for the Casimir effect even in a free theory. Second, the VEV coupling to the mass term implies that the observed mass of the particles in the theory will depend on the size of the system in which the theory is confined.

The surprising extra VEV coupling to the mass term of the path integral formulation of the partition function for the interacting thermal field theory suggests the need to re-evaluate the path integral in the usual in/out formalism used to calculate vacuum scattering amplitudes. Presumably a similar coupling between the VEV and the mass term occurs in this in/out formalism. Since the coupling to the mass term involves the mass, the mass counter term, the coupling, and the coupling counter term, \cref{eq:Final interacting scalar field partition function}, we expect additional infinities that need to be absorbed in both the mass \emph{and} coupling counter terms. Such a change to the theory is likely harmless in the infinite volume limit. However, putting the theory in a finite-sized box will alter the mass of the particles and affect scattering cross sections, \cite{Horowitz:2022rpp}. That an infinity is induced by the presence of the coupling counter term suggests that computing an amplitude at NNLO would be very interesting.

We may straightforwardly extend our results beyond the partition function to the thermal average of an operator $\hat{\mathcal{O}}$ at time $t$:

\begin{align}
    \langle\hat{\mathcal{O}}\rangle(t)
    = & \Tr \left[ \hat{\mathcal{O}}_H(t)e^{-\beta \hat H_{full}[\hat\phi_H(t,\mathbf x_{\mathbf r})]} \right] \nonumber\\
    = & \sum_{n}\bra{n} e^{-\beta \hat{H}_\text{full}[\hat{\phi}_H(t,\mathbf{x_r})]} \hat{\mathcal{O}}_H(t) \ket{n}\nonumber\\
    = & \sum_{n}\bra{n} \hat{U}_I(t_0,t)\hat{U}_I(t,t_0)e^{-\beta \hat{H}_\text{full}[\hat{\phi}_H(t,\mathbf{x_r})]} \nonumber\\
    & \quad \hat{U}_I(t_0,t) \hat{U}_I(t,t_0) \hat{\mathcal{O}}_H(t) \ket{n}\nonumber\\
    = & \sum_{n} \bra{n}\hat{U}_I(t_0,t) \prod_{\mathbf{k} =-\frac{\mathbf{K}}{2}}^\frac{\mathbf{K}}{2} \int d\alpha^*_\mathbf{p_k} d\alpha_\mathbf{p_k} \ket{\boldsymbol{\alpha}(t)}\nonumber\\
    & \quad \bra{\boldsymbol{\alpha}(t)} \hat{U}_I(t,t_0) e^{-\beta \hat{H}_\text{full}[\hat{\phi}_H(t,\mathbf{x_r})]} \nonumber\\
    & \quad \hat{U}_I(t_0,t) \hat{U}_I(t,t_0) \hat{\mathcal{O}}_H(t) \ket{n}\nonumber\\
    = & \prod_{\mathbf{k} =-\frac{\mathbf{K}}{2}}^\frac{\mathbf{K}}{2} \int d\alpha^*_\mathbf{p_k} d\alpha_\mathbf{p_k} \bra{\boldsymbol{\alpha}(t)}\nonumber\\
    & \quad \hat{U}_I(t,t_0)e^{-\beta \hat{H}_\text{full}[\hat{\phi}_H(t,\mathbf{x_r})]} \hat{U}_I(t_0,t) \nonumber\\
    & \quad  \sum_{n}\ketbra{n}{n}\hat{U}_I(t,t_0) \hat{\mathcal{O}}_H(t)\hat{U}_I(t_0,t)\ket{\boldsymbol{\alpha}(t)}\nonumber\\
    = & \prod_{\mathbf{k} =-\frac{\mathbf{K}}{2}}^\frac{\mathbf{K}}{2} \int d\alpha^*_\mathbf{p_k} d\alpha_\mathbf{p_k}\bra{\boldsymbol{\alpha}(t)}\nonumber\\
    & \quad \hat{U}_I(t,t_0) e^{-\beta \hat{H}_\text{full}[\hat{\phi}_H(t,\mathbf{x_r})]}\hat{U}_I(t_0,t)\nonumber\\
    & \quad \hat{U}_I(t,t_0) \hat{\mathcal{O}}_H(t) \hat{U}_I(t_0,t)\ket{\boldsymbol{\alpha}(t)}\nonumber
\end{align}

\begin{align}
    = & \prod_{\mathbf{k} =-\frac{\mathbf{K}}{2}}^\frac{\mathbf{K}}{2} \int d\alpha^*_\mathbf{p_k} d\alpha_\mathbf{p_k}\bra{\boldsymbol{\alpha}(t)} e^{-\beta \hat{H}_\text{full}[\hat{\phi}_I(t,\mathbf{x_r})]}\nonumber\\
    & \quad \hat{\mathcal{O}}_I(t) \ket{\boldsymbol{\alpha}(t)}.\label{eq:Thermal average of operator}
\end{align}

All the objects in \cref{eq:Thermal average of operator} are well defined and one can, in principle, evaluate \cref{eq:Thermal average of operator}. However, to make further progress, one needs the specific form of the operator $\hat{\mathcal{O}}_I(t)$ and the Hamiltonian.

Looking ahead, our results provide a foundation for a controlled computation of higher order corrections to thermodynamic properties such as the pressure, energy density, and entropy density in infinite- and finite-sized systems. Thus we may systematically investigate order-by-order how finite-volume effects modify thermodynamic quantities such as the equation of state and the trace anomaly in scalar field theories \cite{Horowitz:2021dmr,Horowitz:2022rpp}. These findings will serve as a valuable testing ground before extending the framework to fermionic fields and to gauge theories. A natural next step would be to carry out the analogous derivation of the path integral formulation of the partition function in Quantum Electro-Dynamics (QED) and eventually in QCD. With the path integral formulation of the partition function of QCD in hand, one could then compute the finite-volume corrections to the QCD equation of state and its associated trace anomaly, with direct implications for lattice QCD \cite{Umeda:2008zz, Bazavov:2009zn, Huovinen:2009yb} and signals of QGP formation in small systems. 
Such corrections may influence observables sensitive to conformal symmetry breaking \cite{Belitsky:2002su,Attems:2016ugt}, such as flow in small heavy-ion collision systems \cite{Yan:2017jcd} and energy-energy correlators \cite{Hofman:2008ar,Chen:2020vvp,Andres:2022ovj,Budhraja:2024tev}, which are currently being explored as precision probes of the QGP.

\begin{acknowledgments}
All authors wish to thank Charles Gale, Sangyong Jeon, Rob Pisarski, Alexander Rothkopf, and Matt Sievert for insightful discussions. RR thanks Isobel Kolb\'e for her valuable insights related to this research. RR and WAH gratefully acknowledge support from the South African National Research Foundation and the SA-CERN Collaboration. RR further acknowledges support from the Institute for Theoretical Physics Amsterdam (ITFA).
\end{acknowledgments}

\section*{Author Contributions}
\begin{itemize}
    \item RR performed the research and calculations, wrote the first draft of the manuscript, and edited the manuscript. 
    \item MA co-supervised the project, provided crosschecks, and edited the manuscript.
    \item WAH conceived and supervised the project, checked the derivations, and edited the manuscript.
\end{itemize}

\appendix
\section{Kronecker Delta Fourier Decomposition on a Lattice}\label{Kronecker Delta Function Proof}
In this appendix, we derive the identity 
\begin{equation}
    \frac{1}{K+1} \sum_{k=0}^{K} e^{\frac{2\pi i k (r - s)}{K+1}} = \delta_{r,s} = \left\{
    \begin{array}{ll}
    1 & \text{for } r=s \\
    0 & \text{for } r \neq s
    \end{array}
    \right. ,\label{eq:fourier-delta}
\end{equation}
which expresses the Kronecker delta as a discrete Fourier sum. We distinguish two cases:

\begin{itemize}
    \item Case 1: $r = s$
    
    In this case, the exponent becomes zero and every term in the sum is one, such that
    \begin{align}
        \delta_{r,r} = \frac{1}{K+1} \sum_{k=0}^{K} e^0 = \frac{1}{K+1} \cdot (K+1) = 1.\nonumber
    \end{align}
    \item Case 2: $r \neq s$
    
    Define \( \ell \equiv r - s \neq 0 \), and consider the geometric series:
    \begin{align}
        \sum_{k=0}^{K} e^{\frac{2\pi i k \ell}{K+1}} = \frac{1 - e^{2\pi i \ell}}{1 - e^{\frac{2\pi i \ell}{K+1}}}.
    \end{align}
    Since \( \ell \in \mathbb{Z} \setminus \{0\} \), we have $e^{2\pi i \ell} = 1$, so the numerator vanishes
    \begin{align}
        \sum_{k=0}^{K} e^{\frac{2\pi i k \ell}{K+1}} = \frac{1 - 1}{1 - e^{\frac{2\pi i \ell}{K+1}}} = 0.\nonumber
    \end{align}
    Note that the denominator would also vanish if $\frac{\ell}{K+1} \in \mathbb{Z}$. However, since $r,s \in [0, K]$, we have $\abs{r-s} \leq K$, and thus $\frac{\ell}{K+1} < 1$. Hence, the denominator does not vanish, and we find
    \begin{align}
        \delta_{r \neq s} &= 0.
    \end{align}
    \end{itemize}
This completes the proof of the identity stated in \cref{eq:fourier-delta}.

\section{Discrete Spatial Derivative Fourier Decomposition}\label{Discrete Spatial Derivative Fourier Decomposition}
Consider a Fourier exponent: $f(\mathbf{x}_\mathbf{r},\mathbf{p}_\mathbf{k}) \equiv e^{\mathrm{i}\mathbf{p_k} \cdot \mathbf{x}_\mathbf{r}}$. Taking two discrete spatial derivatives yields 
\begin{align}
    &\nabla^2_{\mathbf{x}_\mathbf{r}} f(\mathbf{x}_\mathbf{r},\mathbf{p}_\mathbf{k}) \nonumber\\
    &= \sum_{j=1}^n \frac{f(\mathbf{x}_\mathbf{r} + \Delta x \, \hat{e}_j,\mathbf{p}_\mathbf{k}) - 2f(\mathbf{x}_\mathbf{r},\mathbf{p}_\mathbf{k}) + f(\mathbf{x}_\mathbf{r} - \Delta x\, \hat{e}_j,\mathbf{p}_\mathbf{k})}{(\Delta x\, \hat{e}_j)^2} \nonumber \\
    &= \sum_{j=1}^n \frac{e^{\mathrm{i} \mathbf{p_k} \cdot (\mathbf{x}_\mathbf{r} + \Delta x\, \hat{e}_j)} - 2e^{\mathrm{i} \mathbf{p_k} \cdot \mathbf{x}_\mathbf{r}} + e^{\mathrm{i} \mathbf{p_k} \cdot (\mathbf{x}_\mathbf{r} - \Delta x\, \hat{e}_j)}}{(\Delta x\, \hat{e}_j)^2} \nonumber \\
    &= \sum_{j=1}^n \frac{e^{\mathrm{i} \mathbf{p_k} \cdot \mathbf{x}_\mathbf{r}} \left(e^{\mathrm{i} p_{k_j} \Delta x} + e^{-\mathrm{i} p_{k_j} \Delta x} - 2 \right)}{(\Delta x\, \hat{e}_j)^2} \nonumber \\
    &= \sum_{j=1}^n \frac{e^{\mathrm{i} \mathbf{p_k} \cdot \mathbf{x}_\mathbf{r}} \left(2 \cos(p_{k_j} \Delta x) - 2 \right)}{(\Delta x\, \hat{e}_j)^2}.\nonumber
\end{align}
Applying a Taylor expansion to the cosine in the limit where $K \rightarrow \infty$ and thus $\Delta x \rightarrow 0$, keeping terms up to leading order, we obtain:
\begin{equation}
    2\left(\cos{\left(p_{k_j} \Delta x\, \hat{e}_j\right)}-1\right) \approx -\left(p_{k_j} \Delta x\, \hat{e}_j\right)^2,\nonumber
\end{equation}
such that
\begin{align}
    \nabla^2_{\mathbf{x}_\mathbf{r}} f(\mathbf{x}_\mathbf{r},\,\mathbf{p}_\mathbf{k}) &= -\frac{\left(\mathbf{p_k} \Delta x\right)^2}{(\Delta x)^2} e^{\mathrm{i}\mathbf{p_k} \cdot \mathbf{x}_\mathbf{r}}=-\mathbf{p_k}^2e^{\mathrm{i}\mathbf{p_k} \cdot \mathbf{x}_\mathbf{r}},\nonumber
\end{align}
thus establishing the replacement:\\ $\mathbf{\nabla^2_{x_r}}f(\mathbf{x}_\mathbf{r},\,\mathbf{p}_\mathbf{k}) \rightarrow -\mathbf{p}^2_\mathbf{k}f(\mathbf{x}_\mathbf{r},\,\mathbf{p}_\mathbf{k})$.

\bigskip
\noindent
On the exact same grounds, we may perform the substitution: $\partial_{t_a}^2 f(t_a,\,\omega_m) \rightarrow -\omega_m^2 f(t_a,\,\omega_m)$.

\section{Discretized Canonical Commutation Relations}\label{Discretized Canonical Commutation relations}
We present here a derivation showing that the equal-time canonical commutation relations in \cref{eq:Canonical equal time commutation relations} require that the creation and annihilation operators satisfy the standard commutation relations from
\cref{eq:real scalar ladder operator commutation relation 1,eq:real scalar ladder operator commutation relation 2}. By inserting the field operator mode expansion of \cref{eq:Free Heisenberg picture field mode expansion} and the creation and annihilation operator commutation relations of \cref{eq:real scalar ladder operator commutation relation 1,eq:real scalar ladder operator commutation relation 2} into the commutator, we obtain
\begin{align}
    &[\hat{\phi}_{\mathbf{x}_{\mathbf{r},a}}, \hat{\pi}_{\mathbf{x}_{\mathbf{s},a}}]\nonumber\\
    &= \Delta p^{2n} \sum_{\mathbf{k}, \mathbf{q} = -\frac{\mathbf{K}}{2}}^\frac{\mathbf{K}}{2} \frac{\mathrm{i}}{(2\pi)^{2n} 4 E_\mathbf{k}} \left(e^{-\mathrm{i}(E_\mathbf{k}-E_\mathbf{q})t_a}e^{\mathrm{i}(\mathbf{p_k} \cdot \mathbf{x_r} - \mathbf{p_q} \cdot \mathbf{x_s})} \right. \nonumber\\
    & \left. \qquad \left(\left[\hat{a}_{\mathbf{p_k}}, \hat{a}^\dagger_{\mathbf{p_q}}\right] +\left[\hat{a}_{\mathbf{p_k}}, \hat{a}_{\mathbf{p_q}}\right] \right) - e^{\mathrm{i}(E_\mathbf{k}-E_\mathbf{q})t_a} \right. \nonumber\\
    & \left. \qquad e^{-\mathrm{i}(\mathbf{p_k} \cdot \mathbf{x_r} - \mathbf{p_q} \cdot \mathbf{x_s})} \left(\left[\hat{a}^\dagger_{\mathbf{p_k}}, \hat{a}_{\mathbf{p_q}}\right] + \left[\hat{a}^\dagger_{\mathbf{p_k}}, \hat{a}^\dagger_{\mathbf{p_q}}\right]\right)\right)\nonumber\\
    &= \Delta p^{2n} \sum_{\mathbf{k}, \mathbf{q} = -\frac{\mathbf{K}}{2}}^\frac{\mathbf{K}}{2} \frac{\mathrm{i}}{(2\pi)^{2n} 4 E_\mathbf{k}}\nonumber\\
    & \qquad \left(e^{-\mathrm{i}(E_\mathbf{k}-E_\mathbf{q})t_a}e^{\mathrm{i}(\mathbf{p_k} \cdot \mathbf{x_r} - \mathbf{p_q} \cdot \mathbf{x_s})}(2\pi)^n 2E_{\mathbf{k}} \frac{1}{\Delta p^n} \delta_{\mathbf{k}, \mathbf{q}} \right. \nonumber\\
    & \qquad \left.+ e^{\mathrm{i}(E_\mathbf{k}-E_\mathbf{q})t_a}e^{-\mathrm{i}(\mathbf{p_k} \cdot \mathbf{x_r} - \mathbf{p_q} \cdot \mathbf{x_s})} (2\pi)^n 2E_{\mathbf{k}} \frac{1}{\Delta p^n} \delta_{\mathbf{k}, \mathbf{q}} \right)\nonumber\\
    &= \frac{\mathrm{i}\Delta p^n}{2(2\pi)^n} \sum_{\mathbf{k}=-\frac{\mathbf{K}}{2}}^\frac{\mathbf{K}}{2}  \left(e^{\mathrm{i}\mathbf{p_k} \cdot (\mathbf{x_r} -\mathbf{x_s})} + e^{-\mathrm{i}\mathbf{p_k} \cdot ( \mathbf{x_r} - \mathbf{x_s})} \right)\nonumber\\
    &= \frac{\mathrm{i}N}{V} \delta_{\mathbf{r},\mathbf{s}}\nonumber\\
    &= \frac{\mathrm{i}}{\Delta x^n}\delta_{\mathbf{r},\mathbf{s}}.\label{eq:Cannonical commutation relations proof}
\end{align}

\section{Klein Gordon Hamiltonian}\label{Klein Gordon Hamiltonian}
In this appendix, we express the Hamiltonian operator in terms of creation and annihilation operators using \cref{eq:real scalar ladder operator commutation relation 2,eq:real scalar ladder operator commutation relation 1,eq:Free Heisenberg picture field mode expansion}:
\begin{widetext}
\begin{align}
    \hat{H}&= \frac{\Delta x^n}{2}\sum_{\mathbf{r}=\mathbf{0}}^\mathbf{K} \left[\hat{\pi}_{\mathbf{x}_{\mathbf{r},a}}^2 + \left(\mathbf{\nabla}_\mathbf{x_r}\cdot\hat{\phi}_{\mathbf{x}_{\mathbf{r},a}}\right)^2 + m^2\hat{\phi}_{\mathbf{x}_{\mathbf{r},a}}^2\right]\nonumber\\
    &= \frac{\Delta x^n}{2} \left(\frac{ \Delta p}{2\pi}\right)^{2n}  \sum_{\mathbf{r}=\mathbf{0}}^\mathbf{K} \sum_{\mathbf{q},\mathbf{k}=-\frac{\mathbf{K}}{2}}^\frac{\mathbf{K}}{2}\left[-\frac{1}{4}\left( \hat{a}^{\dagger}_{\mathbf{p_q}} e^{\mathrm{i}p_q \cdot x_r} - \hat{a}_{\mathbf{p_q}} e^{-\mathrm{i}p_q \cdot x_r} \right)\left( \hat{a}^{\dagger}_{\mathbf{p_k}} e^{\mathrm{i}p_k \cdot x_r} - \hat{a}_{\mathbf{p_k}} e^{-\mathrm{i}p_k \cdot x_r} \right)\right.\nonumber\\
    &\left.\qquad+\frac{\mathrm{i}\left(-\mathrm{i}\right)\left(\mathbf{p_q}\cdot\mathbf{p_k}+m^2\right)}{4E_{\mathbf{k}}E_{\mathbf{q}}} \left( \hat{a}^{\dagger}_{\mathbf{p_q}} e^{\mathrm{i}p_q \cdot x_r} + \hat{a}_{\mathbf{p_q}} e^{-\mathrm{i}p_q \cdot x_r} \right)\left( \hat{a}^{\dagger}_{\mathbf{p_k}} e^{\mathrm{i}p_k \cdot x_r} + \hat{a}_{\mathbf{p_k}} e^{-\mathrm{i}p_k \cdot x_r} \right)\right]\nonumber\\
    &=\frac{\Delta x^n}{8} \left(\frac{ \Delta p}{2\pi}\right)^{2n} \sum_{\mathbf{r}=\mathbf{0}}^\mathbf{K} \sum_{\mathbf{q},\mathbf{k}=-\frac{\mathbf{K}}{2}}^\frac{\mathbf{K}}{2} \left[\hat{a}^\dagger_{\mathbf{p_q}}\hat{a}_{\mathbf{p_k}} e^{\mathrm{i}(p_q-p_k) \cdot x_r} + \hat{a}_{\mathbf{p_q}}\hat{a}^\dagger_{\mathbf{p_k}} e^{-\mathrm{i}(p_q-p_k) \cdot x_r}\nonumber- \hat{a}^\dagger_{\mathbf{p_q}}\hat{a}^\dagger_{\mathbf{p_k}} e^{\mathrm{i}(p_q+p_k) \cdot x_r} \right. \nonumber\\
    & \qquad - \hat{a}_{\mathbf{p_q}}\hat{a}_{\mathbf{p_k}} e^{-\mathrm{i}(p_q+p_k) \cdot x_r}
    + \frac{\mathbf{p_q} \cdot \mathbf{p_k}+m^2}{E_\mathbf{q}E_\mathbf{k}}\left(\hat{a}^\dagger_{\mathbf{p_q}}\hat{a}_{\mathbf{p_k}} e^{\mathrm{i}(p_q-p_k) \cdot x_r} + \hat{a}_{\mathbf{p_q}}\hat{a}^\dagger_{\mathbf{p_k}} e^{-\mathrm{i}(p_q-p_k) \cdot x_r} \right. \nonumber\\
    & \qquad \left. \left. + \hat{a}^\dagger_{\mathbf{p_q}}\hat{a}^\dagger_{\mathbf{p_k}} 
    e^{\mathrm{i}(p_q+p_k) \cdot x_r} + \hat{a}_{\mathbf{p_q}}\hat{a}_{\mathbf{p_k}} e^{-\mathrm{i}(p_q+p_k) \cdot x_r}\right)\right]\nonumber\\
    &= \frac{1}{8} \left(\frac{ \Delta p}{2\pi}\right)^{2n} \left(\frac{2\pi}{\Delta p}\right)^n \sum_{\mathbf{q},\mathbf{k}=-\frac{\mathbf{K}}{2}}^\frac{\mathbf{K}}{2} \delta_{\mathbf{q},\mathbf{k}}\left[\hat{a}^\dagger_{\mathbf{p_q}}\hat{a}_{\mathbf{p_k}} + \hat{a}_{\mathbf{p_q}}\hat{a}^\dagger_{\mathbf{p_k}}-\hat{a}^\dagger_{\mathbf{p_q}}\hat{a}^\dagger_{-\mathbf{p_k}} - \hat{a}_{\mathbf{p_q}}\hat{a}_{-\mathbf{p_k}}\right.\nonumber \\
    & \qquad + \left.\frac{\mathbf{p_q} \cdot \mathbf{p_k}+m^2}{E_\mathbf{q}E_\mathbf{k}}\left(\hat{a}^\dagger_{\mathbf{p_q}}\hat{a}_{\mathbf{p_k}}+ \hat{a}_{\mathbf{p_q}}\hat{a}^\dagger_{\mathbf{p_k}}+\hat{a}^\dagger_{\mathbf{p_q}}\hat{a}^\dagger_{-\mathbf{p_k}}+ \hat{a}_{\mathbf{p_q}}\hat{a}_{-\mathbf{p_k}}\right)\right]\nonumber\\
    &=\frac{1}{2}\Delta p^n \sum_{\mathbf{k}=-\frac{\mathbf{K}}{2}}^\frac{\mathbf{K}}{2} \frac{1}{(2\pi)^n2E_\mathbf{k}}E_\mathbf{k}\left[\hat{a}^\dagger_{\mathbf{p_k}}\hat{a}_{\mathbf{p_k}}+\hat{a}_{\mathbf{p_k}}\hat{a}^\dagger_{\mathbf{p_k}}+\left[\hat{a}^\dagger_{\mathbf{p_k}},\hat{a}^\dagger_{\mathbf{p_k}}\right]+\left[\hat{a}_{\mathbf{p_k}},\hat{a}_{\mathbf{p_k}}\right]\right]\nonumber\\
    &= \Delta p^n \sum_{\mathbf{k}=-\frac{\mathbf{K}}{2}}^\frac{\mathbf{K}}{2} \frac{1}{(2\pi)^n E_\mathbf{k}}E_\mathbf{k} \left[\hat{a}^\dagger_{\mathbf{p_k}}\hat{a}_{\mathbf{p_k}}+\frac{1}{2}\left[\hat{a}_{\mathbf{p_k}},\hat{a}^\dagger_{\mathbf{p_k}}\right]\right]\nonumber\\
    &= \Delta p^n \sum_{\mathbf{k}=-\frac{\mathbf{K}}{2}}^\frac{\mathbf{K}}{2} \frac{1}{(2\pi)^n 2E_\mathbf{k}}E_\mathbf{k}\left[\hat{a}^\dagger_{\mathbf{p_k}}\hat{a}_{\mathbf{p_k}}+\left(\frac{2\pi}{\Delta p}\right)^n E_\mathbf{k}\right].\label{eq:Hamiltonian in terms of ladder operators}
\end{align}
\end{widetext}
The second term in the parentheses of \cref{eq:Hamiltonian in terms of ladder operators} represents the vacuum energy, arising from the zero-point energy. This vacuum term reflects the constant energy of the ground state and in field theory is commonly discarded; we will explicitly keep the term in our work.

\section{Partition Function Reality Condition}\label{Partition Function Reality Condition}
The partition function is a real quantity, which means it is self conjugate, $Z^* = Z$. We will show that the reality condition of the partition function results in the following condition:
\begin{align}
    \sum_{i=1}^{N}  \sum_{\mathbf{k}=-\frac{\mathbf{K}}{2}}^\frac{\mathbf{K}}{2}\abs{\alpha_{\mathbf{p}_{\mathbf{k},i}}}^2 &= \sum_{i=1}^{N}  \sum_{\mathbf{k}=-\frac{\mathbf{K}}{2}}^\frac{\mathbf{K}}{2} \alpha^*_{\mathbf{p}_{\mathbf{k},i}}\alpha_{\mathbf{p}_{\mathbf{k},i-1}}\nonumber\\
    &= \sum_{i=1}^{N}  \sum_{\mathbf{k}=-\frac{\mathbf{K}}{2}}^\frac{\mathbf{K}}{2} \alpha_{\mathbf{p}_{\mathbf{k},i}}\alpha^*_{\mathbf{p}_{\mathbf{k},i-1}}.\label{eq:Partition function reality condition}
\end{align}
We can obtain \cref{eq:Partition function reality condition} by considering the complex conjugate of $Z$ in \cref{eq:Finite size discrete real free scalar field partitition function alpha}, which yields
\begin{align}
    Z^* =&\left(\prod_{i=1}^{N} \prod_{\mathbf{k}=-\frac{\mathbf{K}}{2}}^\frac{\mathbf{K}}{2}\frac{1}{2\pi VE_\mathbf{k}}\int_{-\infty}^\infty da_{\mathbf{p}_{\mathbf{k},i}}\int_{-\infty}^\infty db_{\mathbf{p}_{\mathbf{k},i}}\right) \nonumber\\
    & \ \exp\bigg\{-\Delta \tau \Delta p^n  \sum_{i=1}^{N}  \sum_{\mathbf{k}=-\frac{\mathbf{K}}{2}}^\frac{\mathbf{K}}{2}\left[\frac{1}{(2\pi)^n 2E_\mathbf{k}} \right.\nonumber\\
    & \ \left. \left(\alpha_{\mathbf{p}_{\mathbf{k},i}}(\alpha^*_{\mathbf{p}_{\mathbf{k},i}})'+E_\mathbf{k}\alpha_{\mathbf{p}_{\mathbf{k},i}}\alpha^*_{\mathbf{p}_{\mathbf{k},i-1}}\right)+\frac{1}{\Delta p^n} \frac{E_\mathbf{k}}{2}\right]\bigg\}.\label{eq:Complex conjugate partition function}
\end{align}
For $Z = Z^*$, the exponential term in the partition function of \cref{eq:Finite size discrete real free scalar field partitition function alpha} and its complex conjugate of \cref{eq:Complex conjugate partition function} must satisfy the condition
\begin{align}
    \Delta\tau  \Delta p^n  \sum_{i=1}^{N} & \sum_{\mathbf{k}=-\frac{\mathbf{K}}{2}}^\frac{\mathbf{K}}{2}\bigg[\alpha^*_{\mathbf{p}_{\mathbf{k},i}}\frac{\alpha_{\mathbf{p}_{\mathbf{k},i}}-\alpha_{\mathbf{p}_{\mathbf{k},i-1}}}{\Delta\tau}\nonumber\\
    & + E_\mathbf{k} \alpha^*_{\mathbf{p}_{\mathbf{k},i}}\alpha_{\mathbf{p}_{\mathbf{k},i-1}}\bigg] \nonumber\\
    = \Delta \tau \Delta p^n  \sum_{i=1}^{N} & \sum_{\mathbf{k}=-\frac{\mathbf{K}}{2}}^\frac{\mathbf{K}}{2}\bigg[\alpha_{\mathbf{p}_{\mathbf{k},i}}\frac{\alpha^*_{\mathbf{p}_{\mathbf{k},i}}-\alpha^*_{\mathbf{p}_{\mathbf{k},i-1}}}{\Delta \tau}\nonumber\\
    & + E_\mathbf{k} \alpha_{\mathbf{p}_{\mathbf{k},i}}\alpha^*_{\mathbf{p}_{\mathbf{k},i-1}}\bigg].\label{eq:Reality condition partition function identity 2}
\end{align}
Rewriting \cref{eq:Reality condition partition function identity 2} yields
\begin{align}
    & \sum_{i=1}^{N}  \sum_{\mathbf{k}=-\frac{\mathbf{K}}{2}}^\frac{\mathbf{K}}{2}\left[\abs{\alpha_{\mathbf{p}_{\mathbf{k},i}}}^2+(\Delta \tau E_\mathbf{k}-1)\alpha^*_{\mathbf{p}_{\mathbf{k},i}}\alpha_{\mathbf{p}_{\mathbf{k},i-1}}\right]\nonumber\\
    &= \sum_{i=1}^{N}  \sum_{\mathbf{k}=-\frac{\mathbf{K}}{2}}^\frac{\mathbf{K}}{2} \left[\abs{\alpha_{\mathbf{p}_{\mathbf{k},i}}}^2+(\Delta \tau E_\mathbf{k}-1)\alpha_{\mathbf{p}_{\mathbf{k},i}}\alpha^*_{\mathbf{p}_{\mathbf{k},i-1}}\right],\nonumber
\end{align}
which holds if and only if
\begin{align}
    \sum_{i=1}^{N}  \sum_{\mathbf{k}=-\frac{\mathbf{K}}{2}}^\frac{\mathbf{K}}{2} \alpha^*_{\mathbf{p}_{\mathbf{k},i}}\alpha_{\mathbf{p}_{\mathbf{k},i-1}} &= \sum_{i=1}^{N}  \sum_{\mathbf{k}=-\frac{\mathbf{K}}{2}}^\frac{\mathbf{K}}{2} \alpha_{\mathbf{p}_{\mathbf{k},i}}\alpha^*_{\mathbf{p}_{\mathbf{k},i-1}}.\label{eq:Reality condition partition function identity 1}
\end{align}
To prove \cref{eq:Partition function reality condition}, we rely on the discrete analogue of integration by parts, known as summation by parts. Summation by parts, along with some of its consequences, is presented in \cref{Summation by Parts with Periodic Boundaries}. By combining \cref{eq:Reality condition partition function identity 1} with \cref{eq:Summation by parts}, we are now in a position to demonstrate \cref{eq:Partition function reality condition}:

\begin{align}
    \sum_{i=1}^{N} & \sum_{\mathbf{k}=-\frac{\mathbf{K}}{2}}^\frac{\mathbf{K}}{2} \alpha^*_{\mathbf{p}_{\mathbf{k},i}}\alpha_{\mathbf{p}_{\mathbf{k},i-1}}\nonumber\\
    =& \frac{1}{2} \sum_{i=1}^{N}  \sum_{\mathbf{k}=-\frac{\mathbf{K}}{2}}^\frac{\mathbf{K}}{2} \left[\alpha_{\mathbf{p}_{\mathbf{k},i}}\alpha^*_{\mathbf{p}_{\mathbf{k},i-1}}+\alpha^*_{\mathbf{p}_{\mathbf{k},i}}\alpha_{\mathbf{p}_{\mathbf{k},i-1}}\right]\nonumber\\
    =& \frac{1}{2} \sum_{i=1}^{N}  \sum_{\mathbf{k}=-\frac{\mathbf{K}}{2}}^\frac{\mathbf{K}}{2} \left[\alpha_{\mathbf{p}_{\mathbf{k},i}}\left(\alpha^*_{\mathbf{p}_{\mathbf{k},i}}-\Delta\tau\frac{\alpha^*_{\mathbf{p}_{\mathbf{k},i}}-\alpha^*_{\mathbf{p}_{\mathbf{k},i-1}}}{\Delta \tau}\right)\right.\nonumber\\
    & \left. \quad + \alpha^*_{\mathbf{p}_{\mathbf{k},i}}\left(\alpha_{\mathbf{p}_{\mathbf{k},i}}-\Delta\tau\frac{\alpha_{\mathbf{p}_{\mathbf{k},i}}-\alpha_{\mathbf{p}_{\mathbf{k},i-1}}}{\Delta \tau}\right)\right]\nonumber\\
    =& \sum_{i=1}^{N}  \sum_{\mathbf{k}=-\frac{\mathbf{K}}{2}}^\frac{\mathbf{K}}{2} \left[\abs{\alpha_{\mathbf{p}_{\mathbf{k},i}}}^2 - \frac{\Delta\tau}{2} \alpha_{\mathbf{p}_{\mathbf{k},i}} \left(\frac{\alpha^*_{\mathbf{p}_{\mathbf{k},i}}-\alpha^*_{\mathbf{p}_{\mathbf{k},i-1}}}{\Delta \tau} \right. \right. \nonumber\\
    & \left. \left. \quad -\frac{\alpha^*_{\mathbf{p}_{\mathbf{k},i}}-\alpha^*_{\mathbf{p}_{\mathbf{k},i-1}}}{\Delta \tau}\right)\right] + \alpha^*_{\mathbf{p}_{\mathbf{k},N}}\alpha_{\mathbf{p}_{\mathbf{k},N}}-\alpha^*_{\mathbf{p}_{\mathbf{k},0}}\alpha_{\mathbf{p}_{\mathbf{k},0}}\nonumber\\
    =& \sum_{i=1}^{N}  \sum_{\mathbf{k}=-\frac{\mathbf{K}}{2}}^\frac{\mathbf{K}}{2} \abs{\alpha_{\mathbf{p}_{\mathbf{k},i}}}^2,
\end{align}
where in the third line, summation by parts was applied to the final term from the previous line, using that the boundary terms cancel due to periodic boundary conditions imposed by the trace of the partition function.

\section{Summation by Parts with Periodic Boundaries}\label{Summation by Parts with Periodic Boundaries}
We review the discrete analogue of integration by parts, known as summation by parts, and present some consequences of summation by parts that will be used throughout this manuscript. 

The summation by parts identity takes the form
\begin{align}
    \sum_{i=1}^N a_i(b_i - b_{i-1}) 
    &= a_1 b_1 - a_1 b_0 + \cdots + a_N b_N - a_N b_{N-1} \nonumber\\
    &= a_N b_N - a_0 b_0 - \sum_{i=1}^N b_{i-1}(a_i - a_{i-1}) \nonumber\\
    \therefore \sum_{i=1}^N a_i(b_i - b_{i-1}) &+ \sum_{i=1}^N b_{i-1}(a_i - a_{i-1}) 
    = a_N b_N - a_0 b_0. \label{eq:Summation by parts}
\end{align}
An important consequence of summation by parts is that, for any arbitrary function $f_i$, expressions involving the product of said function and their discrete derivatives (i.e. $\sum_{i=1}^N  f_i(f_i - f_{i-1})$) reduce entirely to boundary terms. Specifically, the interior structure of the sum cancels, and only the contributions at the boundaries remain.
\begin{align}
    \underbrace{\sum_{i=1}^N f_i (f_i - f_{i-1})}_{\text{Backward finite difference}} 
    + \underbrace{\sum_{i=1}^N f_{i-1}(f_i - f_{i-1})}_{\text{Forward finite difference}} 
    = f_N^2 - f_0^2. \label{eq:Forward and backward finite difference}
\end{align}
Taking $N$ to infinty, and thus $\Delta \tau \to 0$, we approach the continuum limit. In this limit, the forward and backward finite differences become identical, allowing us to write
\begin{align}
    \sum_{i=1}^N f_i(f_i-f_{i-1}) &= \frac{1}{2}\left(f_N^2-f_0^2\right) = 0,\label{eq:Summation by parts vanishing boundaries}
\end{align}
Since the thermal trace implies periodicity in the index $i$, any function \( f_i \) appearing under the trace satisfies \cref{eq:Summation by parts vanishing boundaries}.

Using \cref{eq:Summation by parts vanishing boundaries} it can be shown that the following expression vanishes:
\begin{align}
    &\left(\frac{\Delta p}{2\pi}\right)^n \sum_{i=1}^N\sum_{\mathbf{k}=-\frac{\mathbf{K}}{2} }^\frac{\mathbf{K}}{2} \phi^*_{\mathbf{p}_{\mathbf{k},i}}\phi'_{\mathbf{p}_{\mathbf{k},i}}\nonumber\\
    & \ = \left(\frac{\Delta p}{2\pi}\right)^n \Delta x^{2n} \sum_{i=1}^N \sum_{\mathbf{k}=-\frac{\mathbf{K}}{2} }^\frac{\mathbf{K}}{2} \sum_{\mathbf{r},\mathbf{s}=\mathbf{0}}^\mathbf{K} \phi_{\mathbf{x}_{\mathbf{r},i}}\phi'_{\mathbf{x}_{\mathbf{s},i}} e^{\mathrm{i}\mathbf{p_k}(\mathbf{x_r}-\mathbf{x_s})} \nonumber\\
    & \ = \Delta x^n \sum_{i=1}^N \sum_{\mathbf{r},\mathbf{s}=\mathbf{0}}^\mathbf{K} \phi_{\mathbf{x}_{\mathbf{r},i}}\phi'_{\mathbf{x}_{\mathbf{s},i}} \delta_{\mathbf{r},\mathbf{s}} \nonumber\\
    & \ = \Delta x^n \sum_{i=1}^N \sum_{\mathbf{r}=\mathbf{0}}^\mathbf{K} \phi_{\mathbf{x}_{\mathbf{r},i}}\phi'_{\mathbf{x}_{\mathbf{r},i}}\nonumber\\
    & \ =\Delta x^n \sum_{\mathbf{r}=\mathbf{0}}^\mathbf{K} \left(\phi_{\mathbf{x}_{\mathbf{r},N}}^2-\phi_{\mathbf{x}_{\mathbf{r},0}}^2\right)\nonumber\\
    & \ =0.\label{eq:Total field derivative vanish}
\end{align}
Similarly, $\left(\frac{\Delta p}{2\pi}\right)^n  \sum_{i=1}^N\sum_{\mathbf{k}=-\frac{\mathbf{K}}{2} }^\frac{\mathbf{K}}{2} \pi^*_{\mathbf{p}_{\mathbf{k},i}}\pi'_{\mathbf{p}_{\mathbf{k},i}}$ vanishes by applying the same argument.

\begin{widetext}
\section{Coherent States \& Displacement Operators}\label{Coherent states}

\subsection{Coherent State Relations}\label{Coherent State Relations}
To ensure our coherent states from \cref{eq:Coherent state definition} are well defined, we aim to prove the following coherent state relations. For the sake of readability, we will leave the limits of the sums implicit throughout this appendix.
\begin{align}
    \hat{a}_{\mathbf{p_k}}\ket{\boldsymbol{\alpha}}&=\alpha_{\mathbf{p_k}}\ket{\boldsymbol{\alpha}}\nonumber\\ 
    \braket{\tilde{\boldsymbol{\alpha}}}{\boldsymbol{\alpha}}&=e^{\Delta p^n \sum_\mathbf{k} \frac{1}{(2\pi)^n2E_\mathbf{k}}\tilde{\alpha}^*_{\mathbf{p_k}}\alpha_{\mathbf{p_k}}}\nonumber\\ 
    \mathbbm{1}&=\prod_\mathbf{k} \int_{-\infty}^\infty \frac{da_{\mathbf{p_k}} db_{\mathbf{p_k}}}{2\pi VE_\mathbf{k}} e^{\Delta p^n \sum_\mathbf{k} \frac{1}{(2\pi)^n2E_\mathbf{k}}\alpha^*_{\mathbf{p_k}}\alpha_{\mathbf{p_k}}}\ketbra{\boldsymbol{\alpha}}{\boldsymbol{\alpha}}.\nonumber
\end{align}
\begin{enumerate}
    \item Claim: The coherent state is an eigenstate of the annihilation operator
    \begin{align}
        \hat{a}_{\mathbf{p_k}}\ket{\boldsymbol{\alpha}}&=\alpha_{\mathbf{p_k}}\ket{\boldsymbol{\alpha}},\label{eq:Coherent state claim 1}\\
        \bra{\boldsymbol{\alpha}} \hat{a}_{\mathbf{p_k}}^\dagger &= \alpha_{\mathbf{p_k}}^* \bra{\boldsymbol{\alpha}}.\label{eq:Coherent state claim 1 conjugate}
    \end{align}
    Proof:
    \begin{align}
        \hat{a}_{\mathbf{p_k}}\ket{\boldsymbol{\alpha}}&=\hat{a}_{\mathbf{p_k}}e^{\Delta p^n \sum_{\mathbf{q}=-\frac{\mathbf{K}}{2}}^\frac{\mathbf{K}}{2}\frac{1}{(2\pi)^n 2E_\mathbf{q}}\alpha_{\mathbf{p_q}}\hat{a}^\dagger_{\mathbf{p_q}}}\ket{0}\nonumber\\
        &=\hat{a}_{\mathbf{p_k}}\sum_{i=0}^{\infty} \frac{1}{i!}\left(\Delta p^n \sum_{\mathbf{q}=-\frac{\mathbf{K}}{2}}^\frac{\mathbf{K}}{2}\frac{1}{(2\pi)^n 2E_\mathbf{q}}\alpha_{\mathbf{p_q}}\hat{a}^\dagger_{\mathbf{p_q}}\right)^i \ket{0}.\label{eq:relativistic Coherent State expansion}
    \end{align}
    Using $\hat{a}_{\mathbf{p_k}}\ket{0} = 0 \text{ and } \left[\hat{a}_{\mathbf{p_k}},\hat{a}^\dagger_{\mathbf{p_q}}\right] = (2\pi)^n 2E_\mathbf{k}\frac{1}{\Delta p^n}\delta_{\mathbf{k},\mathbf{q}}$, we can write
    \begin{align}
        \hat{a}_{\mathbf{p_k}}\left(\hat{a}^\dagger_{\mathbf{p_q}}\right)^i &= \left(\hat{a}^\dagger_{\mathbf{p_q}}\right)^i\hat{a}_{\mathbf{p_k}}+ \sum_{j=1}^i (2\pi)^n 2E_{\mathbf{q}_j} \frac{1}{\Delta p^n} \delta_{\mathbf{k},\mathbf{q}_j}\left(\hat{a}^\dagger_{\mathbf{p_q}}\right)^{i-1},\label{eq:discretized momentum ladder commutator n times}
    \end{align} 
    and insert \cref{eq:discretized momentum ladder commutator n times} into \cref{eq:relativistic Coherent State expansion} to obtain
    \begin{align}
        \hat{a}_{\mathbf{p_k}}\ket{\boldsymbol{\alpha}} &= \sum_{i=1}^{\infty} \frac{1}{i!} \left[\sum_{j=1}^i\sum_{\mathbf{q}_j} (2\pi)^n  2E_{\mathbf{q}_j} \frac{1}{\Delta p^n}\delta_{\mathbf{k},\mathbf{q}_j} \Delta p^n \frac{\alpha_{\mathbf{p}_{\mathbf{q}_j}}}{(2\pi)^n2E_{\mathbf{q}_j}} \left(\Delta p^n \sum_{\mathbf{q}=-\frac{\mathbf{K}}{2}}^\frac{\mathbf{K}}{2}\frac{1}{(2\pi)^n 2E_{\mathbf{q}}}\alpha_{\mathbf{p_q}}\hat{a}_{\mathbf{p_q}}^\dagger\right)^{i-1}\right] \ket{0}\nonumber\\
        &= \alpha_{\mathbf{p_k}} \sum_{i=1}^{\infty} \frac{i}{i!}\left(\Delta p^n \sum_{\mathbf{q}=-\frac{\mathbf{K}}{2}}^\frac{\mathbf{K}}{2}\frac{1}{(2\pi)^n 2E_{\mathbf{q}}}\alpha_{\mathbf{p_q}}\hat{a}_{\mathbf{p_q}}^\dagger\right)^{i-1} \ket{0} \nonumber\\
        &= \alpha_{\mathbf{p_k}} e^{\Delta p^n \sum_{\mathbf{q}=-\frac{\mathbf{K}}{2}}^\frac{\mathbf{K}}{2}\frac{1}{(2\pi)^n 2E_{\mathbf{q}}}\alpha_{\mathbf{p_q}}\hat{a}_{\mathbf{p_q}}^\dagger} \ket{0} = \alpha_{\mathbf{p_k}} \ket{\boldsymbol{\alpha}}.\nonumber
    \end{align}
    Note that in the first line, the sum over $\mathbf{q}_j$ collapsed due to the Kronecker delta. Similarly,
    \begin{align}
        \bra{\boldsymbol{\alpha}} &\equiv \left( \ket{\boldsymbol{\alpha}}\right)^\dagger\nonumber\\
        &= \bra{0} e^{\Delta p^n \sum_\mathbf{k} \frac{1}{(2\pi)^n2E_\mathbf{k}}\alpha_{\mathbf{p_k}}^* \hat{a}_{\mathbf{p_k}}},\nonumber\\
        \bra{\boldsymbol{\alpha}} \hat{a}_{\mathbf{p_k}}^\dagger &= \bra{\boldsymbol{\alpha}} \alpha_{\mathbf{p_k}}^*.\nonumber
    \end{align}
    \item Claim: The coherent state orthogonality relation is
    \begin{align}
        \braket{\tilde{\boldsymbol{\alpha}}}{\boldsymbol{\alpha}} &= e^{\Delta p^n \sum_\mathbf{k} \frac{1}{(2\pi)^n2E_\mathbf{k}}\tilde{\alpha}_{\mathbf{p_k}}^* \alpha_{\mathbf{p_k}}}.\label{eq:Real Coherent state 3}
    \end{align}
    Proof:
    \begin{align}
            \braket{\tilde{\boldsymbol{\alpha}}}{\boldsymbol{\alpha}} &=\bra{0} e^{\Delta p^n \sum_\mathbf{k} \frac{1}{(2\pi)^n2E_\mathbf{k}}\tilde{\alpha}_{\mathbf{p_k}}^* \hat{a}_{\mathbf{p_k}}} e^{\Delta p^n \sum_\mathbf{q} \frac{1}{(2\pi)^n 2E_{\mathbf{q}}}\alpha_{\mathbf{p_q}}\hat{a}_{\mathbf{p_q}}^\dagger} \ket{0} \nonumber\\
            & =\bra{0}  \left(1+\Delta p^n \sum_\mathbf{k} \frac{\tilde{\alpha}_{\mathbf{p_k}}^* \hat{a}_{\mathbf{p_k}}}{(2\pi)^n2E_\mathbf{k}}+\frac{1}{2!}\left(\Delta p^n \sum_\mathbf{k} \frac{\tilde{\alpha}_{\mathbf{p_k}}^* \hat{a}_{\mathbf{p_k}}}{(2\pi)^n2E_\mathbf{k}}\right)^{2}+\cdots\right)\nonumber\\
            & \quad \left(1+ \Delta p^n \sum_\mathbf{q} \frac{\alpha_{\mathbf{p_q}}\hat{a}_{\mathbf{p_q}}^\dagger}{(2\pi)^n 2E_{\mathbf{q}}}+\frac{1}{2!}\left(\Delta p^n \sum_\mathbf{q} \frac{\alpha_{\mathbf{p_q}}\hat{a}_{\mathbf{p_q}}^\dagger}{(2\pi)^n 2E_{\mathbf{q}}}\right)^{2}+\cdots\right) \ket{0}\nonumber\\
            &= \braket{0}{0} + \Delta p^{2n} \sum_{{\mathbf{k}_1},{\mathbf{q}_1}}\frac{\tilde{\alpha}_{\mathbf{p}_{\mathbf{k}_1}}^* \alpha_{\mathbf{p}_{\mathbf{q}_1}}}{(2\pi)^{2n} 4E_{\mathbf{k}_1}E_{\mathbf{q}_1}} \underbrace{\bra{0}\hat{a}_{\mathbf{p}_{\mathbf{k}_1}}\hat{a}_{\mathbf{p}_{\mathbf{q}_1}}^\dagger \ket{0}}_{\left[\hat{a}_{\mathbf{p}_{\mathbf{k}_1}},\hat{a}_{\mathbf{p}_{\mathbf{q}_1}}^\dagger\right]= (2\pi)^n 2E_{\mathbf{p}_{\mathbf{k}_1}} \frac{1}{\Delta p^n} \delta_{\mathbf{k}_1,\mathbf{q}_1}}\nonumber\\
            & \quad + \frac{1}{2!} \Delta p^{4n} \sum_{{\mathbf{k}_1},{\mathbf{k}_2},{\mathbf{q}_1},{\mathbf{q}_2}}\frac{\tilde{\alpha}_{\mathbf{p}_{\mathbf{k}_1}}^*\tilde{\alpha}_{\mathbf{p}_{\mathbf{k}_2}}^*\alpha_{\mathbf{p}_{\mathbf{q}_1}}\alpha_{\mathbf{p}_{\mathbf{q}_2}}}{(2\pi)^{4n} 16E_{\mathbf{k}_1}E_{\mathbf{k}_2}E_{\mathbf{q}_1}E_{\mathbf{q}_2}}\bra{0}\hat{a}_{\mathbf{p}_{\mathbf{k}_1}}\hat{a}_{\mathbf{p}_{\mathbf{k}_2}}\hat{a}_{\mathbf{p}_{\mathbf{q}_1}}^\dagger\hat{a}_{\mathbf{p}_{\mathbf{q}_2}}^\dagger \ket{0} + \cdots\nonumber\\
            &= 1+\Delta p^n \sum_\mathbf{k} \frac{\tilde{\alpha}_{\mathbf{p_k}}^* \hat{a}_{\mathbf{p_k}}}{(2\pi)^n2E_\mathbf{k}}+\frac{1}{2!}\left(\Delta p^n \sum_\mathbf{k} \frac{\tilde{\alpha}_{\mathbf{p_k}}^* \hat{a}_{\mathbf{p_k}}}{(2\pi)^n2E_\mathbf{k}}\right)^{2}+\cdots \nonumber\\
            &= e^{\Delta p^n \sum_\mathbf{k} \frac{1}{(2\pi)^n2E_\mathbf{k}}\tilde{\alpha}_{\mathbf{p_k}}^* \alpha_{\mathbf{p_k}}}.
\end{align}
Our coherent states are thus not pair-wise orthogonal. 
\item Claim:
\begin{align}
    \mathbbm{1} &= N\prod_\mathbf{k}\int_{\alpha_{\mathbf{p_k}}\mathcal{2}\mathbbm{C}} d \alpha_{\mathbf{p_k}}^* d \alpha_{\mathbf{p_k}} e^{-\Delta p^n \sum_\mathbf{k} \frac{1}{(2\pi)^n2E_\mathbf{k}}\tilde{\alpha}_{\mathbf{p_k}}^* \alpha_{\mathbf{p_k}}} \ketbra{\boldsymbol{\alpha}}{\boldsymbol{\alpha}}\label{eq:Real Coherent State 4.1}\\
    &\equiv \prod_\mathbf{k} \frac{1}{2\pi V E_\mathbf{k}}\int_{-\infty}^\infty da_{\mathbf{p_k}}\int_{-\infty}^\infty db_{\mathbf{p_k}} \ e^{-\Delta p^n \sum_\mathbf{k} \frac{1}{(2\pi)^n2E_\mathbf{k}}\alpha_{\mathbf{p_k}}^*\alpha_{\mathbf{p_k}}} \ketbra{\boldsymbol{\alpha}}{\boldsymbol{\alpha}},\nonumber
\end{align}
where $a_{\mathbf{p_k}}, b_{\mathbf{p_k}} \in \mathbbm{R}$ are the real and imaginary parts of $\alpha_{\mathbf{p_k}}$, respectively,
\begin{align}
    \alpha_{\mathbf{p_k}}=a_{\mathbf{p_k}}+\mathrm{i}b_{\mathbf{p_k}}, \quad \alpha_{\mathbf{p_k}}^*=a_{\mathbf{p_k}}-\mathrm{i}b_{\mathbf{p_k}}.\nonumber
\end{align}
Proof: Maintaining a general normalization $N$, our identity element becomes the well-defined integral
\begin{align}
    \mathbbm{1} &= N \left(\prod_\mathbf{k} \int_{-\infty}^\infty da_{\mathbf{p_k}}\int_{-\infty}^\infty db_{\mathbf{p_k}}\right) \ e^{-\Delta p^n \sum_\mathbf{k} \frac{\alpha_{\mathbf{p_k}}^*\left(a_{\mathbf{p_k}},b_{\mathbf{p_k}}\right)\alpha_{\mathbf{p_k}}\left(a_{\mathbf{p_k}},b_{\mathbf{p_k}}\right)}{(2\pi)^n2E_\mathbf{k}}} \ketbra{\boldsymbol{\alpha}(\boldsymbol{a},\boldsymbol{b})}{\boldsymbol{\alpha}(\boldsymbol{a},\boldsymbol{b})}\nonumber\\
    &=N \left(\prod_\mathbf{k} \int_{-\infty}^\infty da_{\mathbf{p_k}} \int_{-\infty}^\infty db_{\mathbf{p_k}}\right) \ e^{-\Delta p^n \sum_\mathbf{k} \frac{\left(a_{\mathbf{p_k}}^2+b_{\mathbf{p_k}}^2\right)}{(2\pi)^n2E_\mathbf{k}}} \ketbra{\boldsymbol{a}+\mathrm{i}\boldsymbol{b}}{\boldsymbol{a}+\mathrm{i}\boldsymbol{b}}\nonumber\\
    &=N \left(\prod_\mathbf{k} \int_{-\infty}^\infty da_{\mathbf{p_k}}\int_{-\infty}^\infty db_{\mathbf{p_k}}\right) \ e^{-\Delta p^n \sum_\mathbf{k} \frac{\left(a^2_{\mathbf{p_k}}+b^2_{\mathbf{p_k}}\right)}{(2\pi)^n2E_\mathbf{k}}} e^{\Delta p^n \sum_\mathbf{q} \frac{\left(a_{\mathbf{p_q}}+\mathrm{i}b_{\mathbf{p_q}}\right)}{(2\pi)^n 2E_{\mathbf{q}}}\hat{a}_{\mathbf{p_q}}^\dagger}\ketbra{0}{0}e^{\Delta p^n \sum_\mathbf{l}\frac{\left(a_{\mathbf{p}_\mathbf{l}}-\mathrm{i}b_{\mathbf{p}_\mathbf{l}}\right)}{(2\pi)^n 2E_{\mathbf{l}}}\hat{a}_{\mathbf{p}_\mathbf{l}}}\nonumber
\end{align}
Since $\mathbf{k}, \mathbf{q},$ and $\mathbf{l}$ are all integer values spanning the same range in each summation, we can factor out the sums and rewrite the expression as a product over a single integer variable. By combining multiple sums into a single product, we simplify the notation and reveal the underlying structure of the finite-sized system. This reformulation is essential in a finite system, as it enables us to retrieve the completeness relation in a more natural way.
\begin{align}
    \mathbbm{1} &= N \prod_{\mathbf{k},\mathbf{q},\mathbf{l}} \int_{-\infty}^\infty da_{\mathbf{p_k}}\int_{-\infty}^\infty db_{\mathbf{p_k}} \ e^{-\Delta p^n \frac{1}{(2\pi)^n 2E_{\mathbf{k}}}\left(a^2_{\mathbf{p_k}}+b^2_{\mathbf{p_k}}\right)}\nonumber\\
    & \qquad e^{\Delta p^n \frac{1}{(2\pi)^n 2E_{\mathbf{q}}}\left(a_{\mathbf{p_q}}+\mathrm{i}b_{\mathbf{p_q}}\right)\hat{a}_{\mathbf{p_q}}^\dagger}\ketbra{0}{0}e^{\Delta p^n \frac{1}{(2\pi)^n 2E_{\mathbf{l}}}\left(a_{\mathbf{p}_\mathbf{l}}-\mathrm{i}b_{\mathbf{p}_\mathbf{l}}\right)\hat{a}_{\mathbf{p}_\mathbf{l}}}\nonumber\\
    &= N \prod_{\mathbf{k}} \int_{-\infty}^\infty da_{\mathbf{p_k}}\int_{-\infty}^\infty db_{\mathbf{p_k}} \ e^{-\Delta p^n \frac{1}{(2\pi)^n 2E_{\mathbf{k}}}\left(a^2_{\mathbf{p_k}}+b^2_{\mathbf{p_k}}\right)}\nonumber\\
    & \qquad e^{\Delta p^n \frac{1}{(2\pi)^n 2E_{\mathbf{k}}}\left(a_{\mathbf{p_k}}+\mathrm{i}b_{\mathbf{p_k}}\right)\hat{a}_{\mathbf{p_k}}^\dagger}\ketbra{0}{0}e^{\Delta p^n \frac{1}{(2\pi)^n 2E_{\mathbf{k}}}\left(a_{\mathbf{p_k}}-\mathrm{i}b_{\mathbf{p_k}}\right)\hat{a}_{\mathbf{p_k}}}.\nonumber
\end{align}
Expanding the coherent state exponential subsequently results in
\begin{align}
    \mathbbm{1} &= N \prod_\mathbf{k} \int_{-\infty}^\infty da_{\mathbf{p_k}}\int_{-\infty}^\infty db_{\mathbf{p_k}} \ e^{-\Delta p^n \frac{a^2_{\mathbf{p_k}}+b^2_{\mathbf{p_k}}}{(2\pi)^n 2E_{\mathbf{k}}}} \nonumber\\
    & \qquad \sum_{i=0}^\infty\frac{1}{i!}\left(\Delta p^n \frac{\left(a_{\mathbf{p_k}}+\mathrm{i}b_{\mathbf{p_k}}\right)\hat{a}_{\mathbf{p_k}}^\dagger}{(2\pi)^n 2E_{\mathbf{k}}}\right)^i\ketbra{0}{0}\sum_{j=0}^\infty\frac{1}{j!}\left(\Delta p^n \frac{\left(a_{\mathbf{p_k}}-\mathrm{i}b_{\mathbf{p_k}}\right)\hat{a}_{\mathbf{p_k}}}{(2\pi)^n 2E_{\mathbf{k}}}\right)^j\nonumber\\
    &=N \prod_\mathbf{k} \int_{-\infty}^\infty da_{\mathbf{p_k}} \int_{-\infty}^\infty db_{\mathbf{p_k}} \ e^{-\Delta p^n \frac{a^2_{\mathbf{p_k}}+b^2_{\mathbf{p_k}}}{(2\pi)^n 2E_{\mathbf{k}}}} \nonumber\\
    & \quad \sum_{i,j=0}^{\infty} \frac{1}{\sqrt{i!j!}} \frac{\left(a_{\mathbf{p_k}}+\mathrm{i}b_{\mathbf{p_k}}\right)^i \left(a_{\mathbf{p_k}}-\mathrm{i}b_{\mathbf{p_k}}\right)^j}{\left(\left(\frac{2\pi}{\Delta p}\right)^n 2E_{\mathbf{k}}\right)^{i+j}}\ketbra{i}{j}.\nonumber
\end{align}
We use radial decomposition to write
\begin{align}
    \alpha_{\mathbf{p_k}} = \rho_{\mathbf{p_k}} e^{\mathrm{i}\theta_{\mathbf{p_k}}}, \ \text{where} \ \rho_{\mathbf{p_k}} \ \mathcal{2} \ [0, \infty), \ \text{and} \ \theta_{\mathbf{p_k}} \ \mathcal{2} \ [0, 2\pi).\nonumber
\end{align}
Our integral then becomes
\begin{align}
    \mathbbm{1} &= N \prod_\mathbf{k} \int_0^\infty d\rho_{\mathbf{p_k}} \int_0^{2\pi} d\theta_{\mathbf{p_k}} \rho_{\mathbf{p_k}} e^{-\left(\frac{\Delta p}{2\pi}\right)^n \frac{\rho^2_{\mathbf{p_k}}}{2E_{\mathbf{k}}}}\sum_{i,j=0}^\infty \frac{\rho^{i+j}_{\mathbf{p_k}}e^{\mathrm{i}\left(i-j\right)\theta_{\mathbf{p_k}}}}{\sqrt{i!j!}\left(\left(\frac{2\pi}{\Delta p}\right)^n 2E_{\mathbf{k}}\right)^{i+j}}\ketbra{i}{j}\nonumber\\
    &= N\prod_\mathbf{k} 2\pi \int_0^\infty d\rho_{\mathbf{p_k}} e^{-\left(\frac{\Delta p}{2\pi}\right)^n \frac{\rho^2_{\mathbf{p_k}}}{2E_{\mathbf{k}}}}\sum_{j=0}^\infty \frac{\rho^{2j+1}_{\mathbf{p_k}}}{j!\left(\left(\frac{2\pi}{\Delta p}\right)^n  2E_{\mathbf{k}}\right)^{2j}}\ketbra{j}{j},\label{eq:Intermediate 1 relativistic completeness}
\end{align}
where we used the following
\begin{align}
    I(x) &= \int_0^\infty d\rho \rho e^{-x\rho^2}=\left[-\frac{1}{2x}e^{-x\rho^2}\right]_{\rho=0}^{\rho=\infty}=\frac{1}{2x}\nonumber\\
    \therefore \int_0^\infty d\rho \rho^{2j+1} e^{-x\rho^2} &= (-1)^j \frac{d^j}{dx^j}I(x)\left|_{x=\left(\frac{\Delta p}{2\pi}\right)^n \frac{1}{2E_{\mathbf{k}}}}\right. = \frac{j!\left(\left(\frac{2\pi}{\Delta p}\right)^n  2E_{\mathbf{k}}\right)^{j+1}}{2},\label{eq:Identity element trick}
\end{align}
which is dimensionally correct, as
\begin{align}
    \left[x\rho^2\right]&=\left[\frac{\rho^2}{\left(\frac{2\pi}{\Delta p}\right)^n 2E_\mathbf{k}}\right]=1\nonumber\\
    \left[d\rho \rho^{2j+1}\right]&=E^{\left(\frac{1-n}{2}\right)(2j+2)}=\left[\left(\left(\frac{2\pi}{\Delta p}\right)^n  2E_\mathbf{k}\right)^{j+1}\right]=E^{(1-n)(j+1)}.\nonumber
\end{align}
By inserting \cref{eq:Identity element trick} into \cref{eq:Intermediate 1 relativistic completeness}, we obtain
\begin{align}
    \mathbbm{1} &= N\prod_\mathbf{k} 2\pi \sum_{j=0}^\infty \frac{j!\left(\left(\frac{2\pi}{\Delta p}\right)^n  2E_{\mathbf{k}}\right)^{j+1}}{2j!\left(\left(\frac{2\pi}{\Delta p}\right)^n  2E_{\mathbf{k}}\right)^{2j}}\ketbra{j}{j}\nonumber\\
    &=N \prod_\mathbf{k} \pi \left(\frac{2\pi}{\Delta p}\right)^n  2E_{\mathbf{k}}\underbrace{\sum_{j=0}^\infty \frac{1}{\left(\left(\frac{2\pi}{\Delta p}\right)^n  2E_{\mathbf{k}}\right)^j}\ketbra{j}{j}}_\mathbbm{1} \Rightarrow N = \prod_\mathbf{k} \frac{1}{2\pi VE_\mathbf{k}}.\nonumber
\end{align}
Note that the number states in our case have dimensions, which must be taken into account when defining the completeness relation as shown above.
The coherent state completeness relation thus becomes
\begin{align}
     \mathbbm{1} &= \left(\prod_\mathbf{k} \frac{1}{2\pi V E_\mathbf{k}}\int_{-\infty}^\infty da_{\mathbf{p_k}}\int_{-\infty}^\infty db_{\mathbf{p_k}}\right) \ e^{-\Delta p^n \sum_\mathbf{k} \frac{a^2_{\mathbf{p_k}}+b^2_{\mathbf{p_k}}}{(2\pi)^n 2E_\mathbf{k}}} \ketbra{\boldsymbol{a}+\mathrm{i}\boldsymbol{b}}{\boldsymbol{a}+\mathrm{i}\boldsymbol{b}}.\nonumber
\end{align}
Since the above expression will be inserted into the partition function many times, we will use the shorthand notation leaving the $a_{\mathbf{p_k}}$ and $b_{\mathbf{p_k}}$ dependence implicit in the exponential.
\begin{align}
     \mathbbm{1} &= \left(\prod_\mathbf{k} \frac{1}{2\pi V E_\mathbf{k}}\int_{-\infty}^\infty da_{\mathbf{p_k}}\int_{-\infty}^\infty db_{\mathbf{p_k}}\right) \ e^{-\Delta p^n \sum_\mathbf{k} \frac{1}{(2\pi)^n2E_\mathbf{k}}\alpha_{\mathbf{p_k}}^*\alpha_{\mathbf{p_k}}} \ketbra{\boldsymbol{\alpha}}{\boldsymbol{\alpha}}.\label{eq:Real Coherent State 4.2}
\end{align}
\end{enumerate}
\end{widetext}
\subsection{Coherent State Representation using Displacement Operator}\label{Coherent State Representation using Displacement Operator}
A coherent state can be generated by acting with the displacement operator on the ground state. Given proper normalization, this leads to the following definition. For clarity and readability, we leave the limits of the sums implicit throughout this appendix.
\begin{align}
    \ket{\boldsymbol{\alpha}_{i,a}} \equiv& e^{\frac{1}{2} \Delta p^n \sum_{\mathbf{k}} \frac{\alpha^*_{\mathbf{p}_{\mathbf{k},i}}\alpha_{\mathbf{p}_{\mathbf{k},i}}}{(2\pi)^n2E_\mathbf{k}}}\nonumber\\
    & \ e^{\Delta p^n \sum_\mathbf{k} \frac{1}{(2\pi)^n 2E_\mathbf{k}} \left[\alpha_{\mathbf{p}_{\mathbf{k},i}}\hat{a}^\dagger_{\mathbf{p}_{\mathbf{k},a}}-\alpha^*_{\mathbf{p}_{\mathbf{k},i}}\hat{a}_{\mathbf{p}_{\mathbf{k},a}}\right]}\ket{0}.\label{eq:Displacement operator definition of coherent state definition from appendix} 
\end{align}
Using the Baker-Campbell-Hausdorff (BCH) formula, 
\begin{align}
    e^Xe^Y &= e^{X+Y+\frac{1}{2}[X,Y]+\cdots},\nonumber
\end{align}
we can prove the equivalence of the coherent state in \cref{eq:Displacement operator definition of coherent state definition from appendix} to that of \cref{eq:Coherent state definition}. The relevant commutators that arise in our use of the BCH formula are:
\begin{align}
    [\hat{a}_{\mathbf{p}_{\mathbf{k},a}},\hat{a}^\dagger_{\mathbf{p}_{\mathbf{q},a}}] &= (2\pi)^n 2E_\mathbf{k} \frac{1}{\Delta p^n}\delta_{\mathbf{k},\mathbf{q}},\nonumber\\
    [\hat{a}_{\mathbf{p}_{\mathbf{k},a}},\hat{a}_{\mathbf{p}_{\mathbf{q},a}}] &=[\hat{a}^\dagger_{\mathbf{p}_{\mathbf{k},a}},\hat{a}^\dagger_{\mathbf{p}_{\mathbf{q},a}}]=0,\nonumber\\
    [\hat{a}_{\mathbf{p}_{\mathbf{k},a}},[\hat{a}_{\mathbf{p}_{\mathbf{q},a}},\hat{a}^\dagger_{\mathbf{p}_{\mathbf{l},a}}]]&=[\hat{a}^\dagger_{\mathbf{p}_{\mathbf{k},a}},[\hat{a}_{\mathbf{p}_{\mathbf{q},a}},\hat{a}^\dagger_{\mathbf{p}_{\mathbf{l},a}}]]=0.\nonumber
\end{align}
Having established the contributions to the BCH formula, we now turn to the proof:
\begin{align}
    \ket{\boldsymbol{\alpha}_{i,a}} =& e^{\frac{1}{2}\Delta p^n\sum_\mathbf{k} \frac{\alpha^*_{\mathbf{p}_{\mathbf{k},i}}\alpha_{\mathbf{p}_{\mathbf{k},i}}}{(2\pi)^n 2E_\mathbf{k}}}\nonumber\\
    & \quad e^{\Delta p^n \sum_\mathbf{k}\frac{1}{(2\pi)^n 2E_\mathbf{k}} \left[\alpha_{\mathbf{p}_{\mathbf{k},i}}\hat{a}^\dagger_{\mathbf{p}_{\mathbf{k},a}}-\alpha^*_{\mathbf{p}_{\mathbf{k},i}}\hat{a}_{\mathbf{p}_{\mathbf{k},a}}\right]}\ket{0}\nonumber\\
    =& e^{\frac{1}{2}\Delta p^n\sum_\mathbf{k} \frac{\alpha^*_{\mathbf{p}_{\mathbf{k},i}}\alpha_{\mathbf{p}_{\mathbf{k},i}}}{(2\pi)^n 2E_\mathbf{k}}}\nonumber\\
    & \quad e^{-\frac{1}{2}\Delta p^{2n}\sum_{\mathbf{k},\mathbf{q}}\frac{\alpha^*_{\mathbf{p}_{\mathbf{k},i}}\alpha_{\mathbf{p}_{\mathbf{q},i}}}{(2\pi)^{2n} 2E_\mathbf{k}E_\mathbf{q}}[\hat{a}_{\mathbf{p}_{\mathbf{k},a}},\hat{a}_{\mathbf{p}_{\mathbf{q},a}}^\dagger]}\nonumber\\
    & \quad e^{\Delta p^n \sum_\mathbf{k} \frac{\alpha_{\mathbf{p}_{\mathbf{k},i}}\hat{a}^\dagger_{\mathbf{p}_{\mathbf{k},a}}}{(2\pi)^n 2E_\mathbf{k}} } e^{-\Delta p^n \sum_\mathbf{k} \frac{\alpha^*_{\mathbf{p}_{\mathbf{k},i}}\hat{a}_{\mathbf{p}_{\mathbf{k},a}}}{(2\pi)^n 2E_\mathbf{k}} }\ket{0}\nonumber\\
    =& e^{\frac{1}{2}\Delta p^n\sum_\mathbf{k} \frac{\alpha^*_{\mathbf{p}_{\mathbf{k},i}}\alpha_{\mathbf{p}_{\mathbf{k},i}}}{(2\pi)^n 2E_\mathbf{k}}}\nonumber\\
    & \quad e^{-\frac{1}{2}\Delta p^{2n}\sum_{\mathbf{k},\mathbf{q}}\frac{\alpha^*_{\mathbf{p}_{\mathbf{k},i}}\alpha_{\mathbf{p}_{\mathbf{q},i}}}{(2\pi)^{2n} 2E_\mathbf{k}E_\mathbf{q}} (2\pi)^n 2E_\mathbf{k} \frac{1}{\Delta p^n} \delta_{\mathbf{k},\mathbf{q}}}\nonumber\\
    & \quad e^{\Delta p^n \sum_\mathbf{k} \frac{\alpha_{\mathbf{p}_{\mathbf{k},i}}\hat{a}^\dagger_{\mathbf{p}_{\mathbf{k},a}}}{(2\pi)^n 2E_\mathbf{k}}  }\ket{0}\nonumber\\
    =& e^{\frac{1}{2}\Delta p^n \sum_\mathbf{k} \frac{\alpha^*_{\mathbf{p}_{\mathbf{k},i}}\alpha_{\mathbf{p}_{\mathbf{k},i}}}{(2\pi)^n 2E_\mathbf{k}}-\frac{1}{2}\Delta p^n\sum_{\mathbf{k},\mathbf{q}}\frac{\alpha^*_{\mathbf{p}_{\mathbf{k},i}}\alpha_{\mathbf{p_q},i}}{(2\pi)^n 2E_\mathbf{k}}\delta_{\mathbf{k},\mathbf{q}}}\nonumber\\
    & \quad e^{\Delta p^n \sum_\mathbf{k} \frac{\alpha_{\mathbf{p}_{\mathbf{k},i}}\hat{a}^\dagger_{\mathbf{p}_{\mathbf{k},a}}}{(2\pi)^n 2E_\mathbf{k}} }\ket{0}\nonumber\\
    =& e^{\Delta p^n \sum_\mathbf{k} \frac{1}{(2\pi)^n 2E_\mathbf{k}} \alpha_{\mathbf{p}_{\mathbf{k},i}}\hat{a}^\dagger_{\mathbf{p}_{\mathbf{k},a}}}\ket{0},
\end{align}
which is exactly the same coherent state we defined in \cref{eq:Coherent state definition}, but with added temporal dependence.

\subsection{Coherent State in the Field Basis}\label{Coherent State in the Field Basis}
In this appendix, we demonstrate how the coherent state defined in \cref{eq:Coherent state definnition displacement operator} can be expressed in terms of the variables and operators $\phi_{\mathbf{x}_{\mathbf{r},i}}, \pi_{\mathbf{x}_{\mathbf{r},i}}, \ \hat{\phi}_{\mathbf{x}_{\mathbf{r},a}}$, and $ \hat{\pi}_{\mathbf{x}_{\mathbf{r},a}}$. 
\subsubsection{Change of Operator Variables}\label{Change of Operator Variables}
We are working in the Heisenberg picture, as explained in \cref{Path Integral from Field Operator Hamiltonian}, where the position space field operators are defined as:
\begin{align}
    \hat{\phi}_{\mathbf{x}_{\mathbf{r},a}} &= \Delta p^n \sum_{\mathbf{k}=-\frac{\mathbf{K}}{2}}^\frac{\mathbf{K}}{2} \frac{1}{(2\pi)^n 2E_{\mathbf{k}}} \left[ \hat{a}_{\mathbf{p}_{\mathbf{k},a}} e^{\mathrm{i}\mathbf{p_k} \cdot \mathbf{x}_\mathbf{r}} \right. \nonumber\\
    & \left. \quad + \hat{a}^\dagger_{\mathbf{p}_{\mathbf{k},a}} e^{-\mathrm{i}\mathbf{p_k} \cdot \mathbf{x}_\mathbf{r}} \right] \Bigg|_{p^0_k=E_\mathbf{k}=\sqrt{\mathbf{p_k}^2+m^2}}.\label{eq:Discretized field operator mode expansion}
\end{align}
The \cref{eq:Discretized field operator mode expansion} is the same as \cref{eq:Free Heisenberg picture field mode expansion}, but \cref{eq:Discretized field operator mode expansion} includes the time dependence within the creation and annihilation operators, which aligns with the time dependent coherent state definitions. Keeping the time dependence inside the ladder operators is also crucial for expressing $\hat{a}_{\mathbf{p}_{\mathbf{k},a}}$ in terms of $\hat{\phi}_{\mathbf{p}_{\mathbf{k},a}}$ and $\hat{\pi}_{\mathbf{p}_{\mathbf{k},a}}$. We have that
\begin{widetext}
\begin{align}
    E_{\mathbf{k}}\hat{\phi}_{\mathbf{p}_{\mathbf{k},a}}+\mathrm{i}\hat{\pi}_{\mathbf{p}_{\mathbf{k},a}} =& \Delta x^n \sum_{\mathbf{r}=\mathbf{0}}^\mathbf{K} e^{-\mathrm{i}\mathbf{p_k}\cdot\mathbf{x}_\mathbf{r}} \left(E_{\mathbf{k}} \hat{\phi}_{\mathbf{x}_{\mathbf{r},a}}+\mathrm{i}\hat{\pi}_{\mathbf{x}_{\mathbf{r},a}}\right)\nonumber\\
    =& \Delta x^n \sum_{\mathbf{r}=\mathbf{0}}^\mathbf{K} e^{-\mathrm{i}\mathbf{p_k}\cdot\mathbf{x}_\mathbf{r}} \Delta p^n  \sum_{\mathbf{q}=-\frac{\mathbf{K}}{2} }^\frac{\mathbf{K}}{2}\frac{1}{(2\pi)^n 2E_\mathbf{q}}\left[E_\mathbf{k}\left(\hat{a}_{\mathbf{p}_{\mathbf{q},a}}e^{\mathrm{i}\mathbf{p_q}\cdot \mathbf{x}_\mathbf{r}}+\hat{a}^\dagger_{\mathbf{p}_{\mathbf{q},a}}e^{-\mathrm{i}\mathbf{p_q}\cdot \mathbf{x}_\mathbf{r}}\right)\right.\nonumber\\
    & \quad \left. +E_\mathbf{q}\left(-\hat{a}^\dagger_{\mathbf{p}_{\mathbf{q},a}}e^{-\mathrm{i}\mathbf{p_q}\cdot \mathbf{x_r}}+\hat{a}_{\mathbf{p}_{\mathbf{q},a}}e^{\mathrm{i}\mathbf{p_q}\cdot \mathbf{x}_\mathbf{r}}\right)\right]\nonumber\\
    =& \Delta x^n \Delta p^n \sum_{\mathbf{r}=\mathbf{0}}^\mathbf{K} \sum_{\mathbf{q}=-\frac{\mathbf{K}}{2} }^\frac{\mathbf{K}}{2}\frac{1}{(2\pi)^n 2E_\mathbf{q}}\left[E_\mathbf{k}\left(\hat{a}_{\mathbf{p}_{\mathbf{q},a}}e^{\mathrm{i}\left(\mathbf{p_q}-\mathbf{p_k}\right)\cdot \mathbf{x}_\mathbf{r}}+\hat{a}^\dagger_{\mathbf{p}_{\mathbf{q},a}}e^{-\mathrm{i}\left(\mathbf{p_q}+\mathbf{p_k}\right)\cdot \mathbf{x}_\mathbf{r}}\right)\right.\nonumber\\
    & \quad \left. + E_\mathbf{q} \left(-\hat{a}^\dagger_{\mathbf{p}_{\mathbf{q},a}}e^{-\mathrm{i}\left(\mathbf{p_q}+\mathbf{p_k}\right)\cdot \mathbf{x}_\mathbf{r}}+\hat{a}_{\mathbf{p}_{\mathbf{q},a}}e^{\mathrm{i}\left(\mathbf{p_q}-\mathbf{p_k}\right)\cdot \mathbf{x}_\mathbf{r}}\right)\right]\nonumber\\
    =&\Delta p^n \sum_{\mathbf{q}=-\frac{\mathbf{K}}{2} }^\frac{\mathbf{K}}{2}\frac{1}{(2\pi)^n2E_\mathbf{q}}\left[E_\mathbf{k} \left(\frac{2\pi}{\Delta p}\right)^n \left(\hat{a}_{\mathbf{p}_{\mathbf{q},a}} \delta_{\mathbf{k},\mathbf{q}}+\hat{a}^\dagger_{\mathbf{p}_{\mathbf{q},a}} \delta_{\mathbf{k},-\mathbf{q}}\right) + E_\mathbf{q} \left(-\hat{a}^\dagger_{\mathbf{p}_{\mathbf{q},a}}\delta_{\mathbf{k},-\mathbf{q}}+\hat{a}_{\mathbf{p}_{\mathbf{q},a}}\delta_{\mathbf{k},\mathbf{q}}\right)\right]\nonumber\\
    =& \left(\frac{\Delta p}{2\pi}\right)^n\left(\frac{2\pi}{\Delta p}\right)^n\frac{E_\mathbf{k}}{2E_\mathbf{k}}\left[\hat{a}_{\mathbf{p}_{\mathbf{k},a}}+\hat{a}^\dagger_{-\mathbf{p}_{\mathbf{k},a}}-\hat{a}^\dagger_{-\mathbf{p}_{\mathbf{k},a}}+\hat{a}_{\mathbf{p}_{\mathbf{k},a}}\right].\nonumber
\end{align}
\end{widetext}
Therefore
\begin{align}
    \hat{a}_{\mathbf{p}_{\mathbf{k},a}}&=E_{\mathbf{k}}\hat{\phi}_{\mathbf{p}_{\mathbf{k},a}}+\mathrm{i}\hat{\pi}_{\mathbf{p}_{\mathbf{k},a}}.\label{eq:Change of operator variables}
\end{align}
For the creation operator, we find a similar relation,
\begin{align}
    \hat{a}^\dagger_{\mathbf{p}_{\mathbf{k},a}} &= \left(\hat{a}_{\mathbf{p}_{\mathbf{k},a}}\right)^\dagger=E_{\mathbf{k}}\hat{\phi}_{-\mathbf{p}_{\mathbf{k},a}}-\mathrm{i}\hat{\pi}_{-\mathbf{p}_{\mathbf{k},a}}.\label{eq:Change of conjugate operator variables}
\end{align}
\subsubsection{Rewriting Coherent States in the Field Basis}\label{Rewriting the Coherent State in the Field Basis}
Starting from the definition of the displacement operator in \cref{eq:Displacement operator}, we perform a change of variables by substituting \cref{eq:Coherent eigenvalue change of variables,eq:Conjugate coherent eigenvalue change of variables,eq:Change of operator variables,eq:Change of conjugate operator variables}
\begin{widetext}
\begin{align}
    \hat{D}(\boldsymbol{\alpha}_{i,a}) &= \exp\bigg\{\Delta p^n \sum_{\mathbf{k}=-\frac{\mathbf{K}}{2}}^\frac{\mathbf{K}}{2} \frac{1}{(2\pi)^n 2E_\mathbf{k}} \left[\alpha_{\mathbf{p}_{\mathbf{k},i}}\hat{a}^\dagger_{\mathbf{p}_{\mathbf{k},a}}-\alpha^*_{\mathbf{p}_{\mathbf{k},i}}\hat{a}_{\mathbf{p}_{\mathbf{k},a}}\right]\bigg\}\nonumber\\
    &= \exp\bigg\{\Delta p^n \sum_{\mathbf{k}=-\frac{\mathbf{K}}{2}}^\frac{\mathbf{K}}{2} \frac{1}{(2\pi)^n 2E_\mathbf{k}} \left[\left(E_{\mathbf{k}}\phi_{\mathbf{p}_{\mathbf{k},i}}+\mathrm{i}\pi_{\mathbf{p}_{\mathbf{k},i}}\right)\left(E_{\mathbf{k}}\hat{\phi}_{-\mathbf{p}_{\mathbf{k},a}}-\mathrm{i}\hat{\pi}_{-\mathbf{p}_{\mathbf{k},a}}\right) \right. \nonumber\\
    & \quad \left.-\left(E_{\mathbf{k}}\phi_{-\mathbf{p}_{\mathbf{k},i}}-\mathrm{i}\pi_{-\mathbf{p}_{\mathbf{k},i}}\right)\left(E_{\mathbf{k}}\hat{\phi}_{\mathbf{p}_{\mathbf{k},a}}+\mathrm{i}\hat{\pi}_{\mathbf{p}_{\mathbf{k},a}}\right)\right]\bigg\}\nonumber\\
    &= \exp\Bigg\{\Delta p^n \sum_{\mathbf{k}=-\frac{\mathbf{K}}{2}}^\frac{\mathbf{K}}{2} \frac{1}{(2\pi)^n 2E_\mathbf{k}} \bigg[E_\mathbf{k}^2\left(\phi_{\mathbf{p}_{\mathbf{k},i}}\hat{\phi}_{-\mathbf{p}_{\mathbf{k},a}}-\phi_{-\mathbf{p}_{\mathbf{k},i}}\hat{\phi}_{\mathbf{p}_{\mathbf{k},a}}\right) + \left(\pi_{\mathbf{p}_{\mathbf{k},i}}\hat{\pi}_{-\mathbf{p}_{\mathbf{k},a}}-\pi_{-\mathbf{p}_{\mathbf{k},i}}\hat{\pi}_{\mathbf{p}_{\mathbf{k},a}}\right) \nonumber\\
    & \quad  -iE_{\mathbf{k}}\left(\phi_{\mathbf{p}_{\mathbf{k},i}}\hat{\pi}_{-\mathbf{p}_{\mathbf{k},a}}-\pi_{\mathbf{p}_{\mathbf{k},i}}\hat{\phi}_{-\mathbf{p}_{\mathbf{k},a}} + \phi_{-\mathbf{p}_{\mathbf{k},i}}\hat{\pi}_{\mathbf{p}_{\mathbf{k},a}}-\pi_{-\mathbf{p}_{\mathbf{k},i}}\hat{\phi}_{\mathbf{p}_{\mathbf{k},a}}\right)\bigg]\Bigg\}.\nonumber
\end{align}
\end{widetext}
Due to the symmetric nature of the sum, we can take $\mathbf{k} \to -\mathbf{k}$ for each independent term without changing the expression, which causes the first two terms in the brackets to cancel. We are thus left with a simplified expression to which we will apply a DFT to transition to position space:
\begin{align}
    \hat{D}(\boldsymbol{\alpha}_{i,a}) =& \exp\bigg\{\Delta p^n \sum_{\mathbf{k}=-\frac{\mathbf{K}}{2}}^\frac{\mathbf{K}}{2} \frac{1}{(2\pi)^n 2E_\mathbf{k}}(-2 \mathrm{i} E_\mathbf{k})\nonumber\\
    & \quad \left(\phi_{\mathbf{p}_{\mathbf{k},i}}\hat{\pi}_{-\mathbf{p}_{\mathbf{k},a}}-\pi_{\mathbf{p}_{\mathbf{k},i}}\hat{\phi}_{-\mathbf{p}_{\mathbf{k},a}}\right)\bigg\}\nonumber\\
    =& \exp\bigg\{-\mathrm{i}\left(\frac{\Delta p}{2\pi}\right)^n \sum_{\mathbf{k}=-\frac{\mathbf{K}}{2}}^\frac{\mathbf{K}}{2} \Delta x^{2n}\nonumber\\
    & \quad \sum_{\mathbf{r},\mathbf{s}=\mathbf{0}}^{\mathbf{K}}\left(\phi_{\mathbf{x}_{\mathbf{r},i}}\hat{\pi}_{\mathbf{x}_{\mathbf{s},a}}-\pi_{\mathbf{x}_{\mathbf{r},i}}\hat{\phi}_{\mathbf{x}_{\mathbf{s},a}}\right)e^{\mathrm{i}\mathbf{p_k}(\mathbf{x}_\mathbf{r}-\mathbf{x}_\mathbf{s})}\bigg\}\nonumber\\
    =& \exp\bigg\{-\mathrm{i} \Delta x^n \sum_{\mathbf{r},\mathbf{s}=\mathbf{0}}^{\mathbf{K}} \nonumber\\
    & \quad \left(\phi_{\mathbf{x}_{\mathbf{r},i}}\hat{\pi}_{\mathbf{x}_{\mathbf{s},a}}-\pi_{\mathbf{x}_{\mathbf{r},i}}\hat{\phi}_{\mathbf{x}_{\mathbf{s},a}}\right)\delta_{\mathbf{r},\mathbf{s}}\bigg\}\nonumber\\
    &= \exp\bigg\{-\mathrm{i} (\Delta x)^n \sum_{\mathbf{r}=\mathbf{0}}^{\mathbf{K}} \nonumber\\
    & \quad \left(\phi_{\mathbf{x}_{\mathbf{r},i}}\hat{\pi}_{\mathbf{x}_{\mathbf{r},a}}-\pi_{\mathbf{x}_{\mathbf{r},i}}\hat{\phi}_{\mathbf{x}_{\mathbf{r},a}}\right)\bigg\}.
\end{align}
We may thus define
\begin{align}
    \hat{D}(\phi_{i,a}) &\equiv e^{-\mathrm{i} (\Delta x)^n \sum_{\mathbf{r}=\mathbf{0}}^{\mathbf{K}} \left(\phi_{\mathbf{x}_{\mathbf{r},i}}\hat{\pi}_{\mathbf{x}_{\mathbf{r},a}}-\pi_{\mathbf{x}_{\mathbf{r},i}}\hat{\phi}_{\mathbf{x}_{\mathbf{r},a}}\right)}\nonumber\\
    \therefore \ket{\phi_{i,a}}&= N^{+}_{i,i}\hat{D}(\phi_{i,a})\ket{0}\nonumber\\
    &= N^{+}_{i,i}\hat{D}(\boldsymbol{\alpha}_{i,a})\ket{0}\nonumber\\
    &=\ket{\boldsymbol{\alpha}_{i,a}},
\end{align}
where $\phi_{\mathbf{p}_{\mathbf{k},i}}$ and $\alpha_{\mathbf{p}_{\mathbf{k},i}}$ are related via \cref{eq:Field operator change of variables,eq:Conjugate field operator change of variables}.

\subsection{Field Basis Displacement Operator Commutation Relations} \label{Field Basis Displacement Operator Commutation Relations}
In this appendix, we prove some commutation relations for the displacement operator with the field operators in discretized space that we will use in our derivations. We will utilize the canonical commutation relations from \cref{eq:real scalar field lagrangian discrete,eq:real scalar ladder operator commutation relation 1,eq:real scalar ladder operator commutation relation 2}. Our fields commute with the creation and annihilation operators as follows:
\begin{align}
    \left[\hat{a}_{\mathbf{p}_{\mathbf{k},a}},\hat{\phi}_{\mathbf{x}_{\mathbf{r},a}} \right] =& \Delta p^n \sum_{\mathbf{q}=-\frac{\mathbf{K}}{2}}^\frac{\mathbf{K}}{2} \frac{1}{(2\pi)^n2E_\mathbf{q}} \left(\left[\hat{a}_{\mathbf{p}_{\mathbf{k},a}},\hat{a}_{\mathbf{p}_{\mathbf{q},a}}\right]e^{\mathrm{i}\mathbf{p_q} \cdot \mathbf{x_r}} \right. \nonumber\\
    & \left. \quad  + \left[\hat{a}_{\mathbf{p}_{\mathbf{k},a}},\hat{a}^\dagger_{\mathbf{p}_{\mathbf{q},a}}\right]e^{- \mathrm{i}\mathbf{p_q} \cdot \mathbf{x_r}}\right)\nonumber\\
    =& \Delta p^n \sum_{\mathbf{q}=-\frac{\mathbf{K}}{2}}^\frac{\mathbf{K}}{2} \frac{1}{(2\pi)^n2E_\mathbf{q}} (2\pi)^n 2E_\mathbf{k} \nonumber\\
    & \quad \frac{1}{\Delta p^n}\delta_{\mathbf{k},\mathbf{q}}e^{- \mathrm{i}\mathbf{p_q} \cdot \mathbf{x_r}}\nonumber\\
    =& e^{- \mathrm{i}\mathbf{p_k} \cdot \mathbf{x_r}},\label{eq:a phi commutator}\\
    \left[\hat{\phi}_{\mathbf{x}_{\mathbf{r},a}},\hat{a}^\dagger_{\mathbf{p}_{\mathbf{k},a}}\right] =& e^{\mathrm{i}\mathbf{p_k} \cdot \mathbf{x_r},}\label{eq:phi a dagger commutator}\\
    \left[\hat{a}_{\mathbf{p}_{\mathbf{k},a}},\hat{\pi}_{\mathbf{x}_{\mathbf{r},a}}\right] =& \frac{\mathrm{i}}{2}\left(\frac{\Delta p}{2\pi}\right)^n \sum_{\mathbf{q}=-\frac{\mathbf{K}}{2}}^\frac{\mathbf{K}}{2} \left[\hat{a}_{\mathbf{p}_{\mathbf{k},a}},\hat{a}^\dagger_{\mathbf{p}_{\mathbf{q},a}}\right]e^{- \mathrm{i}\mathbf{p_q} \cdot \mathbf{x_r}} \nonumber\\
    & \quad - \left[\hat{a}_{\mathbf{p}_{\mathbf{k},a}},\hat{a}_{\mathbf{p}_{\mathbf{q},a}}\right]e^{\mathrm{i}\mathbf{p_q} \cdot \mathbf{x_r}}\nonumber\\
    =& \frac{\mathrm{i}}{2}\left(\frac{\Delta p}{2\pi}\right)^n \left(\frac{2\pi}{\Delta p}\right)^n \nonumber\\
    & \quad \sum_{\mathbf{q}=-\frac{\mathbf{K}}{2}}^\frac{\mathbf{K}}{2} 2 E_\mathbf{q} \delta_{\mathbf{q},\mathbf{k}}e^{-\mathrm{i}\mathbf{p_q} \cdot \mathbf{x_r}}\nonumber\\
    =& \mathrm{i} E_\mathbf{k} e^{- \mathrm{i}\mathbf{p_k} \cdot \mathbf{x_r}}, \label{eq:a pi commutator}\\
    \left[\hat{\pi}_{\mathbf{x}_{\mathbf{r},a}},\hat{a}^\dagger_{\mathbf{p}_{\mathbf{k},a}}\right] =& -\mathrm{i} E_\mathbf{k} e^{\mathrm{i}\mathbf{p_k} \cdot \mathbf{x_r}} .\label{eq:pi a dagger commutator}
\end{align}

Using \cref{eq:a phi commutator,eq:phi a dagger commutator,eq:a pi commutator,eq:pi a dagger commutator}, we can derive the commutators between the displacement operator and the field operators. We start by repeating the definition of the displacement operator:
\begin{align}
    \hat{D}(\phi_{i,a}) &\equiv e^{\hat{A}_{i,a}},\nonumber
\end{align}
where
\begin{align}
    \hat{A}_{i,a} &\equiv -\mathrm{i} (\Delta x)^n\sum_{\mathbf{r}=\mathbf{0}}^\mathbf{K} \left(\phi_{\mathbf{x}_{\mathbf{r},i}}\hat{\pi}_{\mathbf{x}_{\mathbf{r},a}}-\pi_{\mathbf{x}_{\mathbf{r},i}}\hat{\phi}_{\mathbf{x}_{\mathbf{r},a}}\right).\nonumber
\end{align}
With the displacement operator well defined, we then continue with the necessary equal time commutation relations,
\begin{align}
    \left[\hat{\phi}_{\mathbf{x}_{\mathbf{r},a}},\hat{A}_{i,a}\right] =& -\mathrm{i}(\Delta x)^n\sum_{\mathbf{s}=\mathbf{0}}^\mathbf{K} \phi_{\mathbf{x}_{\mathbf{s},i}}[\hat{\phi}_{\mathbf{x}_{\mathbf{r},a}},\hat{\pi}_{\mathbf{x}_{\mathbf{s},a}}] \nonumber\\
    & \quad - \pi_{\mathbf{x}_{\mathbf{s},i}} \cancelto{0}{[\hat{\phi}_{\mathbf{x}_{\mathbf{r},a}},\hat{\phi}_{\mathbf{x}_{\mathbf{s},a}}]}\nonumber\\
    =& -\mathrm{i}(\Delta x)^n\sum_{\mathbf{s}=\mathbf{0}}^\mathbf{K} \phi_{\mathbf{x}_{\mathbf{s},i}} \hat{\mathbbm{1}}\frac{\mathrm{i}}{(\Delta x)^n}\delta_{\mathbf{r},\mathbf{s}}\nonumber\\
    =& \phi_{\mathbf{x}_{\mathbf{r},i}}\hat{\mathbbm{1}},\\
    \left[\hat{\phi}_{\mathbf{x}_{\mathbf{r},a}},\hat{D}(\phi_{i,a})\right] =&
    \left[\hat{\phi}_{\mathbf{x}_{\mathbf{r},a}}, \sum_{n=0}^\infty \frac{(\hat{A}_{i,a})^n}{n!}\right]\nonumber\\
    =& \phi_{\mathbf{x}_{\mathbf{r},i}} \sum_{n=1}^\infty \frac{n}{n!}(\hat{A}_{i,a})^{n-1}\nonumber\\
    =& \phi_{\mathbf{x}_{\mathbf{r},i}} \sum_{n=1}^\infty \frac{(\hat{A}_{i,a})^{n-1}}{(n-1)!}\nonumber\\
    =& \phi_{\mathbf{x}_{\mathbf{r},i}} \hat{D}(\phi_{i,a}).
\end{align}
Similarly, for the momentum conjugate field operator, we find
\begin{align}
    \left[\hat{\pi}_{\mathbf{x}_{\mathbf{r},a}},\hat{A}_{i,a}\right] =& -\mathrm{i}(\Delta x)^n\sum_{\mathbf{s}=\mathbf{0}}^\mathbf{K}\phi_{\mathbf{x}_{\mathbf{s},i}} \cancelto{0}{[\hat{\pi}_{\mathbf{x}_{\mathbf{r},a}},\hat{\pi}_{\mathbf{x}_{\mathbf{s},a}}]}\nonumber\\
    & \quad - \pi_{\mathbf{x}_{\mathbf{s},i}} [\hat{\pi}_{\mathbf{x}_{\mathbf{r},a}},\hat{\phi}_{\mathbf{x}_{\mathbf{s},a}}]\nonumber\\
    =& -\mathrm{i}(\Delta x)^n\sum_{\mathbf{s}=\mathbf{0}}^\mathbf{K} \pi_{\mathbf{x}_{\mathbf{r},i}} \hat{\mathbbm{1}}\frac{\mathrm{i}}{\left(\Delta x\right)^n}\delta_{\mathbf{r},\mathbf{s}}\nonumber\\
    =& \pi_{\mathbf{x}_{\mathbf{r},i}}\hat{\mathbbm{1}},\\
    \left[\hat{\pi}_{\mathbf{x}_{\mathbf{r},a}},\hat{D}(\phi_{i,a})\right]=&
    \left[\hat{\pi}_{\mathbf{x}_{\mathbf{r},a}}, \sum_{n=0}^\infty \frac{(\hat{A}_{i,a})^n}{n!}\right]\nonumber\\
    =& \pi_{\mathbf{x}_{\mathbf{r},i}} \sum_{n=0}^\infty \frac{n}{n!}(\hat{A}_{i,a})^{n-1}\nonumber\\
    =& \pi_{\mathbf{x}_{\mathbf{r},i}} \sum_{n=1}^\infty \frac{(\hat{A}_i)^{n-1}}{(n-1)!}\nonumber\\
    =&\pi_{\mathbf{x}_{\mathbf{r},i}}\hat{D}(\phi_{i,a}).
\end{align}

\subsection{Commutation of Displacement Operators in the Path Integral} \label{Commutation of Displacement Operators in the Path Integral}
We aim to remove the displacement operators from the expression in \cref{eq:Displacement operator normalization} by commuting all annihilation operators to the right, where they act on the vacuum and thereby annihilate it. We proceed as follows:
\begin{widetext}
\begin{align}
    \hat{D}^\dagger(\phi_{i,a}) \hat{D}(\phi_{i-1,a})\ket{0} &=\hat{D}^\dagger(\boldsymbol{\alpha}_{i,a}) \hat{D}(\boldsymbol{\alpha}_{i-1,a})\ket{0}\nonumber\\
    &=  \exp\Bigg\{\Delta p^n \sum_{\mathbf{k}=-\frac{\mathbf{K}}{2}}^\frac{\mathbf{K}}{2} \frac{1}{(2\pi)^n 2E_\mathbf{k}}\left[\alpha^*_{\mathbf{p}_{\mathbf{k},i}}\hat{a}_{\mathbf{p}_{\mathbf{k},a}}-\alpha_{\mathbf{p}_{\mathbf{k},i}}\hat{a}^\dagger_{\mathbf{p}_{\mathbf{k},a}}\right]\Bigg\}\nonumber\\
    & \qquad  \exp\Bigg\{\Delta p^n \sum_{\mathbf{q}=-\frac{\mathbf{K}}{2}}^\frac{\mathbf{K}}{2} \frac{1}{(2\pi)^n2E_\mathbf{q}}\left[\alpha_{\mathbf{p}_{\mathbf{q}.{i-1}}}\hat{a}^\dagger_{\mathbf{p}_{\mathbf{q},a}}-\alpha^*_{\mathbf{p}_{\mathbf{q},{i-1}}}\hat{a}_{\mathbf{p}_{\mathbf{q},a}}\right]\Bigg\}\ket{0}\nonumber\\
    &= N^-_{i-1,i-1}  \exp\Bigg\{\Delta p^n \sum_{\mathbf{k}=-\frac{\mathbf{K}}{2}}^\frac{\mathbf{K}}{2} \frac{1}{(2\pi)^n 2E_\mathbf{k}}\left[\alpha^*_{\mathbf{p}_{\mathbf{k},i}}\hat{a}_{\mathbf{p}_{\mathbf{k},a}}-\alpha_{\mathbf{p}_{\mathbf{k},i}}\hat{a}^\dagger_{\mathbf{p}_{\mathbf{k},a}}\right]\Bigg\} \nonumber\\
    & \quad \exp\Bigg\{\Delta p^n \sum_{\mathbf{q}=-\frac{\mathbf{K}}{2}}^\frac{\mathbf{K}}{2} \frac{\alpha_{\mathbf{p}_{\mathbf{q},{i-1}}}\hat{a}^\dagger_{\mathbf{p}_{\mathbf{q},a}}}{(2\pi)^n2E_\mathbf{q}}\Bigg\} \exp\Bigg\{-\Delta p^n \sum_{\mathbf{q}=-\frac{\mathbf{K}}{2}}^\frac{\mathbf{K}}{2} \frac{\alpha^*_{\mathbf{p}_{\mathbf{q},i-1}}\hat{a}_{\mathbf{p}_{\mathbf{q},a}}}{(2\pi)^n2E_\mathbf{q}}\Bigg\}\ket{0}\nonumber\\
    &= N^-_{i-1,i-1} \exp\Bigg\{\Delta p^n \sum_{\mathbf{k}=-\frac{\mathbf{K}}{2}}^\frac{\mathbf{K}}{2} \frac{1}{(2\pi)^n 2E_\mathbf{k}}\left[\alpha^*_{\mathbf{p}_{\mathbf{k},i}}\hat{a}_{\mathbf{p}_{\mathbf{k},a}}-\alpha_{\mathbf{p}_{\mathbf{k},i}}\hat{a}^\dagger_{\mathbf{p}_{\mathbf{k},a}}\right]\nonumber\\
    & \quad + \Delta p^n \sum_{\mathbf{q}=-\frac{\mathbf{K}}{2}}^\frac{\mathbf{K}}{2} \frac{\alpha_{\mathbf{p}_{\mathbf{q},i-1}}\hat{a}^\dagger_{\mathbf{p}_{\mathbf{q},a}}}{(2\pi)^n 2E_\mathbf{q}}+\frac{1}{2}\Delta p^{2n} \sum_{\mathbf{k},\mathbf{q}=-\frac{\mathbf{K}}{2}}^\frac{\mathbf{K}}{2} \frac{\alpha^*_{\mathbf{p}_{\mathbf{k},i}}\alpha_{\mathbf{p}_{\mathbf{q},i-1}}}{(2\pi)^{2n} 2E_\mathbf{q}E_\mathbf{k}}\left[\hat{a}_{\mathbf{p}_{\mathbf{k},a}},\hat{a}^\dagger_{\mathbf{p}_{\mathbf{q},a}}\right]\Bigg\}\ket{0} \nonumber\\
    &= N^-_{i-1,i-1}N^+_{i,i-1} \exp\Bigg\{\Delta p^n \sum_{\mathbf{k}=-\frac{\mathbf{K}}{2}}^\frac{\mathbf{K}}{2} \frac{-\left(\alpha_{\mathbf{p}_{\mathbf{k},i}}-\alpha_{\mathbf{p}_{\mathbf{k},i-1}}\right)\hat{a}^\dagger_{\mathbf{p}_{\mathbf{k},a}}+\alpha^*_{\mathbf{p}_{\mathbf{k},i}}\hat{a}_{\mathbf{p}_{\mathbf{k},a}}}{(2\pi)^n 2E_\mathbf{k}}\Bigg\}\ket{0} \nonumber\\
    &= N^-_{i-1,i-1}N^+_{i,i-1} \exp\Bigg\{\frac{1}{2}\Delta p^{2n} \sum_{\mathbf{k},\mathbf{q}=-\frac{\mathbf{K}}{2}}^\frac{\mathbf{K}}{2} \frac{\alpha^*_{\mathbf{p}_{\mathbf{k},i}}\left(\alpha_{\mathbf{p}_{\mathbf{q},i}}-\alpha_{\mathbf{p}_{\mathbf{q},i-1}}\right)}{(2\pi)^{2n}2E_\mathbf{q}E_\mathbf{k}}\left[\hat{a}^\dagger_{\mathbf{p}_{\mathbf{k},a}},\hat{a}_{\mathbf{p}_{\mathbf{q},a}}\right]\Bigg\}\nonumber\\
    & \quad  \exp\Bigg\{-\Delta p^n \sum_{\mathbf{k}=-\frac{\mathbf{K}}{2}}^\frac{\mathbf{K}}{2} \frac{\left(\alpha_{\mathbf{p}_{\mathbf{k},i}}-\alpha_{\mathbf{p}_{\mathbf{k},i-1}}\right)\hat{a}^\dagger_{\mathbf{p}_{\mathbf{k},a}}}{(2\pi)^n 2E_\mathbf{k}}\Bigg\} \exp\Bigg\{\Delta p^n \sum_{\mathbf{k}=-\frac{\mathbf{K}}{2}}^\frac{\mathbf{K}}{2} \frac{\alpha^*_{\mathbf{p}_{\mathbf{k},i}}\hat{a}_{\mathbf{p}_{\mathbf{k},a}}}{(2\pi)^n 2E_\mathbf{k}}\Bigg\}\ket{0}\nonumber\\
    &= N^-_{i-1,i-1}N^+_{i,i-1}  \exp\Bigg\{-\frac{1}{2}\Delta p^n \sum_{\mathbf{k}=-\frac{\mathbf{K}}{2}}^\frac{\mathbf{K}}{2} \frac{\alpha^*_{\mathbf{p}_{\mathbf{k},i}}\left(\alpha_{\mathbf{p}_{\mathbf{k},i}}-\alpha_{\mathbf{p}_{\mathbf{k},i-1}}\right)}{(2\pi)^n 2E_\mathbf{k}}\Bigg\}\nonumber\\
    & \quad  \exp\Bigg\{-\Delta p^n \sum_{\mathbf{k}=-\frac{\mathbf{K}}{2}}^\frac{\mathbf{K}}{2} \frac{\left(\alpha_{\mathbf{p}_{\mathbf{k},i}}-\alpha_{\mathbf{p}_{\mathbf{k},i-1}}\right)\hat{a}^\dagger_{\mathbf{p}_{\mathbf{k},a}}}{(2\pi)^n 2E_\mathbf{k}}\Bigg\}\ket{0}\nonumber\\
    &= N^-_{i-1,i-1}N^+_{i,i-1}N^-_{i,i}N^+_{i,i-1} \exp\Bigg\{-\Delta p^n \sum_{\mathbf{k}=-\frac{\mathbf{K}}{2}}^\frac{\mathbf{K}}{2} \frac{\left(\alpha_{\mathbf{p}_{\mathbf{k},i}}-\alpha_{\mathbf{p}_{\mathbf{k},i-1}}\right)\hat{a}^\dagger_{\mathbf{p}_{\mathbf{k},a}}}{(2\pi)^n 2E_\mathbf{k}}\Bigg\}\ket{0}.
\end{align}
\clearpage
\end{widetext}

\subsection{Coherent State Normalization in the Field Basis}\label{Coherent State Normalization in the Field Basis}
Consider the exponent of the normalization constant of \cref{eq:Free partition function field basis phi and pi} 
\begin{align}
    \prod_{i=1}^N N^{-2}_{i,i}N^{+2}_{i,i-1} =& \exp\bigg\{-\Delta \tau \Delta p^n \sum_{i=1}^N \sum_{\mathbf{k}=-\frac{\mathbf{K}}{2} }^\frac{\mathbf{K}}{2}\frac{1}{(2\pi)^n2E_\mathbf{k}}\nonumber\\
    & \quad \frac{\alpha^*_{\mathbf{p}_{\mathbf{k},i}}\left(\alpha_{\mathbf{p}_{\mathbf{k},i}}-\alpha_{\mathbf{p_k},i-1}\right)}{\Delta \tau}\bigg\},\nonumber
\end{align}
and apply a change of field variables from \cref{eq:Coherent eigenvalue change of variables,eq:Conjugate coherent eigenvalue change of variables}
\begin{align}
    -\Delta \tau & \Delta p^n \sum_{i=1}^N \sum_{\mathbf{k}=-\frac{\mathbf{K}}{2} }^\frac{\mathbf{K}}{2}\frac{1}{(2\pi)^n2E_\mathbf{k}}\alpha^*_{\mathbf{p}_{\mathbf{k},i}} \alpha'_{\mathbf{p}_{\mathbf{k},i}}\nonumber\\
    =&-\Delta \tau \Delta p^n \sum_{i=1}^N \sum_{\mathbf{k}=-\frac{\mathbf{K}}{2} }^\frac{\mathbf{K}}{2}\frac{1}{(2\pi)^n2E_\mathbf{k}}\nonumber\\
    & \quad \left(E_\mathbf{k}\phi^*_{\mathbf{p}_{\mathbf{k},i}}-\mathrm{i}\pi^*_{\mathbf{p}_{\mathbf{k},i}}\right)\left(E_\mathbf{k}\phi'_{\mathbf{p}_{\mathbf{k},i}}+\mathrm{i}\pi'_{\mathbf{p}_{\mathbf{k},i}}\right)\nonumber\\
    =&-\Delta \tau \Delta p^n \sum_{i=1}^N \sum_{\mathbf{k}=-\frac{\mathbf{K}}{2} }^\frac{\mathbf{K}}{2}\frac{1}{(2\pi)^n2E_\mathbf{k}} \nonumber\\
    & \quad \left[\mathrm{i}E_\mathbf{k} \phi^*_{\mathbf{p}_{\mathbf{k},i}}\pi'_{\mathbf{p}_{\mathbf{k},i}}-\mathrm{i}E_\mathbf{k} \pi^*_{\mathbf{p}_{\mathbf{k},i}} \phi'_{\mathbf{p}_{\mathbf{k},i}} \right. \nonumber\\
    & \left. \qquad +E_\mathbf{k}^2 \phi^*_{\mathbf{p}_{\mathbf{k},i}}\phi'_{\mathbf{p}_{\mathbf{k},i}}+\pi^*_{\mathbf{p}_{\mathbf{k},i}} \pi'_{\mathbf{p}_{\mathbf{k},i}}\right].\label{eq:Action in terms of phi and pi 1}   
\end{align}
Note that the last two terms in \cref{eq:Action in terms of phi and pi 1} vanish according to the identity proven in \cref{eq:Total field derivative vanish}. Consequently, the exponent of the normalization constant reduces to the following form
\newpage
\begin{align}
    -\Delta \tau & \Delta p^n \sum_{i=1}^N \sum_{\mathbf{k}=-\frac{\mathbf{K}}{2} }^\frac{\mathbf{K}}{2} \frac{1}{(2\pi)^n2E_\mathbf{k}} \alpha^*_{\mathbf{p}_{\mathbf{k},i}}\alpha'_{\mathbf{p}_{\mathbf{k},i}}\nonumber\\
    =& -\frac{\mathrm{i}\Delta\tau}{2} \left(\frac{\Delta p}{2\pi}\right)^n \sum_{i=1}^N \sum_{\mathbf{k}=-\frac{\mathbf{K}}{2} }^\frac{\mathbf{K}}{2}\left[\phi_{\mathbf{p}_{\mathbf{k},i}}\pi^{*'}_{\mathbf{p}_{\mathbf{k},i}}-\pi^*_{\mathbf{p}_{\mathbf{k},i}}\phi'_{\mathbf{p}_{\mathbf{k},i}}\right]\nonumber\\
    =& \frac{\mathrm{i}\Delta\tau}{2} \left(\frac{\Delta p}{2\pi}\right)^n \sum_{i=1}^N \sum_{\mathbf{k}=-\frac{\mathbf{K}}{2} }^\frac{\mathbf{K}}{2}\left[\pi_{\mathbf{p}_{\mathbf{k},i}}\phi^{*'}_{\mathbf{p}_{\mathbf{k},i}}+\pi^*_{\mathbf{p}_{\mathbf{k},i}}\phi'_{\mathbf{p}_{\mathbf{k},i}}\right]\nonumber\\
    =& \frac{\mathrm{i} \Delta\tau}{2} \left(\frac{\Delta p}{2\pi}\right)^n \Delta x^{2n} \sum_{i=1}^N \sum_{\mathbf{k}=-\frac{\mathbf{K}}{2} }^\frac{\mathbf{K}}{2} \sum_{\mathbf{r},\mathbf{s}=\mathbf{0}}^\mathbf{K}\left[\pi_{\mathbf{x}_{\mathbf{r},i}}\phi'_{\mathbf{x}_{\mathbf{s},i}} \right. \nonumber\\
    & \left. \quad +\pi_{\mathbf{x}_{\mathbf{s},i}}\phi'_{\mathbf{x}_{\mathbf{r},i}}\right]e^{\mathrm{i} \mathbf{p_k}(\mathbf{x_s}-\mathbf{x_r})}\nonumber\\
    =& \frac{\mathrm{i} \Delta\tau}{2} \Delta x^n \sum_{i=1}^N \sum_{\mathbf{r},\mathbf{s}=\mathbf{0}}^\mathbf{K}\left[\pi_{\mathbf{x}_{\mathbf{r},i}}\phi'_{\mathbf{x}_{\mathbf{s},i}}+\pi_{\mathbf{x}_{\mathbf{s},i}}\phi'_{\mathbf{x}_{\mathbf{r},i}}\right]\delta_{\mathbf{r},\mathbf{s}}\nonumber\\
    =& \mathrm{i}\Delta\tau \Delta x^n \sum_{i=1}^N \sum_{\mathbf{r}=\mathbf{0}}^{\mathbf{K}}\pi_{\mathbf{x}_{\mathbf{r},i}}\phi'_{\mathbf{x}_{\mathbf{r},i}},
\end{align}
where we first used summation by parts form \cref{Summation by Parts with Periodic Boundaries} and subsequently performed a DFT. The normalization constant thus becomes
\begin{align}
    \prod_{i=1}^N N^{-2}_{i,i}N^{+2}_{i,i-1} =& e^{\mathrm{i}\Delta\tau \Delta x^n \sum_{i=1}^N \sum_{\mathbf{r}=\mathbf{0}}^\mathbf{K}\pi_{\mathbf{x}_{\mathbf{r},i}}\phi'_{\mathbf{x}_{\mathbf{r},i}}}.\label{eq:Coherent state normalization in the field basis}
\end{align}

\clearpage
\bibliographystyle{apsrev4-2}
\bibliography{biblio}

\end{document}